\documentclass[twocolumn,superscriptaddress,floatfix,prb,amsmath,aps,showpacs]{revtex4}
\usepackage{graphicx}
\usepackage{bm}
\usepackage{amssymb}
\usepackage[dvips]{color}
\begin{document}
\bibliographystyle{apsrev}
\def\nn{\nonumber}
\def\dag{\dagger}
\def\u{\uparrow}
\def\d{\downarrow}
\def\j{\bm j}
\def\m{\bm m}
\def\l{\bm l}
\def\0{\bm 0}
\def\k{\bm k}
\title{Dynamical spin structure factors
of quantum spin nematic states}
\date{\today}
\author{Ryuichi Shindou}
\affiliation{Physics Department, Tokyo Institute of Technology,
Ookayama, 2-12-1, Meguro-ku, Tokyo  Japan}
\affiliation{Condensed Matter Theory Laboratory, RIKEN,
2-1 Hirosawa, Wako, Saitama 351-0198, Japan}
\author{Seiji Yunoki}
\affiliation{Computational Condensed Matter Laboratory, RIKEN,
2-1 Hirosawa, Wako, Saitama 351-0198, Japan}
\author{Tsutomu Momoi}
\affiliation{Condensed Matter Theory Laboratory, RIKEN,
2-1 Hirosawa, Wako, Saitama 351-0198, Japan}
\begin{abstract}
Dynamical spin structure factors of quantum spin
nematic states are calculated  in a spin-$\frac{1}{2}$ square-lattice
$J_1$--$J_2$ model with ferromagnetic $J_1$ and
competing antiferromagnetic $J_2$ interactions.
To this end, we use a fermion representation, generalizing it to $N$ flavors.
We begin with a spin-triplet
pairing state of fermion fields, called $Z_2$ planar state,
which is a stable saddle-point solution in the large-$N$ limit
in a finite parameter range where the couplings $J_1$ and $J_2$ compete strongly
[R.~Shindou and T.~Momoi, Phys. Rev. B {\bf 80}, 064410 (2009)].
Using a large-$N$ expansion, we take into account fluctuations
around this saddle point up to corrections of order $1/N$.
The dynamical spin structure factors thus obtained signify
the existence of gapless $q$-linear director-wave (spin-wave)
modes at ${\bm q}=(0,0)$ and gapped `gauge-field' like
collective modes at ${\bm q}=(\pi,\pi)$, whose spectral weight vanishes
as a linear and quadratic function of the momentum respectively.
The low-energy collective modes contain fluctuations of nematic-director,
spin, and gauge degrees of freedom.
Associated with the
gapless $q$-linear modes, we evaluate the temperature
dependence of the nuclear spin relaxation rate $1/T_1$ in the
low-temperature regime as $1/T_{1} \propto T^{2d-1}$,
where $d$ is the effective spatial dimension.
\end{abstract}
\maketitle
\section{introduction}

Frustrated magnets are Mott insulators
in which competing exchange interactions
between localized spins bring about
an extensively large degeneracy in the ground state
energetics. In a certain circumstance, such a
frustrated spin system lifts this degeneracy
quantum-mechanically, only to choose as its
ground state a liquid-like state of
matter, dubbed a quantum spin
liquid.~\cite{LBalents-PALee,pwa,wen,fradkin}
Typically, a ground-state wavefunction of quantum spin liquids
consists of what we call spin-singlet valence
bonds. A spin-singlet valence bond  ---a spin-singlet pair of two $S=1/2$ spins---
is energetically favored by
an antiferromagnetic exchange interaction
between the two spins. A ground-state wavefunction of quantum spin liquids is
given by a
quantum-mechanical superposition of different spatial
partitionings of the spin-singlet valence bonds over the
entire lattice, so that the state preserves not only the
spin-rotational symmetry but also the lattice translational
symmetry.~\cite{pwa}

Having no symmetry-breaking
order parameter, quantum spin liquids have
been regarded as a new quantum state of matter, which
should be sharply contrasted from conventional
magnetic phases such as a N\'eel ordered phase and
a valence bond solid phase.\cite{LBalents-PALee}
In fact, owing to its fluid-like feature,
the spin liquid phase has various unconventional
low-energy excitations, such as a `gauge-field' like
collective excitation and a fractionalized (or `individual')
magnetic excitation
called spinon.~\cite{wen,fradkin}
Experimental and theoretical searches
for this new non-magnetic phase
have been intensively carried out
in the past couple of decades
in the field of quantum magnetism.

Another new quantum state of matter recently explored
in localized spin systems is a quantum
spin nematic
phase,~\cite{ag,chubukov,MomoiS,sms,sm,Vekua,Hikihara1,Sudan,ut,
tz,msk,Hagiwara,mef,sy,smfh,shm,um,Takigawa} which is a quantum-spin analogue of nematic
liquid-crystal phases. Spin nematic states neither possess any spin order, i.e.
sublattice magnetization, nor any crystalline solid-like
structure in spin degrees of freedom,
but, unlike spin-rotational symmetric quantum spin
liquids, they exhibit various types of spin anisotropy, whose order parameters are given by
symmetric rank-2 traceless spin tensor operators\cite{ag}
\begin{equation}
Q_{\j \m,\mu\nu}
=\frac{1}{2}
( S_{{\bm j},\mu} S_{{\bm m},\nu} + S_{{\bm j},\nu} S_{{\bm m},\mu} )
- \frac{\delta_{\mu\nu}}{3}
 {\bm S}_{\bm j} \cdot {\bm S}_{\bm m}
\label{op:tensor}
\end{equation}
for $\mu,\nu=x,y,z$. Here ${\bm S}_{\bm j}=(S_{{\bm j},x},S_{{\bm j},y},S_{{\bm j},z})$
denotes the spin-1/2 vector operator on site ${\bm j}$.
The tensor operator (\ref{op:tensor}) consists of two
spin operators defined on different
sites ${\bm j}$ and ${\bm m}$, usually neighboring two sites, so that
the order parameter is defined on bond $({\bm j},{\bm m})$.
This order is hence called a `bond-type'
spin nematic order.
Ground-state
wavefunctions of this phase
can be essentially described as
quantum-mechanical superpositions of
different spatial partitionings of both spin-singlet valence bonds
and a part of spin-triplet valence bonds,~\cite{sm}
so that quantum spin nematics can be regarded as
`cousins' of symmetric quantum spin liquids, possibly sharing many
of their exotic characters.
At the same time, they should have gapless collective modes
--Nambu-Goldstone modes-- due to the broken
spin-rotational symmetries,
which is distinct from the quantum spin liquids.

\begin{figure}[t]
    \includegraphics[width=30mm]{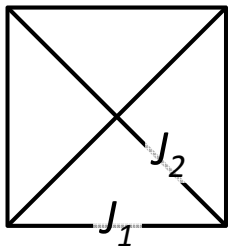}
\caption{Square-lattice $J_1$-$J_2$ frustrated
ferromagnetic model, in which the nearest-neighbor exchange
$J_1$ is ferromagnetic and the next-nearest-neighbor exchange
$J_2$ is antiferromagnetic.}
\label{fig:j1j2}
\end{figure}

Recently, various theoretical investigations have revealed
the appearance of spin nematic
phases in several frustrated spin-1/2 magnets that have both
ferromagnetic couplings and competing
antiferromagnetic couplings.~\cite{chubukov,MomoiS,sms,sm,Vekua,Hikihara1,Sudan,ut,
tz,msk,sy,smfh,shm,um}
Among two-dimensional systems, ground-state properties of
the spin-1/2 $J_1$-$J_2$ model on the square lattice have
been reasonably most studied. The Hamiltonian is given by
\begin{equation}
H=-J_1 \sum_{\langle {\bm i},{\bm j}\rangle} {\bm S}_{\bm i}\cdot {\bm S}_{\bm j}
+ J_2 \sum_{\langle\langle {\bm i}, {\bm j}\rangle\rangle} {\bm S}_{\bm i}\cdot {\bm S}_{\bm j}
\end{equation}
with ferromagnetic $J_1$ and antiferromagnetic $J_2$ ($J_1,J_2>0$),
where the first (second) summation runs over all pairs of nearest-neighbor
(next-nearest-neighbor) sites
(see Fig.\ \ref{fig:j1j2}).
It has been proposed\cite{sms,sym} that a spin nematic phase appears in
a finite parameter range around $J_2/J_1=0.5$.
In contrast to theoretical developments, experimental verifications of spin nematic phases
have just started, especially in quasi-one-dimensional systems,
but they are still very limited.~\cite{Hagiwara,mef,Takigawa}
One of the difficulties for experimentally detecting
this new phase is the absence of any direct probe for the
spin quadratic order parameter Eq.\ (\ref{op:tensor}) and
the lack of theoretical understanding\cite{comment1} of characteristic properties in this
phase.

The purpose of this paper is to clarify dynamical properties of
a quantum spin nematic phase and
to give
a relevant physical characterization to this new
class of quantum spin states.
To this end, we generalize the $J_1$-$J_2$ model
to an $N$-flavor $J_1$-$J_2$ spin model, using a fermion representation.~\cite{sm}
In this approach,
wavefunctions of spin-nematic ground states
are described  with spin-triplet pairings of
fermion fields, whose $d$-vectors specify the
spin-nematic director vectors associated with
the quadrupolar moments.

\begin{figure}[t]
    \includegraphics[width=75mm]{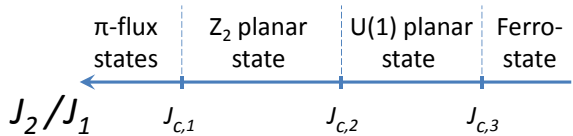}
\caption{Mean-field phase diagram of
the $S=1/2$ square-lattice $J_1$-$J_2$ model
in the large-$N$ limit. In the strong
$J_2$ regime, the mean-field ground state
consists of only two decoupled spin-singlet
pairing states defined on $J_2$ bonds, which are often called $\pi$-flux
states.~\cite{wen,fradkin,am}
When ferromagnetic $J_1$ increases, the mean-field ground state
acquires finite spin-triplet pairing amplitudes on the
ferromagnetic bonds, which connect the decoupled $\pi$-flux
states.~\cite{sm}
These states are called $Z_2$ planar state and
$U(1)$ planar state, depending on the amplitude of singlet pairings,
and both of them are characterized as spin nematic states.
In the strong
$J_1$ regime, the mean-field ground state
contains only spin-triplet pairings,
which corresponds to a fully-polarized ferromagnetic state
(FM state).\cite{sym}
Energetics of the mean-field solutions
conclude that $J_{c,1}=1.325$, $J_{c,2}=1.0448$
and $J_{c,3}=1.02$.}
\label{fig:PD}
\end{figure}

The classical ($S\rightarrow \infty$) phase diagram
of the square-lattice $J_1$-$J_2$
model with ferromagnetic $J_1$
consists of only two phases; a collinear
antiferromagnetic phase in the strong antiferromagnetic $J_2$ regime
and a ferromagnetic phase in the strong ferromagnetic $J_1$ regime.
In the quantum ($S=1/2$) system, it has been
argued that the quantum spin nematic phase emerges in between these two
magnetic ordered phases.~\cite{sms}
The previous saddle point analysis~\cite{sm}
of the spin-1/2 $J_1$-$J_2$ model
concluded that, for large $N$, a certain spin nematic phase dubbed
$Z_2$ planar phase~\cite{BW} stably appears
in ground states in a finite parameter range
where ferromagnetic $J_1$ couplings
strongly compete with antiferromagnetic $J_2$ couplings (see Fig.\ \ref{fig:PD}).
In this pairing state, the spin-triplet $d$-vectors
introduced on ferromagnetic bonds take
a coplanar spatial configuration, which by itself
mimics the pairing symmetry of
a two-dimensional analogue of Balian-Werthamer
state\cite{bw1} in Helium-3 superfluid B phase.
It was demonstrated that
the wavefunction of this $Z_2$ planar state reasonably
reproduces the $d$-wave spin nematic state proposed
by the previous exact diagonalization study.~\cite{sms}
The $Z_2$ planar state possesses the same spatial
configuration of quadrupolar orders on bonds as the
$d$-wave spin nematic state and also both of the
states have the same spatial symmetries.~\cite{sym}
A variational Monte Carlo
study~\cite{sym} based on mean-field solutions
further indicated that, when projected
onto the physical $S=1/2$ spin Hilbert
space, the optimized $Z_2$ planar state achieves
the best optimal energy
in the original ($N=1$) spin-1/2 $J_1$-$J_2$ model
in a finite parameter range, compared
with other competing states.
Among various mean-field ansatze, only the $Z_2$ planar
phase (spin nematic phase) survives, except for the ferromagnetic
and collinear antiferromagnetic phases,
after the projection. The $d$-wave bond spin nematic
phase is hence expected to appear in the
spin-$1/2$ $J_1$-$J_2$ model for any
number $N$.

In this paper, we calculate
dynamical magnetic properties of the
quantum spin nematic phase in a generalized $N$-flavor spin-$1/2$ $J_1$-$J_2$ model
on the square lattice.
Employing a
$1/N$ expansion for large $N$, we take into
account fluctuations around the mean-field (saddle-point) solutions.
The treatment up to order of $1/N$ essentially
corresponds to the so-called random phase
approximation (RPA).  The dynamical spin structure factors Im$\chi_{\mu\mu}({\bm q},\epsilon)$
($\mu=x,y,z$)
thus calculated have two characters; a spin-liquid-like character and
a symmetry-broken-phase character. The former feature
manifests itself as the Stoner continuum of the
individual excitations of free spinons.
The latter character is represented
by gapless collective modes, which
have $q$-linear energy dispersions.
The gapless collective modes are given by long-wavelength fluctuations of
nematic directors. For finite momenta, these director excitations are
accompanied with weak spin excitations, which are measurable through small but
finite spectral weight
in the dynamical spin structure factor.
The spectral weight of Im$\chi_{\mu\mu}({\bm q},\epsilon)$
vanishes as a linear
function of the momentum near the
gapless point, e.g.
Im$\chi_{zz}({\bm q},\epsilon)\simeq a_z v_z|{\bm q}|\delta (\epsilon-v_z|{\bm q}|)+\cdots$,
where $v_z$ denotes the director-wave velocity and $a_z$ is a finite constant.
We further calculated NMR spin relaxation rate $1/T_1$ given by the gapless magnetic modes.
Because the only physical magnetic low-energy modes are
these gapless director-wave excitations around ${\bm q}=(0,0)$, these  excitations can
induce a relatively slow spin relaxation, which has a temperature dependence
$T^{-1}_{1} \propto T^{2d-1}$ in the low-$T$ limit in $d$ dimensions.

We further discuss how the $Z_2$ planar
state changes to different states at the
boundaries to the neighboring
phases in the large $N$ limit,
analyzing the excitation modes.
When the antiferromagnetic coupling $J_2$ decreases,
the mean-field solution transforms from the $Z_2$ planar
state to a $U(1)$ planar state
at $J_2/J_1=J_{c,2}$ (see Fig.\ \ref{fig:PD}).
The calculation to first order in $1/N$ reveals that
two gapped `gauge-field' collective
modes at ${\bm q}=(\pi,\pi)$ become
gapless at the transition point, only to
constitute a compact QED (quantum electrodynamics)
action in the $U(1)$ phase ($J_2/J_1 < J_{c,2}$),
where the space-time instanton
effect associated with this effective action
introduces a strong confining potential
between `free' gapped spinons.
This suggests that the finite mass
at ${\bm q}=(\pi,\pi)$ in the $Z_2$ planar state
quantifies the stability against the confinement effect.
Inside the $U(1)$ planar phase, a couple of other
bosonic modes simultaneously exhibit instabilities.
Due to these instabilities, the $U(1)$ phase would break
various symmetries such as the time-reversal symmetry,
a spin-$\pi$-rotational symmetry and the translational
symmetries.

The organization of this paper is as follows:
In the next section, we first introduce a
generalized $N$-flavor spin-1/2 quantum frustrated ferromagnetic model,
whose large-$N$ limit possesses our previous
mean-field solutions as the exact ground states and whose $N=1$ case
safely reproduces the usual quantum spin-$\frac{1}{2}$ model.
In Sec.~III, we describe the
$1/N$-expansion calculation for the dynamical
spin correlation functions.
In Sec.~IV,
we show the calculated dynamical spin structure factors,
both Im$\chi_{zz}({\bm q},\epsilon)$ and Im$\chi_{+-}({\bm q},\epsilon)$,
and discuss their characteristic features and
physical implications. We also discuss the nature of
the $U(1)$ planar state here.
Section~V contains a calculation of NMR spin relaxation rate $1/T_1$.
Section~VI is devoted to the summary and discussion.

\section{quantum frustrated ferromagnetic model
and its large-$N$ limit on a square lattice}
In this section, we describe our generalized $S=1/2$ $N$-flavor
$J_1$-$J_2$ model. We also briefly review
the $Z_2$ planar ground state, which is a stable saddle-point
solution in the large-$N$ limit of the present model in a finite parameter range
of a strongly competing regime.

\subsection{$N$-flavor spin-1/2 $J_1$-$J_2$ model}
The Hamiltonian for the  generalized $N$-flavor spin-1/2 $J_1$-$J_2$ model is given by~\cite{sm}
\begin{eqnarray}
{\cal H} &=& - \frac{J_1}{N} \sum_{\langle \j,\m\rangle}
\sum^{N}_{a,b=1}  \left(
{\bm S}^{ab}_{\j}\cdot{\bm S}^{ba}_{\m}
+ \psi^{ab}_{\j} \psi^{ba}_{\m} \right)  \nn \\
&& \hspace{-0.3cm}
+ \!\ \frac{J_2}{N} \sum_{\langle\langle \j,\m\rangle \rangle}
\sum_{a,b}
{\bm S}^{ab}_{\j}\cdot {\bm S}^{ba}_{\m}
+ \sum_{\j,a} {\bm h}^{aa}_{\j} \cdot {\bm S}^{aa}_{\j},
\label{largeN}
\end{eqnarray}
where
$\langle \j,\m\rangle$ ($\langle\langle \j,\m\rangle \rangle$)
runs over all nearest-neighbor
(2nd-neighbor) bonds on the square lattice,
and the spin operators
${\bm S}^{ab}_{\j}=(S^{ab}_{\j,1},S^{ab}_{\j,2},S^{ab}_{\j,3})$
and the density operator $\psi^{ab}_{\j}$ are given by
\begin{align}
S^{ab}_{\j,+} &\equiv \frac{1}{2}
\big( f^{a\dag}_{\j,\u} f^{b}_{\j,\d} +
f^{b\dag}_{\j,\u} f^{a}_{\j,\d} \big), \nn \\
S^{ab}_{\j,-} &\equiv \frac{1}{2}
\big( f^{a\dag}_{\j,\d} f^{b}_{\j,\u} +
f^{b\dag}_{\j,\d} f^{a}_{\j,\u} \big), \nn \\
S^{ab}_{\j,3} &\equiv \frac{1}{2}
\big( f^{a\dag}_{\j,\u}f^{b}_{\j,\u}
- f^{b\dag}_{j,\d}f^{a}_{j,\d}\big), \nn \\
\psi^{ab}_{\j} &\equiv \frac{i}{2} \big(
f^{a\dag}_{\j,\alpha}f^{b}_{\j,\alpha}
- f^{b\dag}_{\j,\alpha} f^{a}_{\j,\alpha}\big).
\label{spin}
\end{align}
Here $f_{\j,\alpha}^{a\dagger}$ is a fermion creation operator
with spin $\alpha=\uparrow,\downarrow$
and flavor $a=1,\dots,N$ on site $\j=(j_x,j_y)$.
In this paper, we consider the case $J_1$ is ferromagnetic
and $J_2$ is antiferromagnetic, i.e. $J_1>0$ and $J_2>0$.
We have introduced external magnetic
field ${\bm h}^{aa}_{\j}=(h^{aa}_{\j,1},h^{aa}_{\j,2},h^{aa}_{\j,3})$ to
calculate the spin correlation function.
The physical spin Hilbert space
satisfies the local constraints
\begin{equation}
\sum_{a=1}^N f^{a\dagger}_{{\bm j},\alpha} f^a_{{\bm j},\alpha}=N, \ \ \ \
\sum_{a=1}^N f^{a}_{{\bm j},\alpha} f^a_{{\bm j},\beta}\epsilon_{\alpha\beta}=0
\label{hilbert}
\end{equation}
on each site with $\epsilon_{\uparrow \downarrow}=
-\epsilon_{\downarrow \uparrow}=1$ and
$\epsilon_{\uparrow \uparrow}=
-\epsilon_{\downarrow \downarrow}=0$,
which endows the fermionic Hilbert space
with  the local $SU(2)$ gauge symmetry.
The repeated spin indices imply their summations,
whereas we write the summations for the flavor indices explicitly.

We note that, when $N=1$, the Hamiltonian Eq.~(\ref{largeN})
in the physical Hilbert space reduces to the usual spin-1/2 $J_1$-$J_2$
quantum Heisenberg model.
We regard $N$ to be large when we perform a
$1/N$ expansion.

An equivalent statistical-mechanics problem at temperature $\beta^{-1}$
can be formulated in terms of the path-integral representation.
We decouple four-fermion interaction terms into a quadratic form,
using Hubbard-Stratonovich-type transformation.
Introducing
scalar auxiliary fields ($\chi_{\j\m}$ and $\eta_{\j\m}$) for the antiferromagnetic
interaction and vector auxiliary fields
[${\bm E}_{\j\m}=(E_{\j\m,1},E_{\j\m,2},E_{\j\m,3})$
and ${\bm D}_{\j\m}=(D_{\j\m,1},D_{\j\m,2},D_{\j\m,3})$] for
the ferromagnetic interaction,\cite{sm} we obtain the partition function in the form
\begin{widetext}
\begin{align}
Z[h] =& \int {\cal D}\Psi^{a\dag}
{\cal D}\Psi^{a}{\cal D} {\bm a}_{\tau}
{\cal D}U^{\rm sin} {\cal D}{\bm U}^{\rm tri}
\exp
\left( - \int^{\beta}_{0} d\tau\!\ {\cal L} [h,
U^{\rm sin},{\bm U}^{\rm tri},{\bm a}_{\tau}] \!\ \right),
\label{partition}\\
{\cal L} =& \sum_{a=1}^N \left\{
\frac{1}{2} \sum_{\j} {\rm tr}
\Big[\Psi^{a\dag}_{\j} \Big(\partial_{\tau} +
\sum_{\mu=1}^3 i a^{\mu}_{\j,\tau}
\sigma_{\mu} \Big)\Psi^a_{\j}\Big]
- \frac{J_1}{4} \sum_{\langle j,m\rangle}
 \left(- |{\bm E}_{\j\m}|^2 - |{\bm D}_{\j\m}|^2
+ \sum_{\mu=1}^3{\rm tr}\big[\Psi^{a\dag}_{\j}
U^{\rm tri}_{\j\m,\mu}
\Psi^{a}_{\m} \sigma^T_{\mu}\big]\right) \right. \nn \\
& 
\left.
 - \frac{J_2}{4} \sum_{\langle \langle \j,\m \rangle \rangle}
\left( - |\chi_{\j\m}|^2 - |\eta_{\j\m}|^2
+ {\rm tr}\big[\Psi^{a\dag}_{\j}
U^{\rm sin}_{\j\m}
\Psi^{a}_{\m}\big]\right)
 +
\!\ \frac{1}{4} \!\ \sum_{\j} \sum_{\mu=1}^3
h^{aa}_{\j,\mu} \!\ {\rm tr}\big[\Psi^{a\dag}_{\j}
\Psi^a_{\j} \sigma^T_{\mu} \big] \right\}, \label{lagrangian}
\end{align}
\end{widetext}
where the fermion fields are written in
the $2\times 2$ matrix form
\begin{equation}
\Psi^{a\dag}_{\j} \equiv
\left[\begin{array}{cc}
f^{a\dag}_{\j,\u} & f^a_{\j,\d} \\
f^{a\dag}_{\j,\d} & -f^a_{\j,\u} \\
\end{array}\right]
\end{equation}
and the auxiliary fields are included into the $2\times 2$ matrices $U^{\rm sin}_{\j\m}$ and
${\bm U}^{\rm tri}_{\j\m}
=(U^{\rm tri}_{\j\m,1},U^{\rm tri}_{\j\m,2},U^{\rm tri}_{\j\m,3})$
in the forms 
\begin{eqnarray}
U^{\rm sin}_{\j\m} \equiv \left[\begin{array}{cc}
\chi^{*}_{\j\m} & \eta^{*}_{\j\m}\\
\eta_{\j\m} & -\chi_{\j\m} \\
\end{array}\right], \ \
U^{\rm tri}_{\j\m,\mu} \equiv \left[\begin{array}{cc}
E^{*}_{\j\m,\mu} & D^{*}_{\j\m,\mu}\\
- D_{\j\m,\mu} & E_{\j\m,\mu} \\
\end{array}\right] \nn \\
\end{eqnarray}
($\mu=1,2,3$).
The trace denoted by the symbol ``tr"
is taken over $2 \times 2$ matrices such as
$\Psi^{a}_{\j}$
and the Pauli matrices $\sigma_\mu$.
A Gaussian integral over the auxiliary fields exactly reproduces
the original Hamiltonian given in Eq.~(\ref{largeN}).
Integrating over the temporal gauge fields
${\bm a}_{\j,\tau}=(a_{\j,\tau}^1,a_{\j,\tau}^2,a_{\j,\tau}^3)$ also
strictly imposes the local constraints given by Eq.~(\ref{hilbert}) on every
site and time.

As the Lagrangian is written in a quadratic form of fermion fields,
we can formally rewrite the effective action as
\begin{align}
\int^{\beta}_{0} d\tau {\cal L}
&= N{\cal S}_{\rm I} \nn \\
&\hspace{-1.4cm} + \!\ \frac12 \sum_{\k,n,a}
{\bm f}^{a\dag}_{\k,n} \cdot
{\bm G}^{-1}_{(\k,n|\k',n')}[h^{aa},U^{\rm sin},
{\bm U}^{\rm tri},{\bm a}_{\tau}]
\cdot {\bm f}^a_{\k',n'}, \label{g-g}
\end{align}
where
\begin{align}
{\cal S}_{\rm I} &\equiv  \int^{\beta}_{0} \!\
d\tau \!\ \left[
\frac{J_1}{4} \sum_{\langle \j,\m\rangle}
\big(|{\bm E}_{\j\m}|^2 + |{\bm D}_{\j\m}|^2\big) \right.\nn \\
& \hspace{1.8cm}
\left.+ \frac{J_2}{4} \sum_{\langle \langle \j,\m \rangle \rangle}
\big(|{\chi}_{\j\m}|^2 + |{\eta}_{\j\m}|^2\big) \right] \label{s1-g}
\end{align}
and ${\bm G}$ denotes a $4\times 4$ matrix single-particle
Green function of the fermion field ${\bm f}^{a\dag}_{\k,n}$
written in the Nambu representation
\begin{eqnarray}
{\bm f}^{a\dag}_{\k,n} \equiv
\left(\begin{array}{cccc}
f^{a \dag}_{\k,n,\uparrow}
&  f^{a\dag}_{\k,n,\downarrow}
& f^{a}_{-\k,-n,\uparrow}
&  f^{a}_{-\k,-n,\downarrow} \end{array}\right)
\end{eqnarray}
with
\begin{eqnarray}
f^{a\dag}_{\k,n,\sigma}
\equiv \frac{1}{\sqrt{\beta N_{\Lambda}}}
\sum_{\j} \int^{\beta}_{0}d\tau \!\
e^{i\k\cdot \j+i\omega_{n}\tau} f^{a\dag}_{\j,\sigma}(\tau).
\end{eqnarray}
Here, $N_{\Lambda}$ denotes the total number of
lattice sites, ${\bm k}=(k_x,k_y)$, $\omega_n\equiv (2n+1)\pi\beta^{-1}$,
and $\sigma=\uparrow,\downarrow$.
Note that the functional
${\bm G}^{-1}[h^{aa},U^{\rm sin},{\bm U}^{\rm tri},{\bm a}_{\tau}]$
is a linear function of
the elements of $h^{aa}$, $U^{\rm sin}$,
${\bm U}^{\rm tri}$, and ${\bm a}_{\tau}$.

The integral over the $\Psi$ ($\bm f$) fields in Eq.~(\ref{g-g})
leads to the following partition function,
\begin{align}
Z[h] &= \int {\cal D}U^{\rm sin}
{\cal D}{\bm U}^{\rm tri} {\cal D}{\bm a}_{\tau}
\exp ( - N {\cal S}[h,U^{\rm sin}, {\bm U}^{\rm tri},{\bm a}_{\tau}]), \label{z-g} \\
{\cal S} &\equiv  {\cal S}_{\rm I} + {\cal S}_{\rm II},\\
{\cal S}_{\rm II} &\equiv -\frac{1}{2N} \sum_{a=1}^N {\rm Tr} \big(\ln
{\bm G}^{-1}[h^{aa},U^{\rm sin},{\bm U}^{\rm tri},{\bm a}_{\tau}] \big),
\label{s2}
\end{align}
where the Green function  ${\bm G}$ is diagonal in
the flavor index and the trace of $\ln {\bm G}^{-1}$
is taken over the momentum ($\k$),
the Matsubara frequency ($\omega_n$), and the index of the
$4\times 4$ matrices,
i.e., the spin and particle-hole indices. 

\subsection{Saddle point solution: $Z_2$ planar state}
For large $N$, ${\cal S}_{\rm II}$, as well as ${\cal S}_{\rm I}$,
is a functional of order unity,
and the same is for ${\cal S}$.
In the large $N$ limit, since the prefactor of the action ${\cal S}$ in Eq.~(\ref{z-g})
is proportional to the number $N$,
the partition function (\ref{z-g}) is governed by
the saddle point solution of the fields $U^{\rm sin}$, ${\bm U}^{\rm tri}$, and
${\bm a}_{\tau}$, which satisfies
\begin{eqnarray}
\frac{\delta {\cal S}}{\delta U^{\rm sin}} =
\frac{\delta {\cal S}}{\delta {\bm U}^{\rm tri}} =
\frac{\delta {\cal S}}{\delta {\bm a}_{\tau}} =0,
\end{eqnarray}
and the fluctuations of $U^{\rm sin}$, ${\bm U}^{\rm tri}$, and
${\bm a}_{\tau}$ are weak.

This saddle point solution indeed corresponds to the mean-field solution
derived in Ref.~\onlinecite{sm}.
At the saddle point, the vector auxiliary fields ${\bm D}_{\j\l}$ and ${\bm E}_{\j\l}$,
respectively, relate to
the $d$-vectors of spin-triplet pairing
and spin-triplet hopping, i.e.,
\begin{eqnarray}
\bar{D}_{\j\l,\mu} = i
\langle f_{\j,\alpha} [\sigma_2 \sigma_{\mu}]_{\alpha\beta}
f_{\l,\beta} \rangle, \ \ \
\bar{E}_{\j\l,\mu} =
\langle f^{\dag}_{\j,\alpha} [\sigma_{\mu}]_{\alpha\beta}
f_{\l,\beta} \rangle \label{tri-pairing}
\end{eqnarray}
$(\mu=1,2,3)$,
while the scalar auxiliary fields $\eta_{\j\l}$ and $\chi_{\j\l}$
relate to
the spin-singlet pairing and hopping, i.e.,
\begin{eqnarray}
\bar{\eta}_{\j\l} = -i\langle f_{\j,\alpha}[\sigma_2]_{\alpha\beta}
f_{\l,\beta} \rangle, \ \ \
\bar{\chi}_{\j\l} = \langle f^{\dag}_{\j,\alpha}
f_{\l,\alpha} \rangle.
\label{sin-pairing}
\end{eqnarray}

\begin{figure}
    \includegraphics[width=85mm]{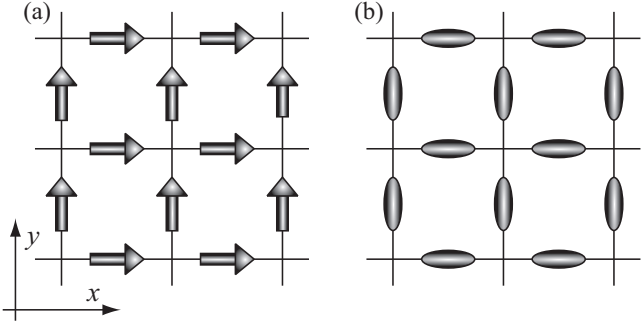}
\caption{Configurations of (a) $d$-vectors for spin triplet pairings and
(b) directors corresponding to the quadrupolar moments
in the $Z_2$ planar state on the square lattice.}
\label{fig:d_config}
\end{figure}

The present authors\cite{sm} previously
investigated various local minima
of the action at zero magnetic field, $h=0$,
assuming that $U^{\rm \sin}$, ${\bm U}^{\rm tri}$,
and ${\bm a}_{\tau}$ are
temporally uniform and also preserve the
translational symmetries of the square lattice.
We found that,
in an intermediate-coupling regime, the saddle
point solution acquires finite spin-triplet pairings
on the ferromagnetic bonds, while singlet pairings
on the antiferromagnetic bonds.
In particular,
a coplanar configuration of orthogonal
$d$-vectors in spin-triplet pairings on the ferromagnetic
bonds [see Fig.~\ref{fig:d_config}(a)]
on top of the `$\pi$-flux'-type singlet pairings~\cite{am} on the antiferromagnetic bonds
realizes the best mean-field
energy among others.~\cite{sm} The
solution is given by
\begin{eqnarray}
&&\hspace{-0.4cm}
\bar{U}^{\rm tri}_{\langle \j,\j+{\bm e}_x\rangle,\mu}
\equiv i\delta_{\mu,1} D \sigma_2, \ \ \
\bar{U}^{\rm tri}_{\langle \j,\j+{\bm e}_y\rangle,\mu}
\equiv i\delta_{\mu,2} D \sigma_2, \nn \\
&& \hspace{-0.4cm}
\bar{U}^{\rm sin}_{\langle \j,\j+{\bm e}_x\pm {\bm e}_y \rangle}
\equiv \chi \sigma_3 \pm \eta \sigma_1, \ \ \
\bar{a}^{\mu}_{\j,\tau} = 0 \label{z2planar}
\end{eqnarray}
with certain real values $D$, $\chi$, and $\eta$,
where ${\bm e}_{x}=(1,0)$ and ${\bm e}_y=(0,1)$.
The presence of the $d$-vectors, $\bar{{\bm D}}_{\langle j,j+{\bm e}_x\rangle}=(D,0,0)$
and
$\bar{{\bm D}}_{\langle j,j+{\bm e}_y\rangle}=(0,D,0)$ [Fig.~\ref{fig:d_config}(a)],
produces a quadrupolar order on bonds; in the mean-field approximation,
we have the relation\cite{sm}
\begin{align}
Q_{{\bm j}{\bm l},\mu\nu}=& -\frac{1}{2} \left(
E_{{\bm j}{\bm l},\mu}E^{*}_{{\bm j}{\bm l},\nu} - \frac{1}{3}\delta_{\mu\nu}
|{\bm E}_{{\bm j}{\bm l}}|^2 \right)
+ {\rm H.c.} \nn \\
&- \frac{1}{2} \left( D_{{\bm j}{\bm l},\mu}D^{*}_{{\bm j}{\bm l},\nu}
- \frac{1}{3}\delta_{\mu\nu}
|{\bm D}_{{\bm j}{\bm l}}|^2\right) + {\rm H.c.} \label{dir}
\end{align}
$(\mu,\nu=1,2,3)$.
Hence the state has an antiferro-quadrupolar order, as shown in
Fig.~\ref{fig:d_config}(b), where all nematic directors are lying in a single plane.
The invariant gauge group~\cite{wen}
dictates that all the gauge excitations
around this mean-field solution have
finite gap [are not required to be
gapless by the local $SU(2)$ gauge
symmetry]. The state has the same
spin-triplet pairing function as a `planar'
type superfluid B-phase of $^3$He.\cite{vollhardt} We
hence dubbed this state the
\emph{$Z_2$ planar state}.

Here we summarize the symmetry of the $Z_2$ planar state.
The wavefunction is invariant under the space translation and
the space reflections with $x$- and $y$-axes.
The state is also invariant under the
time reversal ${\cal T}$; under the operation of ${\cal T}$,
the triplet pairings on the nearest neighbor links change their sign,
but this change sets off by a staggered gauge
transformation $\Psi^{\dag}_{{\bm j}} \rightarrow
(-1)^{j_x+j_y} \Psi^{\dag}_{\bm j}$.
This concludes that this state does not have any spin order
i.e.
$\langle {\bm S}_{\j}\rangle = 0$.~\cite{sm,sym}.
The coplanar ordering of the
$d$-vectors breaks the $SU(2)$ spin-rotational symmetry,
but the state preserves
the spin $\pi$ rotational symmetry around
both $1$-, $2$-, and
$3$-axes. This corresponds to the fact that the ground-state
manifold of the $d$-wave spin nematic state\cite{sms}
has $SU(2)/(Z_2\times Z_2)$ symmetry.
We also note that this pairing state
has a non-trivial {\it staggered}
$U(1)$ spin-rotational symmetry.
That is, the state defined by
Eq.~(\ref{z2planar}) is invariant under
the following staggered spin rotation about $z$ axis
\begin{eqnarray}
\Psi^{\dag}_{\bm j} &\rightarrow
\exp[i(-1)^{j_x+j_y}\theta\sigma_3]
\Psi^{\dag}_{\bm j}, \nn\\
\Psi_{\bm j} &\rightarrow
\Psi_{\bm j}\exp[-i(-1)^{j_x+j_y}\theta\sigma_3]
\label{staggeredU1}
\end{eqnarray}
for any $\theta$.

This $Z_2$ planar state
is shown to be a stable local minimum,
even when projected onto the real spin space.
A variational Monte Carlo study indicates that
the projected BCS wavefunction constructed
from this $Z_2$ planar state
achieves the best optimal energy
in the parameter range $0.42 J_1 \leq J_2 \leq 0.57 J_1$,
which is encompassed by the competing
ferromagnetic phase ($J_2< 0.42 J_1$)
and collinear antiferromagnetic phase
($0.57 J_1<J_2$).~\cite{sym}
Moreover, the wavefunction of the projected $Z_2$ planar state
belongs to
the same space group (including its
irreducible representation)~\cite{sym} as that
of the bond-type spin nematic phase
suggested by the exact diagonalization
analysis~\cite{sms} in the similar parameter regime.
The spin correlation function
calculated with this projected
BCS wavefunction~\cite{sym} exhibits a similar
behavior as those
obtained from the exact diagonalization studies
up to 40 sites.~\cite{richter}
Observing the energetics and the consistencies
with the previous exact diagonalization analyses,
we regard that this projected $Z_2$ planar
phase is indeed realized
as a spin nematic phase
in a certain parameter range around $J_2 \approx 0.5 J_1$
of the present $J_1$-$J_2$ model.
We therefore start from the mean-field
$Z_2$ planar state, to derive the dynamical
magnetic properties of the bond-type
spin nematic phase.

\subsection{Partition function and Green function at the saddle point}

We describe the partition function at the saddle point,
omitting the fluctuations of $U^{\rm sin}$, ${\bm U}^{\rm tri}$,
and ${\bm a}_\tau$ in Eq.~(\ref{partition}).
The Bogoliubov--de Gennes
Hamiltonian for the mean-field $Z_2$ planar state is given by
\begin{align}
{\bm H}^{(0)}_{\k} \equiv&
 \frac{J_1 D}{2} (s_x {\bm \gamma}_3
- s_y {\bm \gamma}_5)
+ J_2 (\chi c_x c_y {\bm \gamma}_4
+ \eta s_x s_y {\bm \gamma}_2)
\label{bdg}
\end{align}
with $s_{\mu} \equiv \sin k_\mu$
and $c_{\mu}\equiv \cos k_\mu$ ($\mu=x,y$).
The $4 \times 4$ $\gamma$-matrices are
defined as
\begin{equation}
\gamma_{1}
= \sigma_2 \otimes \sigma_1
=\left(
 \begin{array}{cc}
  0 & -i \sigma_1  \\
  i \sigma_1 & 0 \\
 \end{array}
\right),
\end{equation}
where the $2 \times 2$ Pauli matrices
$\sigma_{\mu}$ ($\mu=1,2,3$)
in front of the $\otimes$-mark is
for the particle-hole space, while the
other is for the spin space. Using the same notation,
we define the other 4 anti-commutating
$\gamma$-matrices as $\gamma_{2}
= \sigma_2 \otimes \sigma_2$,
$\gamma_{3}
= \sigma_2 \otimes \sigma_3$,
$\gamma_{4}
= \sigma_3 \otimes \sigma_0$,
and $\gamma_{5}
= \sigma_1 \otimes \sigma_0$.

\begin{figure}[tb]
    \includegraphics[width=85mm]{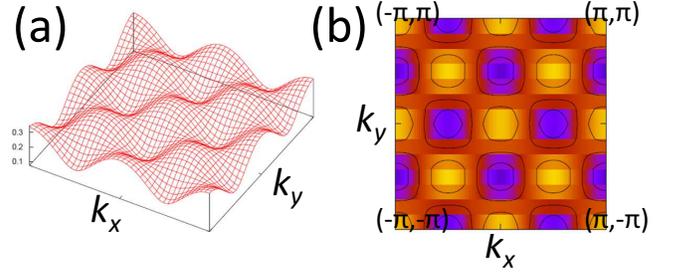}
\caption{(a) Dispersion relation of spinon band
$\xi_{\bm k}$ in the $Z_2$ planar state at $J_2=1.1J_1$. (b) Contour plot of $\xi_{\bm k}$,
showing 8 minima at ${\bm k}=(\pm\frac{\pi}{2},0),
(0,\pm\frac{\pi}{2}),(\pm\frac{\pi}{2},\pi),
(\pi,\pm\frac{\pi}{2})$, and 8 maxima at ${\bm k}=(0,0),
(\pm\frac{\pi}{2},\pm\frac{\pi}{2}),(0,\pi),
(\pi,0),(\pi,\pi)$. }
\label{fig:dis_spinon}
\end{figure}

The partition function at the saddle point takes the form
\begin{align}
Z^{(0)}[h] &=
\exp \left( - N {\cal S}^{(0)}[h]\right), \\
{\cal S}^{(0)}[h] &=  {\cal S}^{(0)}_{\rm I} +
{\cal S}^{(0)}_{\rm II}[h], \label{a-m} \\
{\cal S}^{(0)}_{\rm I} &=  \frac{\beta}{2} \left[
J_1 N_{\Lambda} D^2 +
J_2 N_{\Lambda}
({\chi}^2 + {\eta}^2 ) \right], \label{s1-m} \\
{\cal S}^{(0)}_{\rm II}[h] &= -
\frac{1}{2N} \sum_{a=1}^N {\rm Tr} \left(\ln
 {\bm G}_{0,(\k,n,a|\k',n',a)}^{-1}[h] \right), \label{s2-m}
\end{align}
where the trace 
is over the momentum ($\k$), the Matsubara
frequency ($n$), spin, and particle-hole indices.
The single-particle Green
function ${\bm G}_0[h]$ of the fermion fields
${\bm f}^a_{\k,n}$ at the saddle point
is given by
\begin{align}
&{\bm G}^{-1}_{0,(\k,n,a|\k',n',a)}[h^{aa}] \equiv
\delta_{n,n'}\delta_{\k,\k'} {\bm g}^{-1}_{0}(\k,i\omega_n)
\nn \\
&\hspace{0cm} + \frac{1}{2}
\frac{1}{\sqrt{\beta N_{\Lambda}}}
\sum_{\mu=1,2,3} \sum_{{\bm q},m}
\delta_{\k,\k'+{\bm q}}\delta_{n,n'+m} \!\
h^{aa}_{\mu}({\bm q},m) {\bm u}_{\mu}, \label{g0-m}
\end{align}
where ${\bm g}_{0}$ denotes the $4\times 4$ matrix
single-particle Green function of ${\bm f}^a_{\k,n}$ at zero field,
\begin{align}
&{\bm g}_{0}(\k,i\omega_n)
\equiv \left(i \omega_n {\bm \gamma}_0 -
{\bm H}^{(0)}_{\k}\right)^{-1}, \label{g0-b}
\end{align}
${\bm u}_{\mu}$ $(\mu=1,2,3)$ denote
the $4 \times 4$ Hermite matrices
defined by
\begin{eqnarray}
&& {\bm u}_{1}\equiv
{\bm \gamma}_{15} = -i{\bm \gamma}_1{\bm \gamma}_5,
\ \ \ {\bm u}_2 \equiv {\bm \gamma}_{13}
= -i{\bm \gamma}_1{\bm \gamma}_3, \nn \\
&&\hspace{1.4cm}
{\bm u}_3 \equiv {\bm \gamma}_{35}
= -i{\bm \gamma}_3{\bm \gamma}_5, \label{g-u}
\end{eqnarray}
and
\begin{eqnarray}
h^{aa}_{\mu}({\bm q},m)
\equiv \frac{1}{\sqrt{\beta N_{\Lambda}}}
\sum_{\j} \int^{\beta}_{0} d\tau
\!\ e^{-i{\bm q}\cdot{\j}-i\epsilon_m \tau}
h^{aa}_{\j,\mu}(\tau)
\end{eqnarray}
with $\epsilon_m\equiv 2m\pi\beta^{-1}$.
Diagonalizing the Hamiltonian matrix, we obtain the mean-field
band dispersion relation of spinons,
\begin{eqnarray}
\xi_{\bm k} = \left[\frac{J^2_1 D^2}{4} (s^2_x + s^2_y)
+ J^2_2 (\chi^2 c^2_x c^2_y + \eta^2
s^2_x s^2_y ) \right]^{1/2}, \label{dispersion}
\end{eqnarray}
which has a full gap in the whole Brillouin zone, as shown in Fig.~\ref{fig:dis_spinon}(a).
When we expand the Green function ${\bm G}_{0}$
with small fields $\{ h^{aa}_\mu\}$,
the matrix ${\bm u}_\mu$ corresponds to the
external vertex connecting
the external field $h^{aa}_\mu$
with two single-particle Green functions ${\bm g}_0$.

\section{1/${\bm N}$ expansion for correlation functions}
In the previous section, we have described
the saddle point
solution of a generalized $N$-flavor spin-1/2
frustrated ferromagnetic model.
For large $N$, the fluctuations of the auxiliary fields and the gauge fields
around the saddle point are weak and hence they can be precisely included by
performing a perturbational expansion of the effective action with the fluctuation fields.
This expansion gives a $1/N$ expansion of physical quantities.
In this section, performing this $1/N$ expansion, we
calculate the dynamical spin correlation function~\cite{aa}
\begin{align}
C^{aa}_{\mu\nu}(\j,\tau) &\equiv
\left. \left\langle T_\tau \left[
S^{aa}_{\0,\mu}(0) S^{aa}_{\j,\nu}(\tau)
\right]
\right\rangle\right|_{h=0} \nn\\
&=\left. \frac{\partial^2 Z[h]}{\partial h^{aa}_{\0,\mu}(0) \partial
h^{aa}_{\j,\nu}(\tau)}\right|_{h=0}
\label{corr0}
\end{align}
and the dynamical susceptibility
\begin{align}
\chi^{aa}_{\mu\nu}({\bm q},i\epsilon_m) &=
\sum_{\j}\int^{\beta}_{0} d\tau \!\ e^{-i({\bm q}\cdot\j +
\epsilon_m \tau)}
C^{aa}_{\mu\nu}(\j,\tau),  \label{ft}
\end{align}
where $T_\tau$ denotes the imaginary-time ordering and the flavor index $a$ is fixed to
a certain number.
Specifically, the correlation function of leading order,
$O(1)$, corresponds to the Hartree-Fock contribution
and the correction term of order $1/N$
corresponds to the random phase approximation (RPA) term.
We present a formalism for this large $N$ expansion here and
discuss the dynamical spin structure factors thus obtained
in the next section.

The spin correlation
functions are composed by two parts
\begin{eqnarray}
C^{aa}_{\mu\nu}(\j,\tau)
%
&=& C^{aa}_{\mu\nu,{\rm I}}(\j,\tau)
+ C^{aa}_{\mu\nu,{\rm II}}(\j,\tau)
\end{eqnarray}
with
\begin{align}
C^{aa}_{\mu\nu,{\rm I}}(\j,\tau)
&\equiv - \frac{N}{Z} \int {\cal D}U^{\rm sin} {\cal D}{\bm U}^{\rm tri}
 {\cal D}{\bm a}_{\tau} \nn \\
& \times\frac{\partial^2 {\cal S}}{\partial h^{aa}_{\0,\mu}(0) \partial
h^{aa}_{\j,\nu}(\tau)}
\exp\left( - N {\cal S}\right)\bigg|_{h=0}, \label{corr1} \\
C^{aa}_{\mu\nu,{\rm II}}(\j,\tau) &\equiv
 \frac{N^2}{Z} \int {\cal D}U^{\rm sin} {\cal D}{\bm U}^{\rm tri}
 {\cal D}{\bm a}_{\tau} \nn \\
& \times\frac{\partial {\cal S}}
{\partial h^{aa}_{\0,\mu}(0)} \frac{\partial {\cal S}}
{\partial h^{aa}_{\j,\nu}(\tau)}
\exp\left( - N {\cal S}\right)\bigg|_{h=0},
\label{corr2}
\end{align}
and they relate to corresponding
dynamical susceptibilities
\begin{align}
\chi^{aa}_{\mu\nu,{\rm I}}({\bm q},i\epsilon_m) &\equiv
\sum_{\j}\int^{\beta}_{0} d\tau \!\ e^{-i({\bm q}\cdot\j +
\epsilon_m \tau)}
C^{aa}_{\mu\nu,{\rm I}}(\j,\tau), \nn \\
\chi^{aa}_{\mu\nu,{\rm II}}({\bm q},i\epsilon_m) &\equiv
\sum_{\j}\int^{\beta}_{0} d\tau \!\ e^{-i({\bm q}\cdot\j +
\epsilon_m \tau)}
C^{aa}_{\mu\nu,{\rm II}}(\j,\tau). \nn
\end{align}

\subsection{Correlation functions in the leading order}

\begin{figure}[tb]
    \includegraphics[width=85mm]{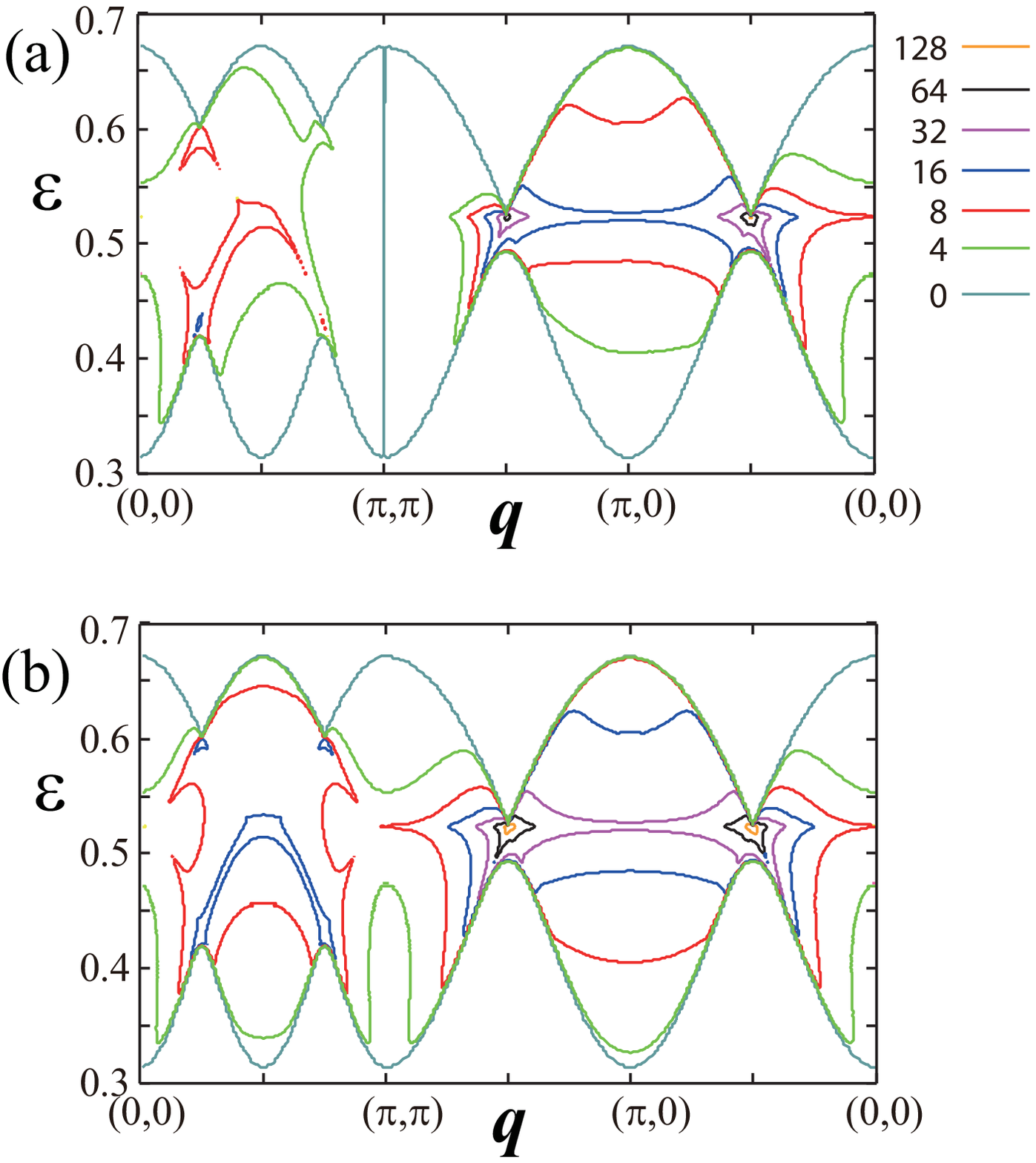}
    \bigskip\\
    \includegraphics[width=85mm]{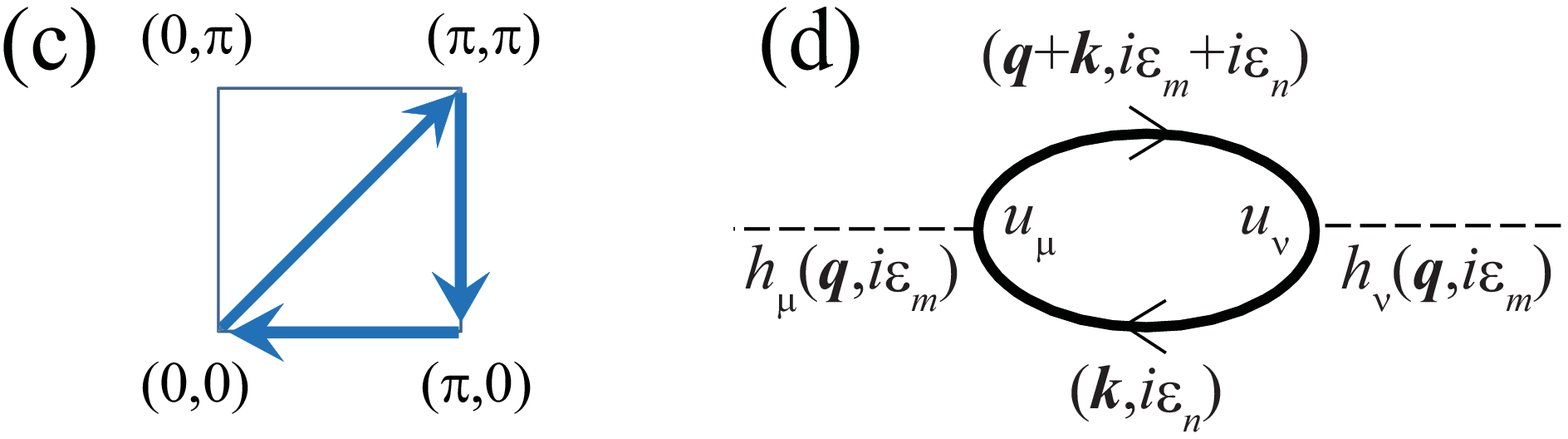}
\caption{Contour plots of (a) Im$\chi^{(0)}_{zz}({\bm q},\epsilon)$
and (b) Im$\chi^{(0)}_{+-}({\bm q},\epsilon)$ at $J_2=1.1 J_1$.
Both of them consist of broad continuum spectra, which have finite
weight only for $\epsilon \ge \omega_c\simeq 0.3$.
(c) Path of the momentum ${\bm q}$ in (a) and (b), which runs from
$(0,0)$ to $(\pi,\pi)$, to $(\pi,0)$ and back to $(0,0)$.
(d) Feynman diagram for Im$\chi^{(0)}_{\mu\mu}({\bm q},i\epsilon_m)$.
Solid lines denote the fermion single-particle Green functions and
dashed lines denote the external magnetic fields. }
\label{fig:HF1}
\end{figure}

Replacing ${\cal S}$ in Eqs.~(\ref{corr1}) and (\ref{corr2})
by ${\cal S}_0$ given in Eqs.~(\ref{a-m}), (\ref{s1-m}), and (\ref{s2-m}),
we obtain the Hartree-Fock contribution to
the spin correlation function,
\begin{align}
C^{aa,(0)}_{\mu\nu}(\j,\tau) =&
\left( - N \frac{\partial^2 {\cal S}^{(0)}}{\partial h^{aa}_{\0,\mu}(0)
\partial h^{aa}_{\j,\nu}(\tau)}  \right. \nn \\
&\ \ \left. \left. + N^2
 \frac{\partial {\cal S}^{(0)}}{\partial h^{aa}_{\0,\mu}(0)}
\frac{\partial {\cal S}^{(0)}}{\partial h^{aa}_{\j,\nu}(\tau)}
\right) \right|_{h=0}, \label{zero1}
\end{align}
or equivalently,
\begin{align}
\lefteqn{\chi^{aa,(0)}_{\mu\nu}({\bm q},i\epsilon_m) } \nn \\
= & - \frac{1}{8 \beta N_{\Lambda}}
\sum_{\k,n}
{\rm Tr} \big[{\bm g}_{0}(\k+{\bm q},i\omega_n+i\epsilon_m)
\!\ {\bm u}_{\mu}\!\
{\bm g}_{0}(\k,i\omega_n) \!\
{\bm u}_{\nu}\big] \nn \\
& + \frac{\delta_{{\bm q},\0}\delta_{m,0}}{16 \beta N_{\Lambda}}
\Big(\sum_{\k,n} {\rm Tr} \big[{\bm g}_{0}(\k,i\omega_n)
\!\ {\bm u}_{\mu}\big] \Big) \nn \\
&\hspace*{1.5cm}\times
\Big(\sum_{\k',n'} {\rm Tr} \big[{\bm g}_{0}(\k',i\omega_{n'})
\!\ {\bm u}_{\nu}\big] \Big),
\label{HF}
\end{align}
where the traces are taken over the
spin and particle-hole indices, i.e., the indices of $4\times 4$ matrices.
From
Eqs.~(\ref{bdg}), (\ref{g0-b}), and (\ref{g-u}), it is clear
that ${\bm g}_0(\k,i\omega_n) \!\ {\bm u}_{\mu}$ is
traceless for each $\mu=1,2,3$, and hence
the second term vanishes. The first term
corresponds to two spinnons propagating with momenta ${\bm k}+{\bm q}$
and $-{\bm k}$ [see Fig.~\ref{fig:HF1}(d)].

After the analytic continuation,
$i\epsilon_n \rightarrow \epsilon +i\delta$,
we obtain the real-time dynamical susceptibilities.
At zero temperature, the imaginary parts of dynamical susceptibilities
at the saddle point are given by
\begin{align}
{\rm Im}\chi^{(0)}_{zz}({\bm q},\epsilon)
&= \frac{\pi}{8N_{\Lambda}}
\sum_{k} \delta(\epsilon-\xi_{+}-
\xi_{-}) \!\ \left[1
- \frac{1}{\xi_{+}\xi_{-}} \right. \nn \\
&\hspace{-1.7cm}
\times(a_{2,+}a_{2,-} - a_{3,+}a_{3,-}
+ a_{4,+}a_{4,-} - a_{5,+}a_{5,-})\Big], \label{imchizz} \\
{\rm Im}\chi^{(0)}_{+-}({\bm q},\epsilon)
&= \frac{\pi}{4N_{\Lambda}}
\sum_{k} \delta(\epsilon-\xi_{+}-
\xi_{-}) \nn \\
&
\times \left[1
- \frac{1}{\xi_{+}\xi_{-}}(a_{2,+}a_{2,-} + a_{4,+}a_{4,-})\right],
\label{imchipm}
\end{align}
where
$a_{2,\pm} \equiv J_2 \eta \!\ s_{x,\pm} s_{y,\pm}$,
$a_{3,\pm} \equiv \frac{J_1 D}{2} s_{x,\pm}$,
$a_{4,\pm} \equiv J_2 \chi \!\ c_{x,\pm} c_{y,\pm}$,
$a_{5,\pm} \equiv -\frac{J_1 D}{2} s_{y,\pm}$,
and $\xi_{\pm} \equiv \sqrt{\sum^5_{j=2} a^2_{j,\pm}}$
with the definitions
$s_{\mu,\pm} \equiv {\rm sin} (k_{\mu} \pm \frac{q_{\mu}}{2})$
and
$c_{\mu,\pm} \equiv {\rm cos} (k_{\mu} \pm \frac{q_{\mu}}{2})$.
These susceptibilities are plotted as a function of
${\bm q}$ and $\epsilon$ in
Figs.~\ref{fig:HF1}(a) and \ref{fig:HF1}(b).
As in the figure,  the Hartree-Fock contributions
consist only of continuum spectra, which
correspond to individual spinon excitations.
Since the mean-field
band dispersion $\xi_{\bm k}$ of the spinon field
is fully gapped [see Fig.~\ref{fig:dis_spinon}(a)],
the continuum spectra appear only above a critical
energy, $\epsilon\ge {\rm max}_{\bm k} (\xi_{+}+\xi_{-})$.
The frequency range of the continuum becomes broadest at
${\bm k}=(0,0)$, $(\pi,0)$, $(\pi,\pi)$, and
$(\frac{\pi}{2},\frac{\pi}{2})$, while
narrow at ${\bm k}=(\pi,\frac{\pi}{2})$ and $(\frac{\pi}{2},0)$.
This feature is because the band dispersion
$\xi_{\bm k}$ has eight minima at
${\bm k}=(\pm\frac{\pi}{2},0)$,
$(0,\pm\frac{\pi}{2})$, $(\pm\frac{\pi}{2},\pi)$, and
$(\pi,\pm\frac{\pi}{2})$, and eight maxima at ${\bm k}=(0,0)$,
$(\frac{\pi}{2},\pm\frac{\pi}{2})$, $(-\frac{\pi}{2},\pm\frac{\pi}{2})$,
$(0,\pi)$,
$(\pi,0)$, and $(\pi,\pi)$ [see Figs.~\ref{fig:dis_spinon}(a) and \ref{fig:dis_spinon}(b)].
One should also notice
that continuum in Im$\chi^{(0)}_{zz}({\bm q},\epsilon)$
has no spectral weight at ${\bm q}=(\pi,\pi)$,
which is attributed to the staggered $U(1)$ spin-rotational
symmetry given by Eq.~(\ref{staggeredU1}).
\begin{figure}
    \includegraphics[width=85mm]{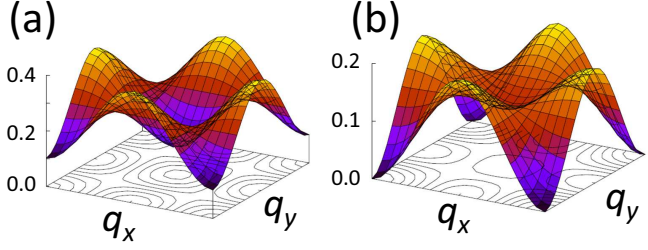}
\caption{ Static structure factors in the mean-field approximation:
(a) $C^{(0)}_{zz}({\bm q},\tau=0)$
and (b) $C^{(0)}_{+-}({\bm q},\tau=0)$ at $J_2=1.1 J_1$.}
\label{fig:static-C}
\end{figure}

Static spin structure factors at the mean-field level 
are given by the frequency integral of
Eqs.~(\ref{imchizz}) and (\ref{imchipm}),
\begin{align}
&C^{(0)}_{zz}({\bm q},\tau=0) =
\int^\infty_0 d\epsilon
{\rm Im}\chi^{(0)}_{zz}({\bm q},\epsilon),
\nn \\
&C^{(0)}_{+-}({\bm q},0) =
\int^\infty_0 d\epsilon
{\rm Im}\chi^{(0)}_{+-}({\bm q},\epsilon),
\nn
\end{align}
both of which exhibit broad peak structures
at ${\bm q}=(0,\pi)$ and $(\pi,0)$, as shown in Fig.~\ref{fig:static-C}.
This indicates the presence of
short-range collinear antiferromagnetic correlations in the $Z_2$ planar state.
This behavior is basically consistent with the
static spin structure factor obtained in the spin nematic phase from the variational
Monte Carlo calculation,~\cite{sym} though the latter one
exhibits relatively stronger collinear antiferromagnetic
correlations.

\subsection{Method of $1/N$ expansion}\label{sec:3B}
To capture the  low-energy collective excitations,
which emerge below the continuum spectra,
we next include fluctuations of
the auxiliary fields ($U^{\rm sin}-{\bar U}^{\rm sin}$ and
${\bm U}^{\rm tri}-{\bar {\bm U}}^{\rm tri}$) and the gauge fields ($i{\bm a}_\tau$)
around their saddle point values.
The fluctuation fields ${\bm r}( \j,\tau)$ for the $Z_2$ planar state are
in total given by the following 35 elements:
\begin{align}
{\bm r}( \j,&\tau) \nn\\
\equiv
 \big(\ & {\rm Re}E_{x,3},\!\
{\rm Im}E_{x,3},\!\ {\rm Re}E_{y,3}, \!\
{\rm Im}E_{y,3},  \nn \\
& 
{\rm Re}D_{x,3},\!\
{\rm Im}D_{x,3},\!\
{\rm Re}D_{y,3},\!\ {\rm Im}D_{y,3},  \nn \\
& 
{\rm Re}E_{x,1},\!\
{\rm Im}E_{x,1},\!\ {\rm Re}E_{y,1},\!\
{\rm Im}E_{y,1}, \nn \\
&
{\rm Re}D_{x,1} - D,\!\ {\rm Im}D_{x,1},\!\
{\rm Re}D_{y,1},\!\ {\rm Im}D_{y,1}, \nn \\
& 
{\rm Re}E_{x,2},\!\
{\rm Im}E_{x,2},\!\ {\rm Re}E_{y,2},\!\
{\rm Im}E_{y,2}, \nn \\
& 
{\rm Re}D_{x,2},\!\ {\rm Im}D_{x,2},\!\
{\rm Re}D_{y,2}-D,\!\ {\rm Im}D_{y,2}, \nn \\
&
{\rm Re}\chi_{x+y}-\chi,\!\ {\rm Re}\chi_{x-y} -\chi,
\!\ {\rm Re}\eta_{x+y}-\eta,\!\ {\rm Re}\eta_{x-y} + \eta, \nn \\
& 
{\rm Im}\chi_{x+y}, \!\ {\rm Im}\chi_{x-y},
\!\ {\rm Im}\eta_{x+y}, {\rm Im}\eta_{x-y},
\!\ ia^1_{\tau}, \!\ i a^{2}_{\tau}, \!\ ia^{3}_{\tau} \
\big).  \label{r-def}
\end{align}
Here the vectors
${\bm D}_{\nu}=(D_{\nu,1},D_{\nu,2},D_{\nu,3})$
and
${\bm E}_{\nu}=(E_{\nu,1},E_{\nu,2},E_{\nu,3})$ with $\nu=x,y$
correspond to
the vector auxiliary fields
given in Eq.~(\ref{lagrangian}),
\begin{eqnarray}
{\bm D}_{\nu}(\j,\tau)
&=&
{\bm D}_{\j-{\bm e}_\nu/2,\j+{\bm e}_\nu/2},\nn\\
%
{\bm E}_{\nu}(\j,\tau)
&=&
{\bm E}_{\j-{\bm e}_\nu/2,\j+{\bm e}_\nu/2},\nn
\end{eqnarray}
where the position vector ${\bm j}$ is defined on the center positions of
the ferromagnetic $J_1$ links,
whereas the fields $\chi_{x\pm y}$ and $\eta_{x\pm y}$ correspond to
the scalar auxiliary fields,
\begin{eqnarray}
\eta_{x\pm y} (\j,\tau )
&=& \eta_{\j-({\bm e}_{x}\pm{\bm e}_{y})/2,
\j+({\bm e}_{x}\pm{\bm e}_{y})/2}, \nn \\
\chi_{x\pm y} (\j,\tau )
&=& \chi_{\j-({\bm e}_{x} \pm{\bm e}_{y})/2,
\j+({\bm e}_{x}\pm{\bm e}_{y})/2}, \nn
\end{eqnarray}
where the position $\bm j$ is
defined on the center positions  of the antiferromagnetic
$J_2$ links.

To obtain quantum corrections to
the Hartree-Fock contribution [Eq.~(\ref{zero1})],
we expand the
action ${\cal S} =  {\cal S}_{\rm I} -
\frac{1}{2N} \sum_{a=1}^N {\rm Tr} \ln {\bm G}^{-1} $ with the small fluctuation fields
${\bm r}( \j,\tau)$ around the saddle point.
Since the functional ${\bm G}^{-1}$ is a linear function of
the elements of $h^{aa}$, $U^{\rm sin}$, ${\bm U}^{\rm tri}$, and ${\bm a}_{\tau}$,
as noted below Eq.~(\ref{g-g}),
the single-particle Green function takes the form
\begin{align}
&{\bm G}^{-1}_{(\k,n,a|\k',n',a)}
= {\bm G}^{-1}_{0,(\k,n,a|\k',n',a)} \nn\\
&+ \frac{1}{\sqrt{\beta N_{\Lambda}}}
\sum_{{\bm q}} \sum_{m}
\delta_{\k,\k'+{\bm q}} \delta_{n,n'+m} 
r_{\alpha}({\bm q},m) {\bm v}_{\alpha}
(\k,\k'),
\label{G-funcwithfluct}
\end{align}
where the bosonic fluctuation fields ${\bm r}$ are transformed as
\begin{align}
r_{\alpha}({\bm q},m) &\equiv
\frac{1}{\sqrt{\beta N_{\Lambda}}}
\sum_{\j} \int^{\beta}_{0} d\tau \!\
e^{-i{\bm j}\cdot{\bm q}-i\epsilon_{m}\tau}
r_{\alpha}(\j,\tau). \nn
\end{align}
Here, the index $\alpha$ ($\alpha=1,\cdots,35$) specifies the element
of the fluctuation fields enumerated in
Eq.~(\ref{r-def}), and
the summation of the position vector ${\bm j}$ runs over
all the center positions of nearest neighbor links for $\alpha=1,\cdots,24$,
all the center positions of second nearest neighbor links for $\alpha=25,\cdots,32$,
and all the lattice sites for $\alpha=33,34,35$.
The summation over the repeated index $\alpha$
is made implicit and will be so henceforth.
In the same sequence as in Eq.~(\ref{r-def}),
the internal vertices ${\bm v}_{\alpha}$
are explicitly given by the $4\times 4$ matrix forms
\begin{align}
{\bm v}(\k,&\k') \nn\\
\equiv
\big(\ &-\overline{c}_x {\bm \gamma}_{35},\!\ -\overline{s}_x
{\bm \gamma}_{12}
, \!\ -\overline{c}_y {\bm \gamma}_{35},\!\ -\overline{s}_y
{\bm \gamma}_{12} , \nn \\
&
\overline{s}_x {\bm \gamma}_{1},\!\
\overline{s}_x
{\bm \gamma}_{14}
, \!\ \overline{s}_y {\bm \gamma}_{1},\!\
\overline{s}_y
{\bm \gamma}_{14} , \nn \\
&
- \overline{c}_x {\bm \gamma}_{15},\!\ -
\overline{s}_x
{\bm \gamma}_{23}
, \!\ - \overline{c}_y {\bm \gamma}_{15},\!\
- \overline{s}_y
{\bm \gamma}_{23}, \nn \\
&
- \overline{s}_x {\bm \gamma}_{3},\!\
- \overline{s}_x
{\bm \gamma}_{34}
, \!\ -\overline{s}_y {\bm \gamma}_{3},\!\
-\overline{s}_y
{\bm \gamma}_{34} ,
 \nn \\
&
\overline{c}_x {\bm \gamma}_{31},\!\
\overline{s}_x
{\bm \gamma}_{25}
, \!\ \overline{c}_y {\bm \gamma}_{31},\!\
\overline{s}_y
{\bm \gamma}_{25}, \nn \\
&
\overline{s}_x {\bm \gamma}_{5},\!\
- \overline{s}_x
{\bm \gamma}_{45}
, \!\ \overline{s}_y {\bm \gamma}_{5},\!\
-\overline{s}_y
{\bm \gamma}_{45}, \nn \\
&
- \overline{c}^{\prime}_{x+y} {\bm \gamma}_{4},\!\
-\overline{c}^{\prime}_{x-y} {\bm \gamma}_{4}, \!\
\overline{c}^{\prime}_{x+y} {\bm \gamma}_{2},\!\
\overline{c}^{\prime}_{x-y} {\bm \gamma}_{2},
\nn \\
& \hspace{-0.2cm}
-\overline{s}^{\prime}_{x+y}
{\bm \gamma}_{0}, \!\
-\overline{s}^{\prime}_{x-y}
{\bm \gamma}_{0}, \!\
\overline{c}^{\prime}_{x+y}
{\bm \gamma}_{24}, \!\
\overline{c}^{\prime}_{x-y}
{\bm \gamma}_{24}, \!\ -{\bm \gamma}_{2} , \!\
{\bm \gamma}_{24}, \!\ {\bm \gamma}_4 \ \big), \label{v-def}
\end{align}
where
\begin{eqnarray}
\overline{c}_{\mu} &=& \frac{J_1}{2}
\cos \Big(\frac{k_{\mu}+k^{\prime}_{\mu}}{2}\Big),
\nn \\
\overline{s}_{\mu} &=& \frac{J_1}{2}
\sin \Big(\frac{k_{\mu}+k^{\prime}_{\mu}}{2}\Big),
\nn \\
\overline{c}^{\prime}_{x\pm y} &=&
\frac{J_2}{2}
\cos \Big(\frac{k_{x}+k^{\prime}_{x}
\pm k_{y} \pm k^{\prime}_{y}}{2}\Big), \nn \\
\overline{s}^{\prime}_{x\pm y} &=&
\frac{J_2}{2}
\sin \Big(\frac{k_{x}+k^{\prime}_{x}
\pm k_{y} \pm k^{\prime}_{y}}{2}\Big). \nn
\end{eqnarray}

Using the expression of Eq.~(\ref{G-funcwithfluct}),
we obtain a series expansion of the
action with the fluctuation fields ${\bm r}$ around the saddle point,
\begin{align}
{\cal S} =&  {\cal S}_{\rm I} -
\frac{1}{2N} \sum_{a=1}^N {\rm Tr} \ln {\bm G}^{-1}  \nn \\
=& {\cal S}^{(0)} +
{\cal S}^{(1)}_{\alpha} r_{\alpha}
+
{\cal S}^{(2)}_{\alpha,\alpha'} r_{\alpha} r_{\alpha'} \nn \\
&+ \sum^{\infty}_{n=3}
 S^{(n)}_{\alpha_1,\alpha_2,\cdots,\alpha_n}
\!\  r_{\alpha_1} r_{\alpha_2}\cdots r_{\alpha_n}, \label{exp}
\end{align}
where the coefficients ${\cal S}^{(n)}_{\alpha_1,\alpha_2,\cdots,\alpha_n}$
are given by
\begin{align}
%
&{\cal S}^{(n)}_{\alpha_1,\alpha_2,\cdots,\alpha_n} =
\left. \frac{1}{n!}\frac{ \partial^n {\cal S}}
{\partial r_{\alpha_1}\cdots \partial r_{\alpha_n}} \right|_{{\bm r}={\bf 0}} \nn\\
&=\left. \frac{ \partial {\cal S}_{\rm I}}
{\partial r_{\alpha_1} } \right|_{{\bm r}={\bf 0}}
\delta_{n,1} +
\left. \frac12 \frac{ \partial^2 {\cal S}_{\rm I}}
{\partial r_{\alpha_1} \partial r_{\alpha_2}}
\right|_{{\bm r}={\bf 0}} \delta_{n,2}
\nn \\
&+ \frac{(-1)^n}{2Nn(\beta N_\Lambda)^{n/2}} \sum_{a=1}^N
{\rm Tr}\big[{\bm G}_0 {\bm v}_{\alpha_1}
{\bm G}_0 {\bm v}_{\alpha_2}\cdots
{\bm G}_0 {\bm v}_{\alpha_n}\big]\label{3rd}
\end{align}
for $n \ge 1$.
The trace here is taken over the momentum,
Matsubara frequency, spin, and particle-hole
indices. The summation of the flavor index is written explicitly.
One finds that all the coefficients ${\cal S}^{(n)}_{\alpha_1,\alpha_2,\cdots,\alpha_n}$
are of order $O(1)$ in the large $N$ limit.
Note that ${\bm v}_{\alpha}$ plays a role of
an internal vertex which connects two Green
functions with the fluctuation field $r_{\alpha}$.

To calculate the correlation functions
Eqs.~(\ref{corr1}) and (\ref{corr2}), we further perform a series expansion of
the action (\ref{exp}) in the small field $h^{aa}$.
We only need the series of $\{h^{aa}_\mu\}$ with a certain flavor $a$,
\begin{align}
{\cal S} &=  \overline{{\cal S}}^{(0,0)} +
\overline{{\cal S}}^{(0,2)}_{\alpha,\alpha'} r_{\alpha} r_{\alpha'}
+ \sum^{\infty}_{n=3}
 \overline{{\cal S}}^{(0,n)}_{\alpha_1,\alpha_2,\cdots,\alpha_n}
\!\  r_{\alpha_1} r_{\alpha_2}\cdots r_{\alpha_n} \nn\\
&
+ \frac{1}{N}\sum_{\mu=1}^3 \sum_{n=1}^\infty
\overline{{\cal S}}^{(1,n)}_{\mu;\alpha_1,\cdots,\alpha_n}
r_{\alpha_1} \cdots r_{\alpha_n}  h^{aa}_\mu \nn \\
&+ \frac{1}{N} \sum_{\mu,\nu=1}^3 \sum^{\infty}_{n=0}
 \overline{{\cal S}}^{(2,n)}_{\mu,\nu;\alpha_1,\cdots,\alpha_n}
r_{\alpha_1} \cdots r_{\alpha_n} h^{aa}_\mu h^{aa}_\nu, \label{h_expand}
\end{align}
where
\begin{align}
&\overline{\cal S}^{(0,n)}_{\alpha_1,\cdots,\alpha_n}\equiv
\left. {\cal S}^{(n)}_{\alpha_1,\cdots \alpha_n}\right|_{h=0},\\
&\overline{\cal S}^{(1,n)}_{\mu;\alpha_1,\cdots,\alpha_n}
\equiv
\left. N \frac{\partial {\cal S}^{(n)}_{\alpha_1,\cdots,\alpha_n}}
{\partial h^{aa}_{\mu}}\right|_{h=0},
\label{pre0} \\
&\overline{\cal S}^{(2,n)}_{\mu,\nu;\alpha_1,\cdots,\alpha_n}
\equiv \left.  \frac{N}{2}
\frac{\partial^2 {\cal S}^{(n)}_{\alpha_1,\cdots,\alpha_n}}
{\partial h^{aa}_{\mu} \partial h^{aa}_{\nu}}\right|_{h=0}
\label{pre}
\end{align}
with $n\ge 0$
and with the definition
${\cal S}^{(n)}_{\alpha_1,\cdots,\alpha_n}={\cal S}^{(0)}$
if $n=0$.
Equation~(\ref{h_expand}) does not have any linear term in ${\bm r}$,
by definition of the saddle point, and
any linear term in $h_\mu^{aa}$, as
$\overline{S}^{(1,0)}_{\mu;}\sim \sum_{k,n}{\rm Tr} [g_0(k,i\omega_n){\bm u}_\mu]=0$. 
The coefficient $\overline{S}^{(2,0)}_{\mu,\nu;}$
corresponds to Hartree-Fock susceptibility $\overline{S}^{(2,0)}_{\mu,\nu;}=-\chi^{aa,(0)}_{\mu\nu}/2$.
Note that the coefficients
$\overline{\cal S}^{(1,n)}_{\mu;\alpha_1,\cdots,\alpha_n}$ and
$\overline{\cal S}^{(2,n)}_{\mu,\nu;\alpha_1,\cdots,\alpha_n}$
are functions of order unity in the large $N$ limit.
This is because $N{\cal S}^{(n)}$ is
composed by a summation over the flavor
indices and the field derivative, selecting
the index $a$, leads to a result of order unity.
All coefficients $\overline{\cal S}^{(0,n)}_{\alpha_1,\cdots,\alpha_n}$,
$\overline{\cal S}^{(1,n)}_{\mu;\alpha_1,\cdots,\alpha_n}$,
and $\overline{\cal S}^{(2,n)}_{\mu,\nu;\alpha_1,\cdots,\alpha_n}$
are order $O(1)$ in the large-$N$ limit.

We regard the quadratic term in ${\bm r}$
as a non-perturbed Gaussian action
for the fluctuation fields and treat the rest of terms
as perturbations.
The coefficient of the quadratic  term  corresponds to
the fluctuation-field propagator
\begin{align}
&\frac{1}{N}
\Big[\Big(\overline{\cal S}^{(0,2)}\Big)^{-1}\Big]_{\alpha_1,\alpha_2}=
\frac{\int d{\bm r} r_{\alpha_1}  r_{\alpha_2}
\exp \Big[ - N \overline{\cal S}^{(0,2)}_{\alpha,\alpha'}
r_{\alpha} r_{\alpha'} \Big] }
{\int d{\bm r}
\exp \Big[ - N \overline{\cal S}^{(0,2)}_{\alpha,\alpha'}
r_{\alpha} r_{\alpha'} \Big] },
\label{r-propagator}
\end{align}
which is given by Eq.~(\ref{3rd}) as
\begin{align}
&\overline{\cal S}^{(0,2)}_{\alpha,\alpha'}
= \left. \frac12 \frac{\partial^2 {\cal S}_{\rm I}}
{\partial r_{\alpha} \partial r_{\alpha'}}\right|_{{\bm r}={\bf 0},h=0} +
 \frac{1}{4\beta N_{\Lambda}}
\sum_{k,n} {\rm Tr}\big[{\bm g}_{0} \!\
{\bm v}_{\alpha} \!\ {\bm g}_{0} \!\ {\bm v}_{\alpha'}\big].
\label{ex-1}
\end{align}

\begin{figure}[tb]
    \includegraphics[width=75mm]{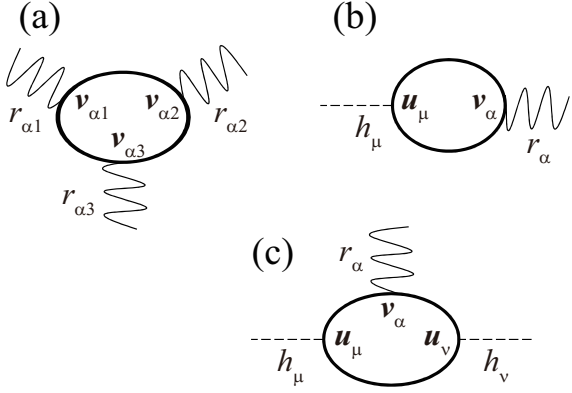}
\caption{Diagrams of interactions in Eq.\ (\ref{h_expand})
containing the
renormalized vertex parts
(a) $\overline{\cal S}^{(0,3)}$,
(b) $\overline{\cal S}^{(1,1)}$,
and (c) $\overline{\cal S}^{(2,1)}$.}
\label{fig:vertex}
\end{figure}

In the perturbation terms of Eq.\ (\ref{h_expand}), each internal vertex
${\bm v}_{\alpha}$ is connected with two single-particle Green Functions (solid lines)
and one fluctuation field (wavy line),
and each external vertex ${\bm u}_{\mu}$ is with two single-particle Green Functions
and one external magnetic field (dashed line),
as shown in Fig.~\ref{fig:vertex}.
The interactions containing $\overline{\cal S}^{(0,n)}$ (with $n>2$),
$\overline{\cal S}^{(1,n)}$,
and $\overline{\cal S}^{(2,n)}$ have single loops composed by multiple
one-particle Green functions and vertices.
See for example
Figs.~\ref{fig:vertex}(a)--(c).
Symbolically,
$\overline{\cal S}^{(1,1)}$ and $\overline{\cal S}^{(2,1)}$
take the following forms
\begin{align}
&\overline{\cal S}^{(1,1)}_{\mu;\alpha}
= \frac{1}{4 \beta N_{\Lambda}}
 \sum_{k,i\omega_n}
{\rm Tr}[{\bm g}_{0} {\bm u}_{\mu}
{\bm g}_{0} {\bm v}_{\alpha} ], \label{ex-???} \\
&\overline{\cal S}^{(2,1)}_{\mu,\nu;\alpha}
= -\frac{1}{16 (\beta N_{\Lambda})^{3/2}}
 \sum_{k,i\omega_n} \Big\{
{\rm Tr}[{\bm g}_{0} {\bm u}_{\nu}
{\bm g}_{0} {\bm u}_{\mu}
{\bm g}_{0} {\bm v}_{\alpha} ] \nn\\
&\hspace{3cm}+{\rm Tr}[{\bm g}_{0} {\bm u}_{\mu}
{\bm g}_{0} {\bm u}_{\nu}
{\bm g}_{0} {\bm v}_{\alpha} ] \Big\}, \label{ex-2}
%
%
\end{align}
where $\k$ and $i\omega_n$ denote
the momentum and the frequency inside of the loops.
The traces in Eqs.~(\ref{ex-1})--(\ref{ex-2})
are only over the particle-hole
and spin indices but not over the flavor index.

Using the series (\ref{h_expand}), we can expand
the spin correlation functions $\chi^{aa}_{\mu\nu}\equiv
\chi^{aa}_{\mu\nu,{\rm I}}+\chi^{aa}_{\mu\nu,{\rm II}}$
as~\cite{aa}
\begin{widetext}
\begin{align}
\chi^{aa}_{\mu\nu,{\rm I}}({\bm q},i\epsilon_n)
&= - \frac{1}{Z} \int {\cal D}{\bm r}  \!\
\Big(\sum^{\infty}_{n=0}
\overline{\cal S}^{(2,n)}_{\mu,\nu;\alpha_1,\cdots,\alpha_n}
r_{\alpha_1} \cdots r_{\alpha_n} \Big)
\sum^{\infty}_{m=0} \frac{(-N)^m}{m !}
\Big(\sum^{\infty}_{l=3}
\overline{\cal S}^{(0,l)}_{\alpha_1,\cdots,\alpha_l}
r_{\alpha_1} \cdots r_{\alpha_l} \Big)^{m}
e^{- N \overline{\cal S}^{(0,2)}_{\alpha,\alpha'}
r_{\alpha} r_{\alpha'}}, \label{c1} \\
\chi^{aa}_{\mu\nu,{\rm II}}({\bm q},i\epsilon_n)
&= \frac{1}{Z} \int {\cal D}{\bm r} \!\
\Big(\sum^{\infty}_{n=1}
\overline{\cal S}^{(1,n)}_{\mu;\alpha_1,\cdots,\alpha_n}
r_{\alpha_1} \cdots r_{\alpha_n} \Big)\!\
\Big(\sum^{\infty}_{l=1}
\overline{\cal S}^{(1,l)}_{\nu;\alpha_1,\cdots,\alpha_l}
r_{\alpha_1} \cdots r_{\alpha_l} \Big) \nn \\
& \hspace{2cm} \times \sum^{\infty}_{m=0}
\frac{(-N)^m}{m !}
\Big(\sum^{\infty}_{j=3}
\overline{\cal S}^{(0,j)}_{\alpha_1,\cdots,\alpha_j}
r_{\alpha_1} \cdots r_{\alpha_j} \Big)^{m}
e^{- N \overline{\cal S}^{(2)}_{\alpha,\alpha'}
r_{\alpha} r_{\alpha'}}, \label{c2}
\end{align}
where $\overline{\cal S}^{(0,0)}$ was
omitted.
The Gaussian integrals over the real-valued fields
${\bm r}$ can be taken, reducing
even numbers of the fields to a sum
over all the possible pairwise contractions among the fields;
\begin{eqnarray}
\int {\cal D}{\bm r} \!\ r_{1} \cdots r_{2k} \!\
\exp \big[ - N \overline{\cal S}^{(0,2)}_{\alpha,\alpha'}
r_{\alpha} r_{\alpha'} \big]
= \sum_{\sigma}
\!\ \frac{1}{N^k} \!\
\big[\overline{\cal S}^{(0,2),-1}\big]_{\sigma(1),\sigma(2)}
\big[\overline{\cal S}^{(0,2),-1}\big]_{\sigma(3),\sigma(4)} \cdots
\big[\overline{\cal S}^{(0,2),-1}\big]_{\sigma(2k-1),\sigma(2k)}.
\label{wick}
\end{eqnarray}
\end{widetext}
The summation over $\sigma$ runs over
the arbitrary permutations among the $2k$ indices.

We can evaluate the $N$ dependence of each term in Eqs.~(\ref{c1}) and
(\ref{c2}), using the facts that all coefficients
$\overline{\cal S}^{(0,n)}_{\alpha_1,\cdots,\alpha_n}$,
$\overline{\cal S}^{(1,n)}_{\mu;\alpha_1,\cdots,\alpha_n}$,
and $\overline{\cal S}^{(2,n)}_{\mu,\nu;\alpha_1,\cdots,\alpha_n}$
are of order unity and one Gaussian integral of
a pair of fluctuation fields leads to one prefactor $1/N$
[see Eq.~(\ref{wick})].
In Eq.~(\ref{c1}), the term
with the summation indices $(l,m,n)$
is of order $O(N^{-n/2-(l-2)m/2})$.
In Eq.~(\ref{c2}), the term
with the summation indices $(l,m,n,j)$
is of order $O(N^{-n/2-l/2-m(j-2)/2})$.
These order estimates give a controlled expansion of physical quantities
with a small parameter $1/N$.

To enumerate all the contractions possible in
Eqs.~(\ref{c1}) and (\ref{c2}), we use Feynman
diagrams.
Connecting one internal
vertex ${\bm v}_{\alpha}$
with another one ${\bm v}_{\beta}$ by a wavy line, which represents
the fluctuation-field 
propagator $[\overline{\cal S}^{(0,2),-1}]_{\alpha,\beta}$,
we obtain all possible diagrams.
As an example, we depict all diagrams of spin correlation functions of order $O(1/N)$ in Fig.~\ref{fig:1-loop}, where
those diagrams which vanish by themselves have
been already omitted.

\subsection{Next-to-leading order corrections to the correlation functions}

\begin{figure}
   \includegraphics[width=70mm]{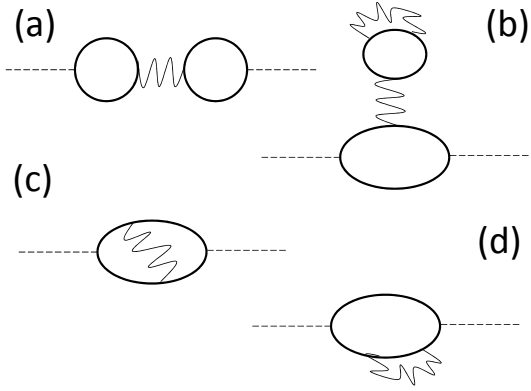}
\caption{$1/N$-contributions to correlation functions,
where dotted lines denote external fields,
wavy lines are the RPA propagators, and solid
lines are single-particle Green functions.
Diagram (a) contributes to $\chi^{aa}_{\mu\mu,{\rm II}}(q,i\epsilon_n)$,
while diagrams (b,c,d) to $\chi^{aa}_{\mu\mu,{\rm I}}(q,i\epsilon_n)$.
They contribute to (a) spin-wave term, (b,d) Hartree-Fock (HF) term with
renormalized single-particle Green functions, and
(c) HF term with a vertex correction. }
\label{fig:1-loop}
\end{figure}

From the order estimation explained in Sec.~\ref{sec:3B},
the leading order
contribution to the spin correlation function
is the Hartree-Fock solution, Eq.~(\ref{HF}),
which is  order unity. The next-to-leading order
corrections of order $O(1/N)$ in the $1/N$ expansion 
are given by the Feynman diagrams 
depicted in Fig.~\ref{fig:1-loop}.

Among these diagrams, only the diagram Fig.~\ref{fig:1-loop}(a)
endows the imaginary part of the
dynamical spin susceptibilities  with finite spectral weight
due to the low-energy collective modes.
Figs.~\ref{fig:1-loop}(b)--(d)
take the same structure as that of the
mean-field diagram Fig.~\ref{fig:HF1}(d).
The difference can be solely attributed to
a proper renormalization of the single-particle Green
function in the cases (b) and (d), and a renormalization
of the external vertex in the case (c).
Thus, their major contribution is more or less modification of
shape and intensity of the Stoner continuum,
which already appears in the leading order.
In Fig.~\ref{fig:1-loop}(a), on the other hand,
the momentum and energy carried by one of
the external lines are transmitted to the other
only through the fluctuation-field propagator, namely RPA propagator, into which
various collective excitations including gapless
Goldstone modes and gapped `gauge-field' like
collective modes are encoded. As a result,
some of low-energy poles of the RPA propagator
show up as coherent bosonic peaks in the imaginary
part of susceptibilities given by Fig.~\ref{fig:1-loop}(a).

We henceforth focus on
Fig.~\ref{fig:1-loop}(a) to discuss low-energy collective modes.
We will see that
these collective modes directly
come from the poles in the RPA propagators which
connect two loops of single-particle Green functions.
To derive its expression,
let us first clarify
possible low-energy poles encoded in the RPA
propagator.
The RPA propagator
defined in Eq.~(\ref{ex-1}) is always a
block-diagonal matrix with respect to the
following four groups of fluctuation fields:
\begin{align}
{\bm R}_{1} \equiv&  {\rm Re}D_{y,3} \!\ {\bm e}^{1}_1
+ {\rm Im}D_{x,3} \!\ {\bm e}^{1}_2 +  {\rm Im}E_{x,3} \!\ {\bm e}^{1}_3
+ {\rm Re}E_{y,3}  \!\ {\bm e}^1_4,  \label{group1} \\
{\bm R}_{2} \equiv&  {\rm Re}D_{x,3} \!\ {\bm e}^2_1
+ {\rm Im}D_{y,3} \!\ {\bm e}^2_2
+ {\rm Im}E_{y,3} \!\ {\bm e}^2_3 +
{\rm Re}E_{x,3} \!\ {\bm e}^2_4,  \label{group2} \\
{\bm R}_{3} \equiv&  {\rm Re}D_{x,2} \!\ {\bm e}^3_1
+ {\rm Re}D_{y,1} \!\ {\bm e}^3_2
+  {\rm Im}D_{y,2} \!\ {\bm e}^3_3  \nn \\
& + {\rm Im}D_{x,1}  \!\ {\bm e}^3_4 +
{\rm Im}E_{y,2} \!\ {\bm e}^3_5  +
{\rm Im}E_{x,1} \!\ {\bm e}^3_6 \nn \\
& \hspace{-0.5cm}
+  {\rm Re}E_{x,2}  \!\ {\bm e}^3_7 +
{\rm Re}E_{y,1} \!\ {\bm e}^3_8 +
\frac{{\rm Im}\chi_{x+y}
+ {\rm Im}\chi_{x-y}}{\sqrt{2}}
\!\ {\bm e}^3_9 \nn \\
& \hspace{-0.5cm}
+  \frac{{\rm Im}\eta_{x+y}
+ {\rm Im}\eta_{x-y}}{\sqrt{2}} \!\ {\bm e}^3_{10}
+ \frac{{\rm Im}\chi_{x+y}
- {\rm Im}\chi_{x-y}}{\sqrt{2}}
 \!\ {\bm e}^3_{11} \nn \\
& \hspace{-0.5cm}
+ \frac{{\rm Im}\eta_{x+y}
- {\rm Im}\eta_{x-y}}{\sqrt{2}} \!\  {\bm e}^3_{12}
 +i a^{3}_{\tau} \!\ {\bm e}^3_{13} +
ia^{1}_{\tau} \!\ {\bm e}^{3}_{14}, \label{group3}  \\
{\bm R}_{4} \equiv&
\big({\rm Re}D_{x,1} - D\big) \!\ {\bm e}^4_1 +
\big({\rm Re}D_{y,2} - D \big) \!\ {\bm e}^{4}_2 \nn \\
&
+ {\rm Im}D_{y,1} \!\ {\bm e}^{4}_3
+ {\rm Im}D_{x,2} \!\ {\bm e}^{4}_{4}
+ {\rm Im}E_{y,1}  \!\ {\bm e}^{4}_5 \nn \\
&
+ {\rm Im}E_{x,2} \!\ {\bm e}^4_{6}
 + {\rm Re}E_{x,1} \!\ {\bm e}^{4}_{7} +
{\rm Re}E_{y,2} \!\ {\bm e}^{4}_{8}  \nn \\
& + ({\rm Re}\chi_{x+y}-\chi)
\!\ {\bm e}^{4}_{9}
 + ({\rm Re}\eta_{x+y} -\eta)
\!\ {\bm e}^{4}_{10}  \nn \\
& \hspace{-1.0cm} + ({\rm Re}\chi_{x+y} - \chi)
\!\ {\bm e}^{4}_{11}
+ ({\rm Re}\eta_{x+y} +\eta)
 \!\ {\bm e}^{4}_{12}
+ ia^2_{\tau} \!\ {\bm e}^4_{13},  \label{group4}
\end{align}
where $\{ {\bm e}^\mu_{\alpha} \}$ denotes the orthonormal basis
of 35-dimensional space ${\mathbb R}^{35}$ and the renamed
fluctuation fields $\{ R_{\mu,\alpha} \}$ are defined through
${\bm R}_{\mu}=\sum_{\alpha} R_{\mu,\alpha}{\bm e}^\mu_\alpha$.
In this definition, for example,
the coefficient of the base ${\bm e}^{1}_1$
corresponds to the $7$-th component of ${\bm r}$
[see Eq.~(\ref{r-def})].
Using this representation, the Gaussian part of the
action is indeed decomposed into four parts,
\begin{eqnarray}
\Big[\overline{\cal S}^{(0,2)}\Big]_{\alpha,\beta}
r_{\alpha} r_{\beta}
= \sum_{\mu=1}^4
\Big[\overline{\cal S}^{(0,2)}_{\mu\mu}\Big]_{\alpha,\beta}
R_{\mu,\alpha} R_{\mu,\beta},
\label{co0}
\end{eqnarray}
where the matrix elements inside of each block are given by
\begin{align}
&\Big[\overline{\cal S}^{(0,2)}_{\mu\mu}
({\bm q},i\epsilon_n)\Big]_{\alpha,\beta}
\equiv \frac12 \frac{\partial^2 {\cal S}_{\rm I}}{\partial r_{\mu,\alpha}
\partial r_{\mu,\beta}} +
 \frac{1}{4\beta  N_{\Lambda}}\sum_{\k,n} \nn \\
& \hspace{0cm}
\times {\rm Tr} \big[{\bm g}_0(\k+{\bm q},i\omega_n+i\epsilon_n)
{\bm v}_{\mu,\beta} \!\ {\bm g}_{0}(\k,i\omega_n)
{\bm v}_{\mu,\alpha}\big] \label{d0}
\end{align}
in the momentum representation.
The internal vertices here are also defined so that
$\sum^{35}_{\alpha=1} r_{\alpha} {\bm v}_{\alpha}
=
\sum^4_{\mu=1}
\sum_{\alpha} R_{\mu,\alpha}
{\bm v}_{\mu,\alpha}$. Explicit expressions
for $[\overline{\cal S}^{(0,2)}_{\mu\mu}]_{\alpha,\beta}$ are given in Appendix
[Eqs.(\ref{a0-1})--(\ref{xi})].

One can also see that, in the vertices $\overline{\cal S}^{(1,1)}_{\mu;\alpha} \!\ {h}_{\mu}
r_{\alpha}$,
the fluctuation fields in ${\bm R}_\mu$ with $\mu=1,2,3$ 
are, respectively, coupled only with the $\mu$-th components of the external magnetic field, $h_\mu$,
in the form
\begin{eqnarray}
\overline{\cal S}^{(1,1)}_{\mu;\alpha} \!\ {h}_{\mu}
r_{\alpha}
= \sum^3_{\mu=1}\Big[\overline{\cal S}^{(1,1)}_{\mu;\mu}\Big]_{\alpha}
h_{\mu} R_{\mu,\alpha},
\label{co1}
\end{eqnarray}
where
\begin{align}
\Big[\overline{\cal S}^{(1,1)}_{\mu;\mu}({\bm q},i\epsilon_n)\Big]_{\alpha}
&\equiv  \frac{1}{4 \beta N_{\Lambda}} \sum_{\k,n} \nn \\
& \hspace{-2cm} \times {\rm Tr} \big[{\bm g}_0(\k+{\bm q},i\omega_n+i\epsilon_n)
{\bm u}_{\mu} \!\ {\bm g}_{0}(\k,i\omega_n)
{\bm v}_{\mu,\alpha}\big], \label{d1}
\end{align}
while
all fluctuation fields enumerated
in ${\bm R}_4$ 
are
disconnected from the external magnetic fields.
Explicit expressions for $\overline{\cal S}^{(1,1)}_{\mu;\alpha}$
($\mu=1,2,3$) are
given in the Appendix [Eqs.(\ref{b0-1})--(\ref{g23})].

Using Eqs.~(\ref{d0}) and (\ref{d1}),
we finally obtain the contribution from Fig.~\ref{fig:1-loop}(a) as
\begin{widetext}
\begin{align}
\chi^{(1)}_{\mu\mu,{\rm II}}({\bm q},i\epsilon_n) =
\Big[\overline{\cal S}^{(1,1)}_{\mu;\mu}({\bm q},i\epsilon_n)
\Big]_{\alpha}
\Big[\Big(\overline{\cal S}^{(0,2)}_{\mu\mu}
({\bm q},i\epsilon_n)\Big)^{-1}
\Big]_{\alpha,\beta}
\Big[\overline{\cal S}^{(1,1)}_{\mu;\mu}
(-{\bm q},-i\epsilon_n)\Big]_{\beta} . \label{co2}
\end{align}
\end{widetext}
Here,
the vertex parts do not give rise to any finite contribution
to ${\rm Im}\chi^{(1)}({\bm q},\epsilon+i\delta)$ below
the Stoner continuum, i.e., for
$|\epsilon| < {\rm min}_{\k} (\xi_{\k+{\bm q}} + \xi_{\k})$.
This is because Eq.~(\ref{d1}) has
the same structure as the Hartree-Fock
contribution, the first term in Eq.~(\ref{HF}),
and it always has the form
\begin{eqnarray}
\Big[\overline{\cal S}^{(1,1)}_{\mu;\mu}
({\bm q},\epsilon)\Big]_{\alpha}
= \sum_{\k}
\frac{A_{\k,{\bm q},\alpha}(\epsilon)}{-\epsilon^2 +
(\xi_{\k+{\bm q}} + \xi_{\k})^2} \label{vertex1}
\end{eqnarray}
for any $\mu$ and $\alpha$, where the numerator is a
regular function of $\epsilon$ with
$[\overline{\cal S}^{(1,1)}_{\mu;\mu}(-{\bm q},-\epsilon)]_{\alpha}
=[\overline{\cal S}^{(1,1)}_{\mu;\mu}({\bm q},\epsilon)]^{*}_{\alpha}$
[see Eqs.~(\ref{fk}), (\ref{gk}), and (\ref{Sg11})--(\ref{g23})].
Hence, we can attribute any finite spectral weight in
Im$\chi^{(1)}_{\mu\mu,{\rm II}}({\bm q},\epsilon)$ below
the Stoner continuum
solely to the poles in the
RPA propagator, i.e., the
zeros of the eigenvalues of Eq.~(\ref{d0}).

Before moving to the discussions of obtained dynamical spin susceptibilities,
we briefly mention about unphysical
zero modes which are encoded in the Gaussian action, i.e., RPA propagator Eq.~(\ref{d0}).
The Gaussian
part in the static limit ($i\epsilon_n=0$) always
has three zero modes at {\it arbitrary} ${\bm q}$,
which comes from the $SU(2)$ local gauge symmetry\cite{sm}
\begin{eqnarray}
\Psi^{\dagger}_{\bm j} \rightarrow \Psi^{\dagger}_{\bm j}
e^{i \phi_{\bm j} \sigma_{\mu}}, \ \ \  \Psi_{\bm j}
\rightarrow e^{-i\phi_{\bm j}\sigma_{\mu}}\Psi_{\bm j}
\label{eqs-local-gauge}
\end{eqnarray}
with $\mu=1,2,3$.
Two zero modes with $\mu=1$ and $3$ belong
to ${\bm R}_3$, while the other ($\mu=2$) belongs to
${\bm R}_4$.
However, these excitations do not change the ground state itself;
all mean-field ansatzes which are
transformed to one another by the local gauge symmetry
should be regarded as an identical state.
Clearly, none of these modes couple with external magnetic
fields at any ${\bm q}$.
We hence regard these three zero modes as unphysical modes.
For example, the gauge transformation
with $\phi_{\bm j}=(-1)^{j_x+j_y}$ for $\mu=3$
requires that ${\bm e}^{3}_{12}$-fluctuation
becomes a zero mode at ${\bm q}=(\pi,\pi)$
in the $Z_2$ planar phase.
From Eq.~(\ref{b0-1}), one finds that
the coupling to the field vanishes as $[{\cal S}^{(1,1)}_{3;3}]_{12} h_3 R_{3,12}=0$.

\section{dynamical structure factors}

In this section, we discuss
the dynamical spin structure factors Im$\chi_{\mu\mu}({\bm q},\epsilon)$ in spin
nematic ground states obtained from
the $1/N$ expansion up to first order in
$1/N$. To find the nature of collective modes,
we analyze the next-to-leading order terms, especially
Im$\chi^{(1)}_{\mu\mu,{\rm II}}({\bm q},\epsilon)$,
at zero temperature
given by Eq.~(\ref{co2}).
For simplicity, the fluctuations of the  temporal
gauge fields, $i{a}^{\mu}_{\tau}$ ($\mu=1,2,3$), are not
included in these calculations, so that collective excitations are
comprised only of the fluctuations of auxiliary fields.
Following a standard convention, we denote the quantity
Im$\chi_{33}({\bm q},\epsilon)$ by Im$\chi_{zz}({\bm q},\epsilon)$
and Im$\chi_{11}({\bm q},\epsilon)$ by Im$\chi_{xx}({\bm q},\epsilon)$
hereafter.

In the first two subsections (Secs.~\ref{sec:4A} and \ref{sec:4B}),
we describe the characters of low-energy
collective modes and associated spectral weight of
the dynamical spin structure factors in the $Z_2$ planar phase.
Typical numerical plots of Im$\chi_{zz}({\bm q},\epsilon)$
and Im$\chi_{+-}({\bm q},\epsilon)$ are shown in
Figs.~\ref{fig:swszz1} and \ref{fig:sws+-1} respectively
for the parameter point $J_2/J_1 =1.1$.
We focus on the collective modes in the vicinity of symmetric momentum points
${\bm q}=(0,0)$, $(\pi,0)$, and $(\pi,\pi)$.
At $\Gamma$ point [${\bm q}=(0,0)$], we find three gapless $q$-linear
collective modes and also gapful modes.
These gapless modes are associated with director-wave excitations,
which are accompanied by
weak spin excitations.

In Sec.~\ref{sec:4C},
we discuss the nature of excitations
when the coupling ratio $J_2/J_1$
is changed to $J_2/J_1= J_{c,2}$, i.e., the
boundary to the neighboring $U(1)$ planar phase.
We find that
two gapped modes in Im$\chi_{zz}({\bm q},\epsilon)$ at ${\bm q}=(\pi,\pi)$
become gapless
(see Fig.~\ref{fig:swszz2}), which
corresponds to the appearance of
gapless gauge excitations (`photon-like' excitation) in the $U(1)$ phase.
We also argue the appearance of a new instability to a certain space symmetry breaking
in the $U(1)$ planar phase.

\subsection{Near $\Gamma$-point}\label{sec:4A}

\begin{figure}
   \includegraphics[width=80mm]{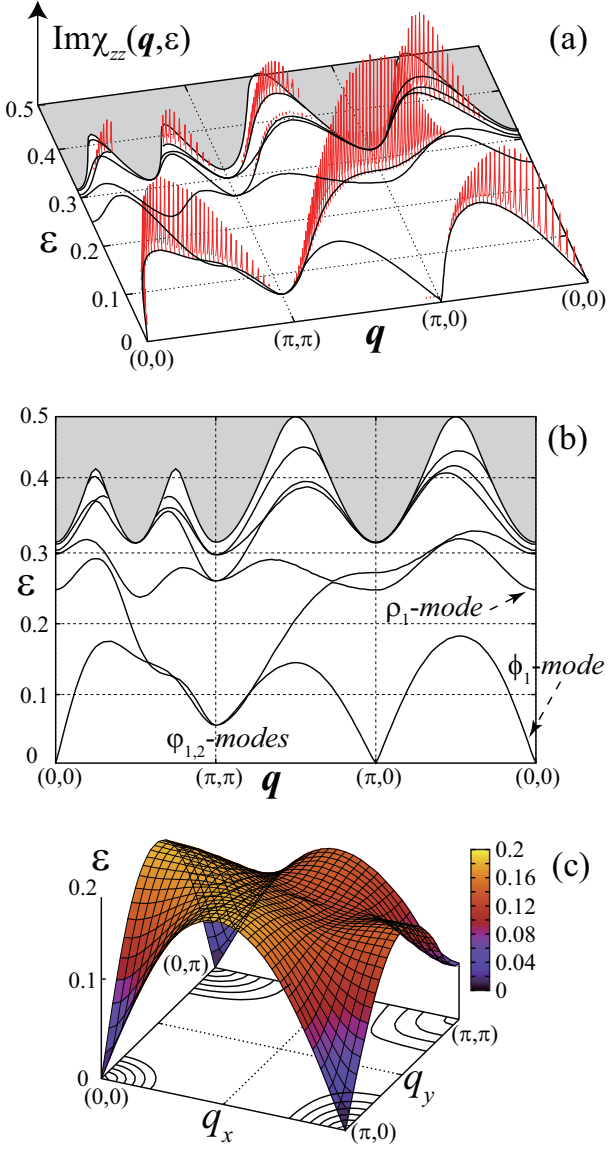}
\caption{ (Color online) Excitation energy spectrum in
the dynamical structure factor ${\rm Im}\chi_{zz}({\bm q},\epsilon)$
in the $Z_2$ planar ground state at $J_2/J_1=1.1$.
(a) Spectral wight of the collective modes.
(b) Momentum-energy dispersion relation, showing the
characteristic collective modes (${\bm \phi}_1$, ${\bm \varphi}_1$,
${\bm \varphi}_2$, ${\bm \rho}_1$) given in the text.
The momentum runs along three high symmetric $q$-points [see
Fig.~\ref{fig:HF1}(c)]. The grey zones denote Stoner continuum.
(c) Momentum-energy dispersion relation for
the lowest collective modes in the
first quadrant of the Brillouin zone.}
\label{fig:swszz1}
\end{figure}

\begin{figure}[tb]
    \includegraphics[width=80mm]{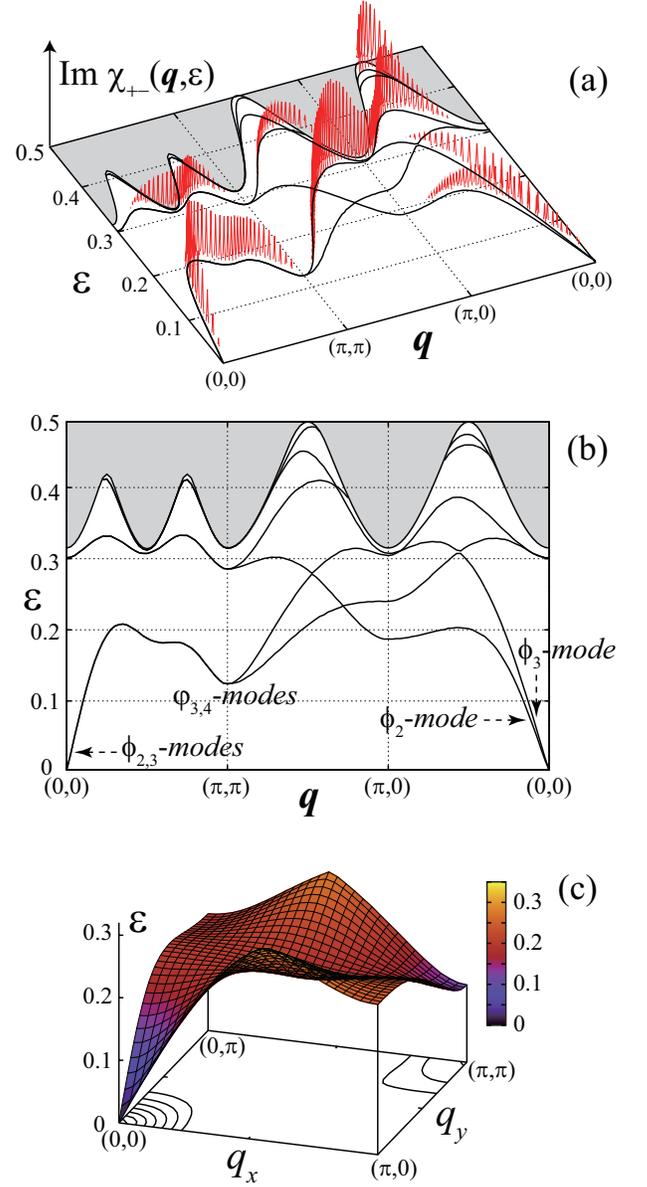}
\caption{(Color online) Excitation energy spectrum in
the dynamical structure factor ${\rm Im}\chi_{+-}({\bm q},\epsilon)$
in the $Z_2$ planar ground state at $J_2/J_1=1.1$.
(a) Spectral wight of the collective modes.
(b) Momentum-energy
dispersion relations, showing the collective modes
(${\bm \phi}_2$, ${\bm \phi}_3$, ${\bm \varphi}_3$,
${\bm \varphi}_4$) given in the text.
The grey zones denote Stoner continuum.
(c) Momentum-energy dispersion relation for
the lowest collective modes in the
first quadrant of the Brillouin zone.}
\label{fig:sws+-1}
\end{figure}

Since the $Z_2$ planar state breaks
all the spin-rotational
symmetries,
there are three gapless Nambu-Goldstone modes at the $\Gamma$-point
corresponding to the long-wavelength director-wave excitations.
The eigenmodes of the corresponding fluctuation fields take the following forms at
${\bm q}=(0,0)$: 
\begin{eqnarray}
&&\hspace{-1cm}
{\bm \phi}_1 \equiv
\frac{1}{\sqrt{2}} \!\ \big( {\bm e}^{3}_{1} - {\bm e}^{3}_2 \big),
\ \ \
{\bm \phi}_2 \equiv {\bm e}^{1}_1,
\ \ \
{\bm \phi}_3 \equiv {\bm  e}^{2}_1. \label{m-NG}
\end{eqnarray}
The ${\bm \phi}_1$-mode and  ${\bm \phi}_{2,3}$-modes appear,
respectively, as the gapless excitations 
in Im$\chi_{zz}({\bm q},\epsilon)$ and Im$\chi_{+-}({\bm q},\epsilon)$
(see Figs.~\ref{fig:swszz1} and \ref{fig:sws+-1}).
These eigenmodes induce
global rotations of nematic directors.
The gapless mode ${\bm \phi}_1$ corresponds to a director rotation
about spin 3-axis, which is generated by
$\sum_{\bm j} S_{{\bm j},3}^{aa}$. The other two gapless modes
${\bm \phi}_2$ and ${\bm \phi}_3$, respectively, correspond to director rotations
about spin 1- and 2-axes,
given by the generators $\sum_{\bm j} S_{{\bm j},\mu}^{aa}$ with $\mu=1$ and 2.
These assignments of the nature of gapless excitations are consistent
with the semi-classical argument given in Ref.~\onlinecite{sm}.
Under the mirror reflection
which exchanges both the spin $1$ and 2 axes and the space $x$ and $y$ axes,
the planar state is symmetric,
and the ${\bm \phi}_{2}$- and ${\bm \phi}_3$-modes
are interchanged.
Thus, these two
modes are energetically degenerate along the line from
$(0,0)$ to $(\pi,\pi)$,
while the degeneracy is lifted away from this
symmetric line (see Fig.~\ref{fig:sws+-1}).

In a finite-momentum regime of these gapless branches,
spin-wave excitations are also induced
together with director-wave excitations.
We can see that a fluctuation field $R_{3,1}{\bm \phi}_1$
(we set $R_{3,2}=-R_{3,1}$) in Im$\chi_{zz}({\bm q},\epsilon)$ induces
both a director rotation about spin 3-axis and a small spin displacement
$\delta S_{{\bm j},3}\sim |{\bm q}|R_{3,1}$ along spin 3-axis.
This spin excitation vanishes at the gapless point since the amplitude is proportional
to $|{\bm q}|$.
In the same way, fluctuations $R_{1,1}{\bm \phi}_2$ and $R_{2,1}{\bm \phi}_3$ in
Im$\chi_{+-}({\bm q},\epsilon)$, respectively,
induce spin displacements $\delta S_{{\bm j},1}\sim |{\bm q}|R_{1,1}$
and $\delta S_{{\bm j},2}\sim |{\bm q}|R_{2,1}$ together with director fluctuations.
These spin excitations give finite spectral weight in
dynamical spin structure factors. Near the $\Gamma$-point,
the spectral weight of these director-wave (spin-wave) modes
vanishes as a linear function of the momentum (or frequency),
\begin{align}
{\rm Im}\chi_{zz}({\bm q},\epsilon)
&= a_z \epsilon \delta (\epsilon - v_z |{\bm q}|) + \cdots,
\label{linear_z} \\
{\rm Im}\chi_{xx}({\bm q},\epsilon)
&= a_x \epsilon \delta \left(\epsilon - \sqrt{v_x^2 q_x^2+v_y^2 q_y^2}\right) + \cdots,
\label{linear_xx}
\end{align}
where $v_\mu$ ($\mu=x,y,z$) denote the director-wave velocities.
Mathematically, this is because the couplings $\bar{\cal S}^{(1,1)}_{\mu;\mu}({\bm q},\epsilon)$ between
spin-wave (director-wave) modes
and external fields have the form $\bar{\cal S}^{(1,1)}_{\mu;\mu}({\bm q},\epsilon)\sim\epsilon$
for small $\epsilon$
and the RPA propagators always  have the form
$[\bar{\cal S}^{(2)}_{\mu\mu}({\bm q},\epsilon)]^{-1}_{\alpha,\beta}
\sim\epsilon^{-1}\delta\textbf{(}\epsilon-\varepsilon_\mu({\bm q})\textbf{)}$, where
$\varepsilon_{\mu}({\bm q})$ denotes the director-wave dispersion relations.

The dispersion relation of the gapless director-wave mode in
Im$\chi_{zz}({\bm q},\epsilon)$ is spatially isotropic in a long-wavelength limit.
The dispersion relation has a $q$-linear form
$\varepsilon_{z}({\bm q})\simeq v_z |{\bm q}|$
near the $\Gamma$ point.
Contrastingly, the gapless mode in
Im$\chi_{xx}({\bm q},\epsilon)$ is anisotropic
even in the long-wavelength limit as
$\epsilon_{x}({\bm q})\simeq (v_x^2 q_x^2+v_y^2 q_y^2)^{1/2}$,
where $v_z \simeq v_x<v_y$;
both the velocity and spectral weight,
$a_x\epsilon=a_x (v_x^2 q_x^2+v_y^2 q_y^2)^{1/2}$,
are spatially anisotropic in the momentum space.
A numerical integral of Im$\chi_{zz}$ and Im$\chi_{xx}$
with respsect to $\epsilon$ for small ${\bm q}$ suggests
that $a_z$ is always greater than $a_x$ in the $Z_2$ planar
phase.

The static magnetic susceptibilities $\chi_\mu$ ($\mu=z,x$) are calculated
from the dynamical spin structure factors using the relation
$\chi_\mu=\lim_{{\bm q}\rightarrow \0}\int_0^\infty d\epsilon
{\rm Im}\chi_{\mu\mu}({\bm q},\epsilon)/\epsilon$.
The collective-mode contribution to the static susceptibility
comes only from the gapless modes in the ${\bm q}\rightarrow \0$ limit.
The result is given by $\chi_{\mu} =a_\mu/\pi$. Thus,
the numerical estimate
concludes that $\chi_z$ is always larger than $\chi_x$.
Hence in a small magnetic field, all nematic directors
are lying on the plane perpendicular to the field.

The low-energy excitations
below the Stoner continuum consist also of collective modes with finite mass
[see Fig.~\ref{fig:swszz1}(b) and
Fig.~\ref{fig:sws+-1}(b)].
In Im$\chi_{zz}({\bm q},\epsilon)$, there are several gapped eigenmodes
near the $\Gamma$-point.
Among them, the lowest gapped eigenmode ${\bm \rho}_1$ at the $\Gamma$-point contains
in-plane antiphase oscillations of two orthogonal directors.
This mode does not induce any spin excitation near $\Gamma$ point,
having no spectral weight in the dynamical spin structure factor Im$\chi_{zz}({\bm q},\epsilon)$.
This antiphase excitation is a direct
analogue of the so-called squashing modes observed
in the superfluid $^3$He-B phase.~\cite{popov,vollhardt,WP-Halperin}
The other gapped modes come from gauge fluctuations or
composite fluctuations of gauge
and director (or spin) degrees of freedom.
In Im$\chi_{+-}({\bm q},\epsilon)$,
there are two gapped eigenmodes near $\Gamma$ point,
which are degenerate at $\Gamma$ point. These
two modes are also composite fluctuations of gauge and
director degrees of freedom.

%
%
%
%

\subsection{Near $(\pi,0)$-point and $(\pi,\pi)$-point}\label{sec:4B}

The vanishing spectral weight of the
spin-wave modes at the $\Gamma$-point
is also expected in a usual antiferromagnetic phase.
In the spin-1/2 $J_1$-$J_2$ model, a collinear antiferromagnetic ordered phase with wave vector
${\bm q}=(\pi,0)$ or $(0,\pi)$ appears in the strong antiferromagnetic $J_2$ regime.
To distinguish
the $Z_2$ planar phase from the
collinear antiferromagnetic phase, we need to look into the
spin structure factor near ${\bm q}=(\pi,0)$ or $(0,\pi)$.
In the antiferromagnetic phase, low-energy excitations
are also composed of gapless spin-wave modes
at either $(\pi,0)$ or $(0,\pi)$, whose spectral weight
remains finite even at these gapless momentum points, e.g.
${\rm Im}\chi_{\mu\mu}\textbf{(}(\pi,0)+{\bm q},\epsilon\textbf{)} \simeq
b'\delta (\epsilon - u' |{\bm q}|) +\cdots$ for $|{\bm q}| \ll 1$
and $\epsilon \ll 1$. By contrast, our calculation
indicates that dynamical spin structure factors in the
spin nematic phase have no finite low-energy
weight near ${\bm q}=(\pi,0)$ and $(0,\pi)$ points, though
there exist two non-magnetic linearly-gapless modes
near these points in Im$\chi_{zz}({\bm q},\epsilon)$ [see Fig.~\ref{fig:swszz1}(a)].
The zero modes at
${\bm q}=(\pi,0)$ and $(0,\pi)$ are, respectively, given
by ${\bm e}^3_8$ and ${\bm e}^3_{7}$ modes, which correspond to Re$E_{y,1}$ and
Re$E_{x,2}$ fields.
These modes are composite
fluctuations of gauge and director degrees of freedom.
Contrary to the director-wave (spin-wave) modes at the $\Gamma$ point,
the existence of these gapless modes at ${\bm q}=(\pi,0)$ and $(0,\pi)$
are required neither by continuous
spin-rotational symmetries nor by local gauge symmetries [see Appendix B],
so that
they could likely acquire finite mass in general situations.
In the present case, the energy of a $Z_2$ planar state
with an additional small staggered mean field
Re$E_{y,1}({\bm j})=re^{i{\bm j}\cdot {\bm Q}_0}$ [${\bm Q}_0=(\pi,0)$]
is expanded as
$E_{\rm MF}(r)/N_\Lambda=E_{\rm MF}(0)/N_\Lambda+cr^4$,
which starts from a quartic term with a positive constant $c$.
We hence expect that a higher-order perturbational expansion to fourth
order in fluctuation fields
opens a gap in the excitation energy at ${\bm q}=(\pi,0)$ and $(0,\pi)$.

The energy gaps of collective modes at ${\bm q}=(\pi,\pi)$
are relevant to the stability of the $Z_2$ planar states.
The lowest gapped excitations in ${\rm Im}\chi_{zz}({\bm q},\epsilon)$
at ${\bm q}=(\pi,\pi)$
are comprised of two fluctuation fields
\begin{align}
{\bm \varphi}_1 &= i\alpha {\bm e}^3_{1}
+ \beta {\bm e}^3_{3}
- \gamma {\bm e}^3_{6}
+ i\delta
{\bm e}^3_{9}, \label{v3} \\
{\bm \varphi}_2 &= i\alpha {\bm e}^3_{2}
+ \beta {\bm e}^3_{4}
- \gamma {\bm e}^3_{5}
+ i\delta
{\bm e}^{3}_{11}. \label{v4}
\end{align}
Their momentum-energy dispersions are degenerate
at ${\bm q}=(\pi,\pi)$ [see Fig.~\ref{fig:swszz1}(b)]
due to the mirror symmetry with respect to the $(x+y)$-axis.
The lowest gapped modes in ${\rm Im}\chi_{+-}({\bm q},\epsilon)$
at ${\bm q}=(\pi,\pi)$ are comprised of two fluctuations
\begin{eqnarray}
{\bm \varphi}_3 = i\epsilon {\bm e}^2_{1} + \zeta {\bm e}^2_{2}, \ \ \
{\bm \varphi}_4 = i\epsilon {\bm e}^1_{1} + \zeta {\bm e}^1_{2}
\label{v5}
\end{eqnarray}
[see Fig.~\ref{fig:sws+-1}(b)]. These two modes are also energetically degenerate.
These four modes are composite fluctuations of gauge
and director degrees of freedom.
When the momentum approaches the symmetric point ${\bm q}=(\pi,\pi)$,
the spectral weight of these four gapped modes
vanishes as a quadratic function of the momentum, i.e.
${\rm Im}\chi_{\mu\mu}\textbf{(}(\pi,\pi)+{\bm k},\epsilon\textbf{)} \simeq
\alpha^{\prime\prime} |{\bm k}|^2
\delta (\epsilon - m-v^{\prime\prime}|{\bm k}|^2) $ for
$|{\bm k}| \ll 1$.
The vanishing of the spectral weight 
is a consequence of the staggered $U(1)$ spin-rotational
symmetry [Eq.~(\ref{staggeredU1})] in the $Z_2$ planar state.
As will be described in the
next subsection, on decreasing $J_2$, these four modes ${\bm \varphi}_i$ ($i=1,2,3,4$)
become gapless at the phase boundary $J_2/J_1=J_{c,2}$
between the adjacent $U(1)$ planar phase.

\subsection{Transition to the $U(1)$ planar state and its instability}\label{sec:4C}
In the rest of this section, we briefly discuss instabilities to the $Z_2$ planar state,
induced by energy-gap closing.
In the saddle-point solution of the $Z_2$ planar phase\cite{sm},
when the antiferromagnetic exchange $J_2$ decreases,
the $d$-wave spin-singlet pairing
amplitude $\eta$ is reduced to zero
at the critical point
$J_2/J_1 = J_{c,2} \simeq 1.0448$, while
the spin-triplet pairing amplitude $D$
and the $s$-wave excitonic pairing amplitude $\chi$
remain finite beyond the boundary,
$J_2/J_1 < J_{c,2}$.
Such a planar state in $J_2/J_1 < J_{c,2}$ is
dubbed the $U(1)$ planar state, since the ansatz
is invariant under the staggered $U(1)$ rotation
around the $3$-axis in the gauge space.

\begin{figure}[t]
\begin{center}
\includegraphics[width=70mm]{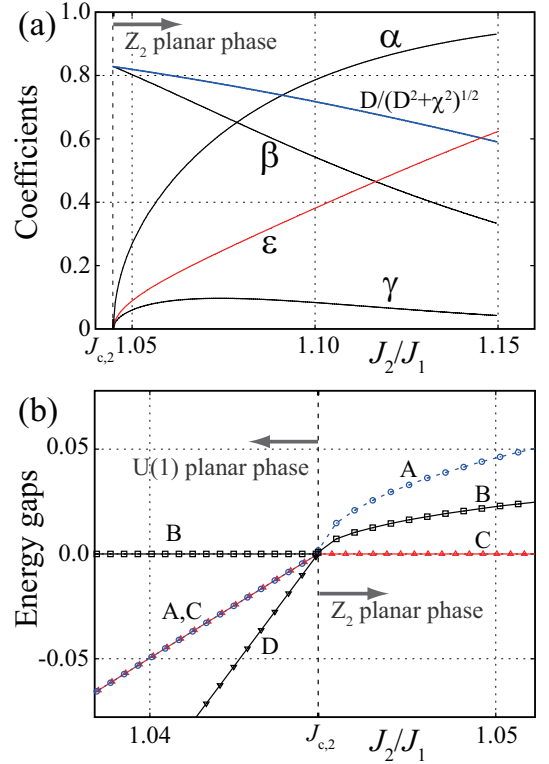}
\caption{(Color online) (a) Coefficients of eigenmodes ${\bm \varphi}_1$,
${\bm \varphi}_2$, and ${\bm \varphi}_3$ [Eqs.~(\ref{v3}), (\ref{v4}), and (\ref{v5})] as
a function of $J_2/J_1$, where
the units are taken as
$\alpha^2+\beta^2+\gamma^2+\delta^2=1$
and $\epsilon^2+\zeta^2=1$.
At the critical point $J_2/J_1=J_{c,2}$ $\alpha$,
$\gamma$ and $\epsilon$ are reduced to
zero, while
$\beta$, $\delta$ and $\zeta$,  respectively, converge to
$D/{\sqrt{D^2+\chi^2}}$,
$\chi/{\sqrt{D^2+\chi^2}}$ and
$1$. (b) Mass of eigenmodes
as a function of $J_2/J_1$. Negative energy gaps indicate the presence of instabilities.
A: Mass of ${\bm \varphi}_3$- and
${\bm \varphi}_4$-modes at ${\bm q}=(\pi,\pi)$ in the $Z_2$ planar phase
and mass of ${\bm e}^2_2$- and
${\bm e}^1_2$-modes at ${\bm q}=(\pi,\pi)$ in the $U(1)$ planar phase.
B: Mass of
${\bm \varphi}_1$- and ${\bm \varphi}_2$-modes at ${\bm q}=(\pi,\pi)$.
C: Mass of ${\bm e}^3_{8}$-mode at ${\bm q}=(\pi,0)$
and mass of ${\bm e}^3_7$-mode
at ${\bm q}=(0,\pi)$.
D: Mass of ${\bm e}^3_{12}$-mode at ${\bm q}=(\pi,\pi)$.}
\label{fig:coef-mass}
\end{center}
\end{figure}

\begin{figure}
   \includegraphics[width=75mm]{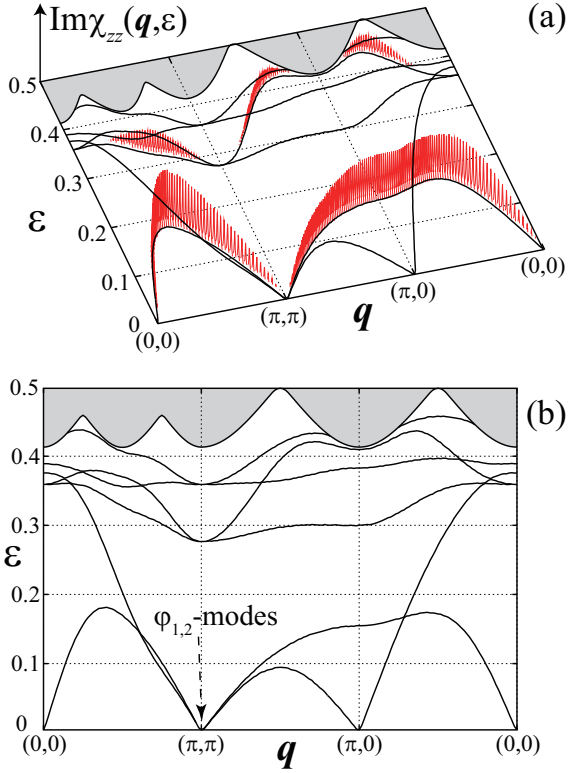}
\caption{(Color online)
Dynamical structure factor ${\rm Im}\chi_{zz}({\bm q},\epsilon)$
at the critical point $J_2/J_1=J_{c,2}= 1.0448$
between the $Z_2$ planar and
$U(1)$ planar phases.
The grey zones denote Stoner continuum.
(a)~Spectral wight of the collective modes.
(b)~Momentum-energy dispersion
relations
for the collective modes, 
showing that the lowest excitations at the $(\pi,\pi)$-point
become gapless.}
\label{fig:swszz2}
\end{figure}

\begin{figure}
   \includegraphics[width=75mm]{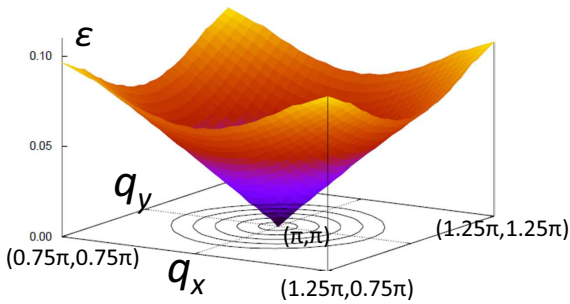}
\caption{(Color online)
Photon-like momentum-energy
dispersion in ${\rm Im}\chi_{zz}({\bm q},\epsilon)$
at the $(\pi,\pi)$ point in the $U(1)$
planar phase at $J_2/J_1=1.025 < J_{c,2}$.}
\label{fig:u1-pht}
\end{figure}

Owing to the restoration of this global $U(1)$ gauge
symmetry, the gapped fluctuations ${\bm \varphi}_i$ ($i=1,2$)
in the $Z_2$ planar state
become massless gauge fluctuations at $J_2/J_1=J_{c,2}$
[see Fig.\ \ref{fig:coef-mass}(b) and Fig.~\ref{fig:swszz2}].
To be specific, at $J_2/J_1=J_{c,2}$, the coefficients $\alpha$
and $\gamma$ in the fields ${\bm \varphi}_1$ and ${\bm \varphi}_2$
[Eqs.~(\ref{v3}) and (\ref{v4})] reduce
to zero and the ratio $\beta/\delta$ converges
to $D/\chi$ [Fig.~\ref{fig:coef-mass}(a)].
The fields have the forms
\begin{align}
{\bm \varphi}^{(c)}_1 &= D
\!\ {\bm e}^3_{3} + i \chi
\!\ {\bm e}^{3}_{9}, \label{vc1} \\
{\bm \varphi}^{(c)}_2 &= D
\!\ {\bm e}^3_{4} + i \chi
\!\ {\bm e}^{3}_{11}. \label{vc2}
\end{align}
Such fluctuation fields are induced
by generators in the gauge space;
\begin{eqnarray}
\int^{\beta}_{0} d\tau \!\
{\cal L} &=& {\cal S}_{\rm I} + \int^{\beta}_{0} d\tau
\Big\{ \!\  \frac{1}{2} \sum_{\j} {\rm Tr}
\big[\Psi^{\dag}_{\j} \partial_{\tau} \Psi_{\j}\big]
\nn \\
&& \hspace{-1cm} \ - \frac{J_1}{4}\!\ \sum_{\j} {\rm Tr}
\big[\Psi^{\dag}_{\j} \!\ D \sigma_2 \!\
e^{i (-1)^{j_x+j_y} a^3_x \!\ \sigma_3}\!\ \Psi_{{\j}+{\bm e}_x} \!\
\sigma^T_{1} \!\ \big] \nn \\
&&\hspace{-1cm}
- \frac{J_1}{4}\!\ \sum_{\j} {\rm Tr}
\big[\Psi^{\dag}_{\j} \!\ D \sigma_2 \!\
 e^{i (-1)^{j_x+j_y} a^3_y \!\ \sigma_3} \!\ \Psi_{\j+{\bm e}_y} \!\
\sigma^{T}_{2} \!\ \big] \nn \\
&& \hspace{-1.3cm}
- \!\  \frac{J_2}{4}\!\ \sum_j {\rm Tr}
\big[\Psi^{\dag}_{\j} \!\ \chi \sigma_3 \!\
e^{i (-1)^{j_x+j_y}(a^3_x + a^3_y) \sigma_3}\!\
\Psi_{\j+{\bm e}_x+{\bm e}_y}\big] \nn \\
&&\hspace{-1.6cm} - \frac{J_2}{4}\!\ \sum_{\j} {\rm Tr}
\big[\Psi^{\dag}_{\j+{\bm e}_x} \!\ \chi \sigma_3 \!\
e^{-i (-1)^{j_x+j_y}(a^3_x - a^3_y) \sigma_3}
 \!\ \Psi_{\j+{\bm e}_y}\big] \Big\}. \nn
\end{eqnarray}
Namely, an expansion with
respect to the slowly-varying gauge fields, $a^3_{x}(\j,\tau)$
and $a^3_y(\j,\tau)$, leads to
\begin{eqnarray}
\int^{\beta}_{0} d\tau \!\
{\cal L} &= & {\cal S}_{\rm I} + \frac{1}{2} \sum_{{\bm k},n}
{\bm f}^{\dagger}_{{\bm k},n}
{\bm g}^{-1}_{0}({\bm k},i\omega_n) {\bm f}_{{\bm k},n}
+ \frac{1}{\sqrt{\beta N_{\Lambda}}} \nn \\
&& \hspace{-2.0cm}
\times \sum^{35}_{\alpha=1} \sum^{|{\bm q}|\ll \pi}_{{\bm q},m}
\sum_{{\bm k},n}
{\bm f}^{\dagger}_{{\bm k},n} \!\
{\bm v}_{\alpha}\textbf{(}{\bm k},{\bm k}-(\pi,\pi)-{\bm q}\textbf{)} \!\
{\bm f}_{{\bm k}-(\pi,\pi)-{\bm q},n-m} \nn \\
&& \hspace{-0.5cm}
\times \!\
\Big\{a^{3}_{x}({\bm q},i\omega_m) \!\
{\bm \varphi}^{(c)}_{1,\alpha} + a^{3}_{y}({\bm q},i\omega_m) \!\
{\bm \varphi}^{(c)}_{2,\alpha} \Big\}, \nn
\end{eqnarray}
where ${\bm \varphi}^{(c)}_{1,\alpha}$ and
${\bm \varphi}^{(c)}_{2,\alpha}$ are exactly given by
Eqs.~(\ref{vc1}) and (\ref{vc2}). An integration over the fermion field
and subsequent expansion of the action in terms of
$a^3_{x}$ and $a^3_{y}$ leads to a quadratic form of the
effective action, $\sum_{{\bm q},i\omega_n,\alpha,\beta}
M_{\alpha\beta} ({\bm q},i\omega_n) a_{\alpha}({\bm q},i\omega_n)
a_{\beta}(-{\bm q},-i\omega_n)$. Now that the $U(1)$ planar state is
invariant under the staggered $U(1)$ rotation around the 3-axis
in the gauge space, the effective action thus obtained
is transformed into $\sum
M_{\alpha\beta}({\bm q},i\omega_n)
(a_{\alpha}+\partial_{\alpha}\theta) ({\bm q},i\omega_n)
(a_{\beta}+\partial_{\beta}\theta)(-{\bm q},-i\omega_n)$ under a
$U(1)$ local gauge transformation;
$\Psi^{\dagger}_{\j} \rightarrow
\Psi^{\dagger}_{\j} e^{i(-1)^{j_x+j_y}\theta({\bm r}) \sigma_3}$
with slowly varying function $\theta({\bm r})$.
On the one end, any physical quantities including the action
should have been invariant under any local gauge transformation,
which enforces $M_{\alpha\beta}({\bm q}=0,i\omega_n=0)$ to be
zero precisely. More accurately, it is required that
$a^3_{x}$ and $a^3_{y}$ in combination with
the staggered component of the temporal gauge field,
i.e. $a^3_{0} \equiv (-1)^{j_x+j_y}\!\ a^{3}_{\j,\tau}$,
must take a gauge invariant quadratic form as their
effective action, which turns out to be the
Maxwell form~\cite{sm}
\begin{eqnarray}
F_{\rm gauge} = \int^{\beta}_{0} d\tau \int d^2 r
\left( u {\bm E}^2 + \frac{K}{2} B^2\right).
\end{eqnarray}
The `emergent' electromagnetic fields are defined as
$E_{\alpha} \equiv \partial_{\tau} a^3_{x} - \partial_{\alpha} a^3_0$,
and  $B \equiv \partial_x a^3_y - \partial_y a^3_x$.

In the $2+1$ dimensional space, this Maxwell form
does not suppress the fluctuations of these gauge fields
efficiently, so that the $U(1)$ planar state is generally
unstable against these fluctuations. That is, the
space-time instanton which is allowed by the
corresponding compact QED action, $\int^{\beta}_0 \int d^2 r
\{ u{\bm E}^2 - K \cos (\epsilon_{\alpha\beta}
\partial_{\alpha}a^3_{\beta}) \}$, proliferate in the $2+1$
dimensional space, only to introduce strong confining
potentials between two
neutral `free' fermions (spinon).~\cite{polyakov}
In the context of spin-singlet quantum spin liquids,
it is known that resulting confining phases are
accompanied by the reduction of the space group
symmetry of original mean-field states.~\cite{read-sachdev}

In the present situation,
this symmetry reduction is driven by the
condensation of ${\bm \varphi}_{3}$-,
${\bm \varphi}_4$-, and
${\bm e}^{3}_{12}$-modes
at ${\bm q}=(\pi,\pi)$, ${\bm e}^3_{8}$-mode
at ${\bm q}=(\pi,0)$, and
${\bm e}^{3}_{7}$-mode at ${\bm q}=(0,\pi)$, where
${\bm e}^{3}_{12}$ originates from the unphysical zero
modes mentioned in the previous section.
As shown in Fig.~\ref{fig:coef-mass}(b), the mass of
these modes are all negative in the $U(1)$ phase,
$J_2/J_1<J_{c,2}$, where ${\bm \varphi}_3$- and
${\bm \varphi}_4$-modes are transformed into
${\bm e}^2_2$ and ${\bm e}^1_2$ respectively
[see Fig.~\ref{fig:coef-mass}(a)].
We thus expect that the
$U(1)$ phase is generally accompanied by
condensations of
Im$D_{x,3}$, Im$D_{y,3}$, Im$\eta_{x+y}-$Im$\eta_{x-y}$
with ${\bm q}=(\pi,\pi)$,
Re$E_{y,1}$ with ${\bm q}=(\pi,0)$, and Re$E_{x,2}$ with
${\bm q}=(0,\pi)$. Such
condensations break the time-reversal symmetry
${\cal T}$, the $\pi$ spin-rotation symmetry around
the $z$-axis ${\cal R}^{\rm spin}_{\pi,z}$ and
the translational symmetries $T_{\mu}$ $(\mu=x,y)$.
These symmetry breakings would possibly endow the
$U(1)$ phase with ferrimagnetic moments.  Having such
magnetic orderings in the background, a pair of spinon
and anti-spinon introduced in the $U(1)$
planar phase generally pay those energy
cost which are proportional to the spatial distance
between these two.~\cite{wen,fradkin,pepin}
Because of this strong confining potential, the
pair is spatially confined to each other in the $U(1)$
planar phase.

We note that this $U(1)$ planar state does not
survive as a stable ground state if the mean-field
solutions are projected to the real spin space\cite{sym}.
It is hence expected that the transition from
the $Z_2$ planar state to the $U(1)$ planar state
appears only in the large-$N$ spin model.

\subsection{Transition to the $\pi$-flux states}
When the antiferromagnetic exchange $J_2$ increases
in the $Z_2$ planar phase,
the spin-triplet pairing field $D$ decreases and vanishes at
$J_2/J_1 = J_{c,1}\simeq 1.325$, while the
other two remains almost constant. In
$J_2/J_1 \ge J_{c,1}$,
the saddle-point solution is a $\pi$-flux state having
$D=0$ and $\chi=\eta \ne 0$,
where the
Stoner excitations become gapless at five
(inequivalent) symmetric momentum
points $(0,0)$,
$(\pi/2,\pi/2)$, $(\pi,\pi)$, $(\pi,0)$ and $(0,\pi)$.
Correspondingly, all the collective excitations
and their spectral weight in the spin structure factor
merge into the lower edge of the Stoner continuum,
when $J_2/J_1$ gets closer to the critical value $J_{c,1}$
from below. 
We note that in the usual $N=1$ $S=1/2$ spin model  this $\pi$-flux
phase becomes a collinear antiferromagnetic phase\cite{sym}.
It is hence expected that the transition from the $Z_2$ planar phase
to the $\pi$-flux phase appears only
in the large-$N$ model.

\section{NMR relaxation time}
In the previous section, we have observed that
the spectral weight of  gapless spin-wave (director-wave)
modes vanishes as a linear
function of the momentum near the $\Gamma$ point.
These modes are the only magnetic low-lying excitations in the $Z_2$
planar state.
This behavior is also observed theoretically
in other kinds of spin nematic
phases~\cite{ta,smeralds} and can be regarded as a common
property of quantum spin nematics in
$d\ge 2$.~\cite{comment1}
In this section, we will calculate
the longitudinal relaxation time $T_1$ of the
nuclear magnetic resonance (NMR) in the spin nematic
phase, which also captures a low-energy property of the dynamical
spin structure factor through the relation
\begin{align}
\frac{1}{T_1} &= \frac{2\gamma^2_n T}{\hbar^2\gamma^2_e}
\lim_{\omega \rightarrow 0} \sum^3_{\mu=1}
\sum_{{\bm q}} A_{\mu}
\frac{{\rm Im}\chi_{\mu\mu}({\bm q},\omega)}{\omega} .
\end{align}
Here, $\gamma_e$ and $\gamma_n$ stand
for the gyromagnetic ratio
of electron spin and nuclear spin, respectively, and
$A_{\mu}$ ($\mu=x,y,z$) denote the form factors, which depend on
the geometry of couplings between
nuclear spins and electron spins.~\cite{moriya,ss}
We argue that the low-temperature
behavior of the NMR relaxation rate $1/T_1$ also exhibits
a characteristic temperature dependence.

When the temperature is sufficiently low,
nuclear spins relaxations are mainly attributed to
the scattering processes involved with
the gapless spin-wave modes.
There are two types of relevant scattering processes:~\cite{moriya}
one is (i) a direct process, in which a nuclear spin
is flipped by either one-magnon
emission or one-magnon absorption.
As in usual magnetic
Mott insulators, the direct process in spin nematic
phases is forbidden in usual experimental
situations, because all magnetic compounds
inevitably contain tiny spin-anisotropy
fields, such as Dzyaloshinsky-Moriya exchange
field and dipolar field, which opens a gap relatively larger than
the nuclear Larmor frequency.
In such cases,
the scattering process is dominated by (ii) the
so-called Raman process, where a nuclear spin flipping
is accompanied by the simultaneous occurrence
of one-magnon  emission and
one-magnon  absorption.

\begin{figure}
   \includegraphics[width=50mm]{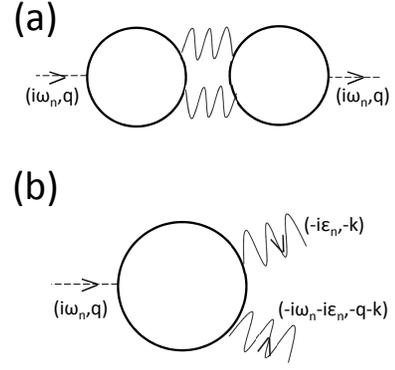}
\caption{(a) Raman scattering process. (b) Each vertex part consists
of two internal lines (wavy lines; gapless director-wave modes)
and one external line (dashed line; magnetic field).
}
\label{fig:2loop}
\end{figure}

In the framework of the $1/N$ expansion, this
Raman process can be captured by a 2-loop diagram of order $1/N^2$
which has simultaneous two fluctuation-field propagators between the two loops,
as depicted in Fig.~\ref{fig:2loop}(a).
Here, two fluctuation-field propagators (wavy lines) correspond to magnon
(director-wave) emission and absorption.
In the following, we will argue that
this Raman contribution results in a characteristic
low-temperature dependence of the
NMR relaxation rate,
\begin{equation}
\frac{1}{T_1} = a T^{2d-1} + \cdots
\label{t1invb0}
\end{equation}
[see Eq.~(\ref{t1invb})], where $d$ denotes the (effective)
spatial dimension.~\cite{comment2}

\begin{table}[btp]
\begin{tabular}{c||c|c|c|c|c|c}
& $h_x$ & $h_y$ & $h_z$ & ${\phi}_1$
& ${\phi}_2$ & ${\phi}_3$ \\ \hline \hline
${\cal T}$ & $-h_x$ & $-h_y$ & $-h_z$ & ${\phi}_1$
& ${\phi}_2$ & ${\phi}_3$ \\ \hline
$\sigma_x$& $h_x$ & $h_y$ & $h_z$ & ${\phi}_1$
& ${\phi}_2$ & ${\phi}_3$ \\ \hline
$\sigma_y$& $h_x$ & $h_y$ & $h_z$ & ${\phi}_1$
& ${\phi}_2$ & ${\phi}_3$ \\ \hline
$R_{\frac{\pi}{2},z}$ & $h_y$ & $-h_x$ & $h_z$
 & ${\phi}_1$ & ${\phi}_3$ & ${\phi}_2$ \\ \hline
$R^{\rm spin}_{\pi,z}$& $-h_x$ & $-h_y$ & $h_z$
& ${\phi}_1$ & $-{\phi}_2$ & $-{\phi}_3$ \\ \hline
 \end{tabular}
\caption{Transformation properties of gapless director-wave modes
under time-reversal ${\cal T}$, mirror operations $\sigma_{\mu}$
with respect to the $\mu$-axis,
and rotations in spin and lattice space.
$R_{\frac{\pi}{2},z}\equiv R^{\rm spin}_{\frac{\pi}{2},z} \!\
R^{\rm lat}_{\frac{\pi}{2},z}$, where $R^{\rm spin/lat}_{\theta,z}$
denotes the spin(lattice)-rotation by $\theta$
around the $z$-axis.
Note that ${\cal T}$, $\sigma_x$, $\sigma_y$,
$R_{\frac{\pi}{2},z}$, and $R^{\rm spin}_{\pi,z}$ are, respectively, accompanied by
proper gauge transformations,
$\Psi^{\dagger}_{\j} \rightarrow \Psi^{\dagger}_{\j} (-1)^{j_x+j_y}$,
$\Psi^{\dagger}_{\j} \rightarrow \Psi^{\dagger}_{\j} \sigma_1(-1)^{j_x}$,
$\Psi^{\dagger}_{\j} \rightarrow \Psi^{\dagger}_{\j} \sigma_1 (-1)^{j_y}$,
$\Psi^{\dagger}_{\j} \rightarrow \Psi^{\dagger}_{\j} \sigma_1 (-1)^{j_y}$,
and $\Psi^{\dagger}_{\j} \rightarrow \Psi^{\dagger}_{\j} (-1)^{j_x+j_y}$.}
\label{table:1}
\end{table}

The Raman process shown in Fig.~\ref{fig:2loop}(a) consists of
two vertex parts ${\cal \bar{S}}^{(1,2)}$, each of which has one
external (dashed) line representing magnetic field and two
internal (wavy) lines representing the fluctuating modes.
Among the fluctuating modes, the gapless director-wave modes (${\bm \phi}_1$,
${\bm \phi}_2$, and ${\bm \phi}_3$) with low frequencies only
contribute to the relaxation rate
at sufficiently low temperature.
We first determine a form of the vertex part $\overline{\cal S}^{(1,2)}$
associated with these gapless modes by using a  symmetry argument.
Under the time-reversal transformation combined with
a proper gauge transformation $\Psi^{\dagger}_{\j}
\rightarrow \Psi^{\dagger}_{\j}(-1)^{j_x+j_y}$ and
$\Psi_{\j} \rightarrow (-1)^{j_x+j_y} \Psi_{\j}$,
none of the director-wave modes changes its sign,
while external magnetic field changes the sign.
Since the interaction ${\cal \bar{S}}^{(1,2)}h_\mu\phi_j\phi_m$
respects the time-reversal symmetry,
the vertex part ${\cal \bar{S}}^{(1,2)}$ is an odd function of frequency.
Expanding the vertex part ${\cal \bar{S}}^{(1,2)}$
of Fig.~\ref{fig:2loop}(b)
with frequencies $i\omega_n$ and $i\epsilon_n$,
we obtain a linear function of
\begin{eqnarray}
i\omega_n h_{\mu}(i\omega_n,{\bm q}) \phi_{j}(i\epsilon_n,{\bm k})
\phi_{m}(-i\epsilon_n-i\omega_n, -{\bm q}-{\bm k}) \label{vertexa}
\end{eqnarray}
and
\begin{eqnarray}
i\epsilon_n h_{\mu}(i\omega_n,{\bm q}) \phi_{j}(i\epsilon_n,{\bm k})
\phi_{m}(-i\epsilon_n-i\omega_n, -{\bm q}-{\bm k}), \label{vertexb}
\end{eqnarray}
where $\mu=x,y,z$ and $j,m=1,2,3$, in the leading term.
Since third- and higher-order time-derivative terms
lead to subleading contributions in the
relaxation rate, we will consider
only the first-order time-derivative terms.
Transformation properties of the
gapless director-wave modes under other
symmetry operations (see Table.~\ref{table:1}) further
restrict allowed
combinations of $\mu$, $j$ and $m$ in
Eqs.~(\ref{vertexa}) and (\ref{vertexb}) to the following sets
\begin{align}
&i\omega_n h_z \phi_1 \phi_1, \ \ i\omega_n h_z \phi_2 \phi_2, \ \
i\omega_n h_z \phi_3 \phi_3, \ \ i\omega_n h_z \phi_2 \phi_3, \nn \\
&i\epsilon_n \big(h_x \phi_1 \phi_2 - h_y \phi_1 \phi_3\big), \ \
(i\epsilon_n +i\omega_n)
\big(h_x \phi_1 \phi_2 - h_y \phi_1 \phi_3\big),  \nn \\
&i\epsilon_n \big(h_x \phi_1 \phi_3 - h_y \phi_1 \phi_2\big), \ \
(i\epsilon_n +i\omega_n) \big(h_x \phi_1 \phi_3 - h_y \phi_1 \phi_2\big), \nn
\end{align}
where we have omitted momentum and frequency arguments in
$h_{\mu}$ and $\phi_i$.

Taking the contractions between the two internal lines that
connects two vertex parts in
Fig.~\ref{fig:2loop}(a),
we can see that this Raman contribution always takes the following
form
\begin{align}
&\chi^{\rm ram}_{\mu\mu}({\bm q},i\omega_n)
= \nn \\
& \ \ \frac{1}{\beta}\sum_{{\bm k}}\sum_{\epsilon_n}
\frac{\alpha_{\mu}\omega^2_n+\beta_{\mu}
(\omega_n+\epsilon_n)\epsilon_n +
\gamma_{\mu}\omega_n\epsilon_n }{[(\omega_n+\epsilon_n)^2
+ \varepsilon^2_1] (\epsilon^2_n + \varepsilon^2_2)}, \nn
\end{align}
where $\varepsilon_1 \equiv \varepsilon_{-{\bm k}-{\bm q}}$ and
$\varepsilon_2 \equiv \varepsilon_{\bm k}$ correspond to
the energies of the gapless director-wave modes.
For simplicity, we assumed that the momentum-energy
dispersions of these three linearly-gapless modes
are all the same, while their difference
does not change the conclusion on the leading temperature dependence, i.e.,
Eq.~(\ref{t1invb0}). The coefficients $\alpha_{\mu}$, $\beta_{\mu}$,
and $\gamma_{\mu}$ ($\mu=x,y,z$) are determined
by specific microscopic evaluations of the vertex part. As far
as the leading-order contribution is
concerned, we can treat these coefficients as constants
independent of momentum and frequency.

With the analytic continuation, $i\omega_n \rightarrow
\omega+i\delta$, the Raman contribution to
the imaginary part of the dynamical susceptibilities
can be calculated as
\begin{align}
{\rm Im}&\chi^{\rm ram}_{\mu\mu}({\bm q},\omega)
= {\rm Im}\chi^{\rm ram}_{\mu\mu}
({\bm q},i\omega_n =\omega+i\delta)  \nn \\
=& \sum_{\bm k}
\left(\frac{\alpha_{\mu} \pi \omega^2}{4 \varepsilon_1 \varepsilon_2}
+\frac{\beta_{\mu} \pi}{4}
+\frac{\gamma_{\mu} \pi \omega}{4 \varepsilon_1}
\right)
[ n^{\rm B}(\varepsilon_1)-n^{\rm B}(\varepsilon_2) ]
 \nn \\
& \times \delta(\omega-\varepsilon_1+\varepsilon_2) \nn \\
& - \sum_{\bm k}
\left(\frac{\alpha_{\mu} \pi \omega^2}{4 \varepsilon_1 \varepsilon_2}
+\frac{\beta_{\mu} \pi}{4}
-\frac{\gamma_{\mu} \pi \omega}{4 \varepsilon_1}\right)
[ n^{\rm B}(\varepsilon_1)-n^{\rm B}(\varepsilon_2) ]
 \nn \\
& \hspace{5mm}\times \delta(\omega+\varepsilon_1-\varepsilon_2), \nn
\end{align}
where $n^{\rm B}(\varepsilon)$ denotes the Bose distribution
function, $n^{\rm B}(\varepsilon)= (e^{\beta \varepsilon}-1)^{-1}$.
Since $\omega$ will be replaced by zero (tiny
nuclear Ramor frequency),  we have dropped those terms which
are proportional to either $\delta(\omega-\varepsilon_1-\varepsilon_2)$
or $\delta(\omega+\varepsilon_1+\varepsilon_2)$, while
keeping those terms which are proportional to either  $\delta(\omega-\varepsilon_1+\varepsilon_2)$
or $\delta(\omega+\varepsilon_1-\varepsilon_2)$.
In terms of density of state for the gapless mode
defined as $N(\omega)=\sum_{\bm k} \delta(\omega -\varepsilon_{{\bm k}})$,
we obtain an expression for the NMR relaxation rate as
\begin{align}
\frac{1}{T_1}
&=\lim_{\omega \rightarrow 0}
\frac{\pi \gamma^2_n T}{\hbar^2 \gamma^2_e \omega} \nn \\
&\times\int^{\Lambda}_{0} d\Omega
\bigg\{
[n^{\rm B}(\omega+\Omega) - n^{\rm B}(\Omega) ] \!\
N(\Omega)N(\Omega+\omega)  \nn \\
& \hspace{1cm} \times
\bigg(\frac{{\bm A}\cdot {\bm \alpha} \!\
 \omega^2}{(\omega+\Omega)\Omega}
+ {\bm A}\cdot {\bm \beta} +
\frac{{\bm A}\cdot {\bm \gamma} \!\ \omega}{ (\omega+\Omega)}\bigg)
\nn \\
&\hspace{1cm} -
[n^{\rm B}(-\omega+\Omega) - n^{\rm B}(\Omega) ] \!\
N(\Omega)N(\Omega-\omega)  \nn \\
& \hspace{1cm} \times
\bigg( \frac{{\bm A}\cdot {\bm \alpha} \!\
 \omega^2}{(-\omega+\Omega)\Omega}
+ {\bm A}\cdot {\bm \beta} -
\frac{{\bm A}\cdot {\bm \gamma} \!\
\omega}{ (-\omega+\Omega)}\bigg)
\bigg\}, \label{t1inv}
\end{align}
where ${\bm A}=(A_{x}, A_{y}, A_{z})$, ${\bm \alpha}=(\alpha_{x}, \alpha_{y}, \alpha_{z})$,
${\bm \beta}=(\beta_{x}, \beta_{y}, \beta_{z})$, and
${\bm \gamma}=(\gamma_{x}, \gamma_{y}, \gamma_{z})$.
We assume that the form factors $A_{\mu}$ $(\mu=x,y,z)$ have
neither temperature dependence nor momentum dependence.
Since $N(\Omega)\propto \Omega^{d-1}$ in $d$ dimensions,
we can evaluate the temperature dependence of
the relaxation rate in the low-temperature limit
as
\begin{eqnarray}
\frac{1}{T_1} = \frac{2\pi \!\ \gamma^2_n \!\ T^{2d-1}}{\hbar^2\gamma^2_e}
\int^{\infty}_{0} dX f(X), \label{t1invb}
\end{eqnarray}
where $f(X)$ is free from the temperature and given by
\begin{align}
f(X) &= \frac{d}{dx} \bigg\{ [\overline{n}^{\rm B}(x+X)
- \overline{n}^{\rm B}(X)] \!\ N(X) N(X+x) \nn \\
&\hspace{0.5cm} \times
\left(\frac{{\bm A}\cdot {\bm \alpha} \!\
 x^2}{(x+X)X}
+ {\bm A}\cdot {\bm \beta} +
\frac{{\bm A}\cdot {\bm \gamma} \!\ x}{(x+X)}
\right)\bigg\}\bigg|_{x=0} \nn
\end{align}
with
\begin{align}
\overline{n}^{\rm B}(x)= \frac{1}{e^{x}-1}. \nn
\end{align}

\section{summary}
We have studied dynamical properties of a spin nematic state called $Z_2$ planar
state in a generalized $N$-flavor spin-1/2 $J_1$-$J_2$ model on the square lattice.
In the large-$N$ limit, the $Z_2$ planar state is the ground state in a finite
parameter range\cite{sm} in which ferromagnetic
coupling $J_1$ competes strongly with the antiferromagnetic coupling $J_2$.
The $Z_2$ planar state has the completely same magnetic properties,\cite{sym}
including an antiferro-quadrupolar order, as
the $d$-wave spin nematic state proposed in the spin-1/2 $J_1$-$J_2$ model
($N=1$) on the square lattice. 
Using the standard $1/N$ expansion,
we have calculated the dynamical spin-structure factors up to order of $1/N$
in this quantum spin nematic state for large $N$.

The obtained dynamical spin structure factors
have two characters; they
have both a spin-liquid like character and a symmetry-broken phase character.
The former feature is represented by the so-called
Stoner continuum of individual excitations of gapped neutral fermions (spinons).
Due to the existence of an antiferro-quadrupolar order,
the dynamical spin structure factors also acquire
coherent peaks below the continuum, which
signifies the existence of gapless director-wave (spin-wave)
collective modes. These director-wave modes have linear dispersions with
respect to the momentum in the long-wavelength limit. These director fluctuations
are accompanied with weak spin excitations and hence
they have a finite spectral weight in the dynamical spin structure factors,
which is proportional to the momentum, e.g.
Im$\chi_{zz}({\bm q},\epsilon)\simeq a_z v_z |{\bm q}|\delta(\epsilon-v_z|{\bm q}|)$.
A careful analysis revealed that these $q$-linear modes are the only magnetic low-energy excitations
in this spin nematic state.
Accordingly, temperature
dependence of (magnetic contributions to)
the specific heat in the present spin
nematic phase can be evaluated as a
quadratic function of temperature, $C_{v} \sim T^2$, in the
low-$T$ regime, while that of
the NMR relaxation rate is evaluated
as $T^{-1}_{1} \sim T^{3}$.
The latter unusual behavior of the NMR relaxation rate
was also discussed in an antiferro-quadrupolar phase
in a spin-1 bilinear-biquadratic model.\cite{ta}

The lowest gapped excitations around ${\bm q}=(\pi,\pi)$ are identified as a
certain kind of Higgs bosons, whose finite
mass quantifies the stability of the present $Z_2$ planar
phase against the `confinement effect'.
Though these massive modes are gauge-like excitations, they have finite spectral
weight in the dynamical {\it spin} structure factor,
once the momentum is deviated from
${\bm q}=(\pi,\pi)$. Thus the mass can
be experimentally measured with
inelastic neutron scattering experiments.
When these Higgs bosons lose their
mass, which is the case near the
ferromagnetic phase boundary for large $N$,~\cite{sm}
a `linear' confining potential should be
introduced between two neutral fermions and
the $Z_2$ planar phase is transformed into another
phase having {\it no} gapped free spinon.
We also found that, at this transition point,
a couple of other gapped bosonic
modes at high symmetric momentum points
simultaneously exhibit instabilities,
which break the time-reversal
symmetry, a spin-$\pi$-rotational
symmetry, and the translational symmetries
of the square lattice.

\begin{acknowledgments}
We acknowledge Masahiro Sato, Kazutaka Takahashi,
Akira Furusaki, Andrey Chubukov,
Leon Balents, Kimitoshi Kono, Nic Shannon, and
Andrew Smerald for helpful discussions. We also thank to
Sebastien Burdin for his insightful comment
on the nature of the $U(1)$ planar phase.
RS was partially supported by the Institute
of Physical and Chemical Research (RIKEN).
This work was supported by Grants-in-Aid for Scientific Research
from MEXT, Japan (No.\ 22014016 and No.\ 23540397).
\end{acknowledgments}

\ \

\appendix
\section{Explicit expressions for Eqs.~(\ref{d0}) and (\ref{d1})}
The RPA propagators $\overline{\cal S}^{(0,2)}_{jj}(q,i\epsilon_n)$
$(j=1,2,3)$, defined in Eq.~(\ref{d0}),
are calculated as follows:
\begin{widetext}
\begin{eqnarray}
\overline{\cal S}^{(0,2)}_{33} \equiv
\bordermatrix{
& {\bm e}^3_{1} & {\bm e}^3_{2} &
{\bm e}^3_{3} & {\bm e}^3_{4} &
{\bm e}^3_{5} & {\bm e}^3_{6} &
{\bm e}^3_{7} & {\bm e}^3_{8} &
{\bm e}^3_{9} & {\bm e}^3_{10} &
{\bm e}^3_{11} & {\bm e}^3_{12} \cr
{\bm e}^{3}_1 & \alpha_1 & \beta_{1,2} & \beta_{1,3} & & \beta_{1,5} &
& \beta_{1,7} & & \beta_{1,9} & & \beta_{1,11} & \cr
{\bm e}^{3}_2 & \beta_{1,2} & \alpha_{2} & & \beta_{2,4}  & & \beta_{2,6}
& & \beta_{2,8} & \beta_{2,9} & & \beta_{2,11} & \cr
{\bm e}^{3}_3 & -\beta_{1,3} & &\alpha_{3} & \beta_{3,4} & \beta_{3,5} &
& \beta_{3,7} & & \beta_{3,9} &\beta_{3,10} & \beta_{3,11} & \beta_{3,12} \cr
{\bm e}^3_4 & & -\beta_{2,4} & \beta_{3,4} & \alpha_4 & & \beta_{4,6} &
& \beta_{4,8} & \beta_{4,9} & \beta_{4,10} & \beta_{4,11} &  \beta_{4,12} \cr
{\bm e}^{3}_5 & -\beta_{1,5} & & \beta_{3,5} & & \alpha_{5} & \beta_{5,6}
&  \beta_{5,7} & & \beta_{5,9} & \beta_{5,10} & \beta_{5,11} & \beta_{5,12} \cr
{\bm e}^3_6 & & -\beta_{2,6}  & & \beta_{4,6} & \beta_{5,6} & \alpha_{6} &
&\beta_{6,8} & \beta_{6,9} &\beta_{6,10} & \beta_{6,11} & \beta_{6,12} \cr
{\bm e}^{3}_7 & -\beta_{1,7} & & \beta_{3,7} &  &\beta_{5,7} & & \alpha_{7}
& \beta_{7,8} &  & \beta_{7,10} & & \beta_{7,12} \cr
{\bm e}^{3}_8 &  & -\beta_{2,8} & & \beta_{4,8}  & & \beta_{6,8} & \beta_{7,8}
& \alpha_{8} & & \beta_{8,10} &  & \beta_{8,12} \cr
{\bm e}^3_{9} & \beta_{1,9} & \beta_{2,9} &-\beta_{3,9} & -\beta_{4,9} & -\beta_{5,9}
& -\beta_{6,9} & & & \alpha_{9} &\beta_{9,10} & \beta_{9,11} & \beta_{9,12} \cr
{\bm e}^{3}_{10} & &  &\beta_{3,10} & \beta_{4,10} & \beta_{5,10} & \beta_{6,10}
& \beta_{7,10} & \beta_{8,10} & -\beta_{9,10} &\alpha_{10} & \beta_{10,11} & \beta_{10,12} \cr
{\bm e}^{3}_{11} & \beta_{1,11} & \beta_{2,11} &-\beta_{3,11} & -\beta_{4,11}
& -\beta_{5,11} & -\beta_{6,11} & & & \beta_{9,11} & -\beta_{10,11} & \alpha_{11} & \beta_{11,12} \cr
{\bm e}^{3}_{12} & &  &\beta_{3,12} & \beta_{4,12} & \beta_{5,12} & \beta_{6,12}
& \beta_{7,12} & \beta_{8,12} & -\beta_{9,12} &\beta_{10,12} & -\beta_{11,12} & \alpha_{12} \cr
},   \label{a0-1}
\end{eqnarray}
\end{widetext}
\begin{eqnarray}
\overline{\cal S}^{(0,2)}_{11} \equiv
\bordermatrix{
& {\bm e}^1_{1} & {\bm e}^1_{2} &
{\bm e}^1_{3} & {\bm e}^1_{4}  \cr
{\bm e}^{1}_1 & \alpha^\prime_1 &
\beta^{\prime}_{1,2} & \beta^{\prime}_{1,3} & \beta^{\prime}_{1,4} \cr
{\bm e}^{1}_2 & -\beta^{\prime}_{1,2}
& \alpha^{\prime}_{2} & \beta^{\prime}_{2,3} & \beta^{\prime}_{2,4} \cr
{\bm e}^{1}_3
& -\beta^{\prime}_{1,3} & \beta^{\prime}_{2,3}
&\alpha^{\prime}_{3} & \beta^{\prime}_{3,4} \cr
{\bm e}^{1}_4 &  -\beta^{\prime}_{1,4}
& \beta^{\prime}_{2,4} & \beta^{\prime}_{3,4}
& \alpha^{\prime}_4  \cr }, \label{a0-2}
\end{eqnarray}
and
\begin{eqnarray}
\overline{\cal S}^{(0,2)}_{22} \equiv
\bordermatrix{
& {\bm e}^2_{1} & {\bm e}^2_{2} &
{\bm e}^2_{3} & {\bm e}^2_{4}  \cr
{\bm e}^{2}_1 & \alpha^{\prime\prime}_1 &
\beta^{\prime\prime}_{1,2} & \beta^{\prime\prime}_{1,3}
& \beta^{\prime\prime}_{1,4} \cr
{\bm e}^{2}_2 & -\beta^{\prime\prime}_{1,2}
& \alpha^{\prime\prime}_{2}
& \beta^{\prime\prime}_{2,3} & \beta^{\prime\prime}_{2,4} \cr
{\bm e}^{2}_3
& -\beta^{\prime\prime}_{1,3} & \beta^{\prime\prime}_{2,3}
&\alpha^{\prime\prime}_{3} & \beta^{\prime\prime}_{3,4} \cr
{\bm e}^{2}_4 &  -\beta^{\prime\prime}_{1,4}
& \beta^{\prime\prime}_{2,4} & \beta^{\prime\prime}_{3,4}
& \alpha^{\prime\prime}_4  \cr }. \label{a0-3}
\end{eqnarray}
Respective matrix elements are calculated as follows:
\begin{align}
\alpha_{1} &= \frac{1}{N_{\Lambda}} \sum_{\bm k}
s^2_x \big\{h_{\bm k} - ({\bm \xi}_{+}{\bm \xi}_{-}
- 2a_{5,+}a_{5,-}) f_{\bm k} \big\}  \nn \\
&\hspace{0.6cm}
- \frac{1}{N_{\Lambda}}
\sum_{\bm k} s^2_x \big\{h^0_{\bm k}
- \xi^2 f^0_{\bm k}\big\}, \label{a1} \\
%
\alpha_{2} &= \frac{1}{N_{\Lambda}} \sum_{\bm k}
s^2_y \big\{h_{\bm k} - ({\bm \xi}_{+}{\bm \xi}_{-}
- 2a_{3,+}a_{3,-}) f_{\bm k} \big\} \nn \\
&\hspace{0.6cm}
- \frac{1}{N_{\Lambda}}\sum_{\bm k} s^2_y \big\{h^0_{\bm k}
- \xi^2 f^0_{\bm k}\big\}, \label{a2}
\end{align}
\begin{align}
\alpha_{3} &= \frac{1}{N_{\Lambda}} \sum_{\bm k}
s^2_y \big\{h_{\bm k} - ({\bm \xi}_{+}{\bm \xi}_{-}
- 2a_{2,+}a_{2,-} \nn \\
&\hspace{0.3cm}
- 2a_{3,+} a_{3,-}) f_{\bm k} \big\}
- \frac{1}{N_{\Lambda}}\sum_{\bm k}
s^2_y \big\{h^0_{\bm k}
- \xi^2 f^0_{\bm k}\big\}, \label{a3} \\
%
\alpha_{4} &= \frac{1}{N_{\Lambda}} \sum_{\bm k}
s^2_x \big\{h_{\bm k} - ({\bm \xi}_{+}{\bm \xi}_{-}
- 2a_{2,+}a_{2,-} \nn \\
&\hspace{0.3cm}
 - 2a_{5,+}a_{5,-}) f_{\bm k} \big\}
- \frac{1}{N_{\Lambda}}\sum_{\bm k} s^2_x
\big\{h^0_{\bm k}
- \xi^2 f^0_{\bm k}\big\}, \label{a4}
\end{align}
\begin{align}
\alpha_{5} &= \frac{1}{N_{\Lambda}} \sum_{\bm k}
s^2_y \big\{h_{\bm k} - ({\bm \xi}_{+}{\bm \xi}_{-}
- 2a_{3,+}a_{3,-} \nn \\
&\hspace{0.3cm}
- 2a_{4,+}a_{4,-}) f_{\bm k} \big\}
- \frac{1}{N_{\Lambda}}\sum_{\bm k} s^2_y
\big\{h^0_{\bm k}
- \xi^2 f^0_{\bm k}\big\}, \label{a5} \\
\alpha_{6} &= \frac{1}{N_{\Lambda}} \sum_{\bm k}
s^2_x \big\{h_{\bm k} - ({\bm \xi}_{+}{\bm \xi}_{-}
- 2a_{4,+}a_{4,-} \nn \\
&\hspace{0.3cm} - 2a_{5,+}a_{5,-}) f_{\bm k} \big\}
- \frac{1}{N_{\Lambda}}\sum_{\bm k} s^2_x \big\{h^0_{\bm k}
- \xi^2 f^0_{\bm k}\big\}, \label{a6}
\end{align}
\begin{align}
\alpha_{7} &= \frac{1}{N_{\Lambda}} \sum_{\bm k}
c^2_x \big\{h_{\bm k} - ({\bm \xi}_{+}{\bm \xi}_{-}
- 2a_{2,+}a_{2,-} - 2a_{4,+} a_{4,-} \nn \\
&\hspace{0.3cm}
- 2a_{5,+}a_{5,-}) f_{\bm k} \big\}
- \frac{1}{N_{\Lambda}}\sum_{\bm k} s^2_x
\big\{h^0_{\bm k}
- \xi^2 f^0_{\bm k}\big\}, \label{a7} \\
\alpha_{8} &= \frac{1}{N_{\Lambda}} \sum_{\bm k}
c^2_y \big\{h_{\bm k} - ({\bm \xi}_{+}{\bm \xi}_{-}
- 2a_{2,+}a_{2,-} - 2a_{3,+} a_{3,-} \nn \\
&\hspace{0.3cm} - 2a_{4,+}a_{4,-}) f_{\bm k} \big\}
- \frac{1}{N_{\Lambda}}\sum_{\bm k} s^2_y \big\{h^0_{\bm k}
- \xi^2 f^0_{\bm k}\big\}, \label{a8}
\end{align}
%
where $\frac{J^2_1}{4}$ were omitted from the overall factors
in the right hand sides.
%
\begin{align}
\alpha_{9} &= \frac{2}{N_{\Lambda}} \sum_{\bm k}
 c^2_y \big\{
s^2_x (h_{\bm k} +  {\bm \xi}_{+}{\bm \xi}_{-}
f_{\bm k} )
-  c^2_x (h^0_{\bm k}
- \xi^2 f^0_{\bm k})\big\}, \label{a9} \\
\alpha_{11} &= \frac{2}{N_{\Lambda}}
\sum_{\bm k}  c^2_x \big\{
s^2_y (h_{\bm k} +  {\bm \xi}_{+}{\bm \xi}_{-}
f_{\bm k} )
-  c^2_y (h^0_{\bm k}
- \xi^2 f^0_{\bm k})\big\}, \label{a10}
\end{align}
\begin{align}
\alpha_{10} &= \frac{1}{N_{\Lambda}} \sum_{\bm k}
2c^2_xc^2_y \big\{h_{\bm k} - ({\bm \xi}_{+}{\bm \xi}_{-}
- 2a_{3,+}a_{3,-} \nn \\
&\hspace{0.3cm}
- 2a_{5,+} a_{5,-}) f_{\bm k} \big\}
- \frac{1}{N_{\Lambda}}\sum_{\bm k} 2c^2_xc^2_y
\big\{h^0_{\bm k}
- \xi^2 f^0_{\bm k}\big\}, \label{a11} \\
\alpha_{12} &= \frac{1}{N_{\Lambda}} \sum_{\bm k}
2s^2_x s^2_y \big\{h_{\bm k} - ({\bm \xi}_{+}{\bm \xi}_{-}
- 2a_{3,+}a_{3,-} \nn \\
&\hspace{0.3cm}
- 2a_{5,+}a_{5,-}) f_{\bm k} \big\}
- \frac{1}{N_{\Lambda}}\sum_{\bm k} 2s^2_x s^2_y \big\{h^0_{\bm k}
- \xi^2 f^0_{\bm k}\big\}, \label{a12}
\end{align}
where $\frac{J^2_2}{4}$ were omitted from the overall factors
in the right hand sides.
\begin{align}
\alpha^{\prime}_{1} &= \frac{1}{N_{\Lambda}} \sum_{\bm k}
s^2_y \big\{h_{\bm k} - {\bm \xi}_{+}{\bm \xi}_{-}
f_{\bm k} - h^0_{\bm k}
+ \xi^2 f^0_{\bm k}\big\}, \label{ad1}
\end{align}
\begin{align}
\alpha^{\prime}_{2} &= \frac{1}{N_{\Lambda}} \sum_{\bm k}
s^2_x \big\{h_{\bm k} - ({\bm \xi}_{+}{\bm \xi}_{-}
- 2a_{2,+}a_{2,-} - 2a_{3,+}a_{3,-} \nn \\
&\hspace{0.3cm}
- 2a_{5,+} a_{5,-} ) f_{\bm k} \big\}
- \frac{1}{N_{\Lambda}}\sum_{\bm k} s^2_x \big\{h^0_{\bm k}
- \xi^2 f^0_{\bm k}\big\}, \label{ad2} \\
\alpha^{\prime}_{3} &= \frac{1}{N_{\Lambda}} \sum_{\bm k}
s^2_x \big\{h_{\bm k} - ({\bm \xi}_{+}{\bm \xi}_{-}
- 2a_{3,+}a_{3,-} - 2a_{4,+}a_{4,-} \nn \\
&\hspace{0.3cm}
- 2a_{5,+} a_{5,-} ) f_{\bm k} \big\}
- \frac{1}{N_{\Lambda}}\sum_{\bm k} s^2_x \big\{h^0_{\bm k}
- \xi^2 f^0_{\bm k}\big\}, \label{ad3}  \\
\alpha^{\prime}_{4} &= \frac{1}{N_{\Lambda}} \sum_{\bm k}
c^2_y \big\{h_{\bm k} - ({\bm \xi}_{+}{\bm \xi}_{-}
- 2a_{2,+}a_{2,-} \nn \\
&\hspace{0.3cm}
- 2a_{4,+}a_{4,-}) f_{\bm k} \big\}
- \frac{1}{N_{\Lambda}}\sum_{\bm k} s^2_y \big\{h^0_{\bm k}
- \xi^2 f^0_{\bm k}\big\}, \label{ad4}
\end{align}
\begin{align}
\alpha^{\prime\prime}_{1} &= \frac{1}{N_{\Lambda}} \sum_{\bm k}
s^2_x \big\{h_{\bm k} - {\bm \xi}_{+}{\bm \xi}_{-}
f_{\bm k} - h^0_{\bm k}
+ \xi^2 f^0_{\bm k}\big\}, \label{add1}
\end{align}
\begin{align}
\alpha^{\prime\prime}_{2} &= \frac{1}{N_{\Lambda}} \sum_{\bm k}
s^2_y \big\{h_{\bm k} - ({\bm \xi}_{+}{\bm \xi}_{-}
- 2a_{2,+}a_{2,-} - 2a_{3,+}a_{3,-} \nn \\
&\hspace{0.3cm}
- 2a_{5,+} a_{5,-} ) f_{\bm k} \big\}
- \frac{1}{N_{\Lambda}}\sum_{\bm k} s^2_y \big\{h^0_{\bm k}
- \xi^2 f^0_{\bm k}\big\}, \label{add2} \\
\alpha^{\prime\prime}_{3} &= \frac{1}{N_{\Lambda}} \sum_{\bm k}
s^2_y \big\{h_{\bm k} - ({\bm \xi}_{+}{\bm \xi}_{-}
- 2a_{3,+}a_{3,-} - 2a_{4,+}a_{4,-}  \nn \\
&\hspace{0.3cm}
- 2a_{5,+} a_{5,-} ) f_{\bm k} \big\}
- \frac{1}{N_{\Lambda}}\sum_{\bm k} s^2_y \big\{h^0_{\bm k}
- \xi^2 f^0_{\bm k}\big\}, \label{add3} \\
\alpha^{\prime\prime}_{4} &= \frac{1}{N_{\Lambda}} \sum_{\bm k}
c^2_x \big\{h_{\bm k} - ({\bm \xi}_{+}{\bm \xi}_{-}
- 2a_{2,+}a_{2,-} \nn \\
&\hspace{0.6cm}
- 2a_{4,+}a_{4,-}) f_{\bm k} \big\}
- \frac{1}{N_{\Lambda}}\sum_{\bm k} s^2_x \big\{h^0_{\bm k}
- \xi^2 f^0_{\bm k}\big\}, \label{add4}
\end{align}
where $\frac{J^1_1}{4}$ were omitted from the overall factors
in the right hand sides.
\begin{align}
\beta_{1,2} &= - \frac{1}{N_{\Lambda}} \sum_{\bm k}
s_x s_y (a_{3,+} a_{5,-} + a_{5,+} a_{3,-}) f_{\bm k}, \label{b12} \\
\beta_{1,3} &= - \frac{i}{N_{\Lambda}} \sum_{\bm k}
s_x s_y (a_{4,-} g^{+}_{\bm k} -
a_{4,+} g^{-}_{\bm k}) , \label{b13}
\end{align}
\begin{align}
\beta_{1,5} &= \frac{i}{N_{\Lambda}} \sum_{\bm k}
s_x s_y (a_{2,-} g^{+}_{\bm k} -
a_{2,+} g^{-}_{\bm k}) , \label{b15} \\
\beta_{1,7} &=  \frac{i}{N_{\Lambda}} \sum_{\bm k}
s_x c_x (a_{2,+} a_{4,-} - a_{2,-} a_{4,+}) f_{\bm k}, \label{b17}
\end{align}
%
\begin{align}
\beta_{2,4} & = - \frac{i}{N_{\Lambda}} \sum_{\bm k}
s_x s_y (a_{4,-} g^{+}_{\bm k} -
a_{4,+} g^{-}_{\bm k}) , \label{b24} \\
\beta_{2,6} &= \frac{i}{N_{\Lambda}} \sum_{\bm k}
s_x s_y (a_{2,-} g^{+}_{\bm k} -
a_{2,+} g^{-}_{\bm k}) , \label{b26}
\end{align}
\begin{align}
\beta_{2,8} &=  \frac{i}{N_{\Lambda}} \sum_{\bm k}
s_y c_y (a_{2,+} a_{4,-} - a_{2,-} a_{4,+}) f_{\bm k}, \label{b28} \\
%
%
\beta_{3,4} &= \frac{1}{N_{\Lambda}} \sum_{\bm k}
s_x s_y (a_{3,+} a_{5,-} +
a_{3,-} a_{5,+} ) f_{\bm k} , \label{b34}
\end{align}
\begin{align}
\beta_{3,5} &= \frac{1}{N_{\Lambda}} \sum_{\bm k}
s^2_y (a_{2,+} a_{4,-} +
a_{2,-} a_{4,+} ) f_{\bm k} , \label{b35} \\
\beta_{3,7} &= - \frac{1}{N_{\Lambda}} \sum_{\bm k}
c_x s_y (a_{2,+} g^{-}_{\bm k} +
a_{2,-} g^{+}_{\bm k}) , \label{b37}
\end{align}
\begin{align}
\beta_{4,6} &= \frac{1}{N_{\Lambda}} \sum_{\bm k}
s^2_x (a_{2,+} a_{4,-} +
a_{2,-} a_{4,+} ) f_{\bm k} , \label{b46} \\
\beta_{4,8} &= - \frac{1}{N_{\Lambda}} \sum_{\bm k}
s_x c_y (a_{2,+} g^{-}_{\bm k} +
a_{2,-} g^{+}_{\bm k}) , \label{b48}
\end{align}
\begin{align}
\beta_{5,6} &= \frac{1}{N_{\Lambda}} \sum_{\bm k}
s_x s_y (a_{3,+} a_{5,-} +
a_{3,-} a_{5,+} ) f_{\bm k} , \label{b56} \\
\beta_{5,7} &= - \frac{1}{N_{\Lambda}} \sum_{\bm k}
c_x s_y (a_{4,+} g^{-}_{\bm k} +
a_{4,-} g^{+}_{\bm k}) , \label{b57}
\end{align}
\begin{align}
\beta_{6,8} &= - \frac{1}{N_{\Lambda}} \sum_{\bm k}
s_x c_y (a_{4,+} g^{-}_{\bm k} +
a_{4,-} g^{+}_{\bm k}) , \label{b68} \\
\beta_{7,8} &= - \frac{1}{N_{\Lambda}} \sum_{\bm k}
c_x c_y (a_{3,+} a_{5,-} +
a_{3,-} a_{5,+} ) f_{\bm k} , \label{b78}
\end{align}
where $\frac{J^2_1}{4}$ were omitted from the overall factors
in the right hand sides.
\begin{align}
\beta_{1,9} &= -\frac{\sqrt{2}}{N_{\Lambda}}\sum_{\bm k}
s^2_x c_y
(a_{5,-}g^{+}_{\bm k} + a_{5,+}g^{-}_{\bm k} ), \label{b19} \\
\beta_{1,11} &= - \frac{\sqrt{2}}{N_{\Lambda}}\sum_{\bm k}
s_x s_y c_x (a_{5,-}g^{+}_{\bm k} + a_{5,+}g^{-}_{\bm k} ),
\label{b111}
\end{align}
\begin{align}
\beta_{2,9} &=  \frac{\sqrt{2}}{N_{\Lambda}}\sum_{\bm k}
s_y s_x c_y (a_{3,-}g^{+}_{\bm k} + a_{3,+}g^{-}_{\bm k} ),
\label{b29} \\
\beta_{2,11} &=  \frac{\sqrt{2}}{N_{\Lambda}}\sum_{\bm k}
s_y s_y c_x (a_{3,-}g^{+}_{\bm k} + a_{3,+}g^{-}_{\bm k} ),
\label{b211}
\end{align}
\begin{align}
\beta_{3,9} &=  \frac{\sqrt{2}i}{N_{\Lambda}}\sum_{\bm k}
s_y s_x c_y (a_{4,+}a_{5,-} - a_{5,+}a_{4,-} ) f_{\bm k},
\label{b39} \\
\beta_{3,11} &=  \frac{\sqrt{2}i}{N_{\Lambda}}\sum_{\bm k}
s_y s_y c_x (a_{4,+}a_{5,-} - a_{5,+}a_{4,-}) f_{\bm k},
\label{b311}
\end{align}
\begin{align}
\beta_{3,10} &=  - \frac{\sqrt{2}}{N_{\Lambda}}\sum_{\bm k}
s_y c_x c_y (a_{2,+}a_{5,-} + a_{5,+}a_{2,-}) f_{\bm k},
\label{b310} \\
\beta_{3,12} &=  \frac{\sqrt{2}}{N_{\Lambda}}\sum_{\bm k}
s_y s_x s_y (a_{2,+}a_{5,-} + a_{5,+}a_{2,-} )f_{\bm k},
\label{b312}
\end{align}
\begin{align}
\beta_{4,9} &= \frac{\sqrt{2}i}{N_{\Lambda}}\sum_{\bm k}
s_x s_x c_y (a_{3,+}a_{4,-} - a_{4,+}a_{3,-} ) f_{\bm k},
\label{b49} \\
\beta_{4,11} &=  \frac{\sqrt{2}i}{N_{\Lambda}}\sum_{\bm k}
s_x s_y c_x (a_{3,+}a_{4,-} - a_{4,+}a_{3,-} )f_{\bm k},\label{b411}
\end{align}
\begin{align}
\beta_{4,10} &=  \frac{\sqrt{2}}{N_{\Lambda}}\sum_{\bm k}
s_x c_x c_y (a_{2,+}a_{3,-} + a_{3,+}a_{2,-} )f_{\bm k},
\label{b410} \\
\beta_{4,12} &=  -\frac{\sqrt{2}}{N_{\Lambda}}\sum_{\bm k}
s_x s_x s_y (a_{2,+}a_{3,-} + a_{3,+}a_{2,-} ) f_{\bm k},
\label{b412}
\end{align}
\begin{align}
\beta_{5,9} &=  - \frac{\sqrt{2}i}{N_{\Lambda}}\sum_{\bm k}
s_y s_x c_y (a_{2,+}a_{5,-} - a_{5,+}a_{2,-} ) f_{\bm k},
\label{b59} \\
\beta_{5,11} &= - \frac{\sqrt{2}i}{N_{\Lambda}}\sum_{\bm k}
s_y c_x s_y (a_{2,+}a_{5,-} - a_{5,+}a_{2,-}) f_{\bm k},
\label{b511}
\end{align}
\begin{align}
\beta_{5,10} &=  - \frac{\sqrt{2}}{N_{\Lambda}}\sum_{\bm k}
s_y c_x c_y ( a_{4,+}a_{5,-} + a_{5,+}a_{4,-}) f_{\bm k},
\label{b510} \\
\beta_{5,12} &= \frac{\sqrt{2}}{N_{\Lambda}}\sum_{\bm k}
s_y s_x s_y (a_{4,+}a_{5,-} + a_{5,+}a_{4,-} ) f_{\bm k},
\label{b512}
\end{align}
\begin{align}
\beta_{6,9} &= \frac{\sqrt{2}i}{N_{\Lambda}}\sum_{\bm k}
s_x s_x c_y (a_{2,+}a_{3,-} - a_{3,+}a_{2,-}) f_{\bm k},
\label{b69} \\
\beta_{6,11} &=  \frac{\sqrt{2}i}{N_{\Lambda}}\sum_{\bm k}
s_x c_x s_y (a_{2,+}a_{3,-} - a_{3,+}a_{2,-} ) f_{\bm k},
\label{b611}
\end{align}
\begin{align}
\beta_{6,10} &=  \frac{\sqrt{2}}{N_{\Lambda}}\sum_{\bm k}
s_x c_x c_y (a_{4,+}a_{3,-} + a_{3,+}a_{4,-}) f_{\bm k},
\label{b610} \\
\beta_{6,12} &= - \frac{\sqrt{2}}{N_{\Lambda}}\sum_{\bm k}
s_x s_x s_y (a_{4,+}a_{3,-} + a_{3,+}a_{4,-}) f_{\bm k},
\label{b612}
\end{align}
\begin{align}
\beta_{7,10} &= \frac{\sqrt{2}}{N_{\Lambda}}\sum_{\bm k}
c_x c_x c_y (a_{5,-}g^{+}_{\bm k} + a_{5,+}g^{-}_{\bm k} ),
\label{b710} \\
\beta_{7,12} &= -\frac{\sqrt{2}}{N_{\Lambda}}\sum_{\bm k}
c_x s_x s_y (a_{5,-}g^{+}_{\bm k} + a_{5,+}g^{-}_{\bm k}),
\label{b712}
\end{align}
\begin{align}
\beta_{8,10} &= -\frac{\sqrt{2}}{N_{\Lambda}}\sum_{\bm k}
c_y c_x c_y (a_{3,-}g^{+}_{\bm k} + a_{3,+}g^{-}_{\bm k} ),
\label{b810} \\
\beta_{8,12} &= \frac{\sqrt{2}}{N_{\Lambda}}\sum_{\bm k}
c_y s_x s_y (a_{3,-}g^{+}_{\bm k} + a_{3,+}g^{-}_{\bm k} ),
\label{b812}
\end{align}
where $\frac{J_1J_2}{4}$ were omitted from the overall factors
in the right hand sides.
\begin{align}
\beta_{9,11} &= \frac{1}{N_{\Lambda}} \sum_{\bm k}
2s_xs_yc_xc_y \big\{ \!\ h_{\bm k} + {\bm \xi}_{+}
{\bm \xi}_{-} f_{\bm k} \!\ \big\}, \label{b911} \\
\beta_{10,12} &= - \frac{1}{N_{\Lambda}} \sum_{\bm k}
2s_x s_yc_xc_y \big\{ h_{\bm k} - ({\bm \xi}_{+}{\bm \xi}_{-}
\nn \\
& \hspace{1.5cm}
- 2 a_{3,+}a_{3,-} - 2 a_{5,+}a_{5,-}) f_{\bm k} \big\},
\label{b1012}
\end{align}
\begin{align}
\beta_{9,10} &= \frac{2i}{N_{\Lambda}} \sum_{\bm k}
\!\ s_x c_x c^2_y ( a_{2,+} a_{4,-}
-a_{4,+}a_{2,-} ) f_{\bm k},  \label{b910} \\
\beta_{9,12} &= -\frac{2i}{N_{\Lambda}} \sum_{\bm k}
\!\ s^2_x s_y c_y ( a_{2,+} a_{4,-}
-a_{4,+}a_{2,-} ) f_{\bm k}, \label{b912}
\end{align}
\begin{align}
\beta_{10,11} &= - \frac{2i}{N_{\Lambda}} \sum_{\bm k}
\!\ c^2_x s_yc_y ( a_{2,+} a_{4,-}
-a_{4,+}a_{2,-} ) f_{\bm k}, \label{b1011} \\
\beta_{11,12} &=  - \frac{2i}{N_{\Lambda}} \sum_{\bm k}
\!\ s_x c_x s^2_y ( a_{2,+} a_{4,-}
-a_{4,+}a_{2,-} ) f_{\bm k}, \label{b1112}
\end{align}
where $\frac{J^2_2}{4}$ were omitted from the overall factors
in the right hand sides.
\begin{align}
\beta^{\prime}_{1,2}&= -\frac{i}{N_{\Lambda}} \sum_{\bm k}
s_x s_y (a_{4,-}g^{+}_{\bm k} - a_{4,+}g^{-}_{\bm k} ),
\label{bd12} \\
\beta^{\prime}_{1,3}&= \frac{i}{N_{\Lambda}} \sum_{\bm k} s_x s_y
(a_{2,-}g^{+}_{\bm k} - a_{2,+}g^{-}_{\bm k} ), \label{bd13}
\end{align}
\begin{align}
\beta^{\prime}_{1,4}&= \frac{i}{N_{\Lambda}} \sum_{\bm k} s_y c_y
(a_{2,+}a_{4,-} - a_{4,+}a_{2,-} ) f_{\bm k} , \label{bd14}  \\
\beta^{\prime}_{2,3} &= \frac{1}{N_{\Lambda}} \sum_{\bm k} s^2_x
(a_{2,+}a_{4,-} + a_{4,+}a_{2,-}) f_{\bm k}, \label{bd23}
\end{align}
\begin{align}
\beta^{\prime}_{2,4}&= -\frac{1}{N_{\Lambda}} \sum_{\bm k}
s_x c_y (a_{2,+}g^{-}_{\bm k} + a_{2,-}g^{+}_{\bm k} ),
\label{bd24} \\
\beta^{\prime}_{3,4}&= -\frac{1}{N_{\Lambda}} \sum_{\bm k} s_x c_y
(a_{4,+}g^{-}_{\bm k} + a_{4,-}g^{+}_{\bm k} ), \label{bd34}
\end{align}
\begin{align}
\beta^{\prime\prime}_{1,2}&= -\frac{i}{N_{\Lambda}} \sum_{\bm k}
s_x s_y (a_{4,-}g^{+}_{\bm k} - a_{4,+}g^{-}_{\bm k} ),
\label{bdd12} \\
\beta^{\prime\prime}_{1,3} &= \frac{i}{N_{\Lambda}} \sum_{\bm k} s_x s_y
(a_{2,-}g^{+}_{\bm k} - a_{2,+}g^{-}_{\bm k} ), \label{bdd13}
\end{align}
\begin{align}
\beta^{\prime\prime}_{1,4}&= \frac{i}{N_{\Lambda}} \sum_{\bm k} s_x c_x
(a_{2,+}a_{4,-} - a_{4,+}a_{2,-}) f_{\bm k} , \label{bdd14} \\
\beta^{\prime\prime}_{2,3} &= \frac{1}{N_{\Lambda}} \sum_{\bm k}
s^2_y  (a_{2,+}a_{4,-} + a_{4,+}a_{2,-}) f_{\bm k}, \label{bdd23}
\end{align}
\begin{align}
\beta^{\prime\prime}_{2,4}&= -\frac{1}{N_{\Lambda}} \sum_{\bm k}
s_y c_x (a_{2,+}g^{-}_{\bm k} + a_{2,-}g^{+}_{\bm k} ),
\label{bdd24} \\
\beta^{\prime\prime}_{3,4} &= -\frac{1}{N_{\Lambda}}
\sum_{\bm k} s_y c_x
(a_{4,+}g^{-}_{\bm k} + a_{4,-}g^{+}_{\bm k} ), \label{bdd34}
\end{align}
where $\frac{J^2_1}{4}$ were omitted from the overall factors
in the right hand sides. In all of these equations, symbols are defined as
$s_{\mu} \equiv \sin \big(k_{\mu}\big),
c_{\mu} \equiv \cos \big(k_{\mu}\big)$ with
$\mu=x,y$, and
\begin{align}
a_{2,\pm} &\equiv J_2 \eta
\sin \Big(k_x \pm \frac{q_{x}}{2}\Big)
\sin \Big(k_y \pm \frac{q_{y}}{2}\Big), \label{a2pm} \\
a_{3,\pm} & \equiv \frac{J_1 D}{2} \sin
\Big(k_x \pm \frac{q_{x}}{2}\Big), \label{a3pm} \\
a_{4,\pm} & \equiv J_2 \chi \cos\Big(k_x \pm \frac{q_{x}}{2}\Big)
\cos \Big(k_y \pm \frac{q_{y}}{2}\Big), \label{a4pm} \\
a_{5,\pm} & \equiv - \frac{J_1 D}{2}
\sin \Big(k_y \pm \frac{q_{y}}{2}\Big), \label{a5pm}
\end{align}
\begin{align}
a_{2} &\equiv J_2 \eta
\sin k_x \sin k_y  \ \ \
a_{3}  \equiv \frac{J_1 D}{2} \sin k_x , \label{a23} \\
a_{4} & \equiv J_2 \chi \cos k_x
\cos k_y , \ \ \
a_{5}  \equiv
- \frac{J_1 D}{2} \sin k_y , \label{a45}
\end{align}
\begin{align}
f_{\bm k} &\equiv \frac{1}{2}\frac{\xi_{+} + \xi_{-}}{\xi_{+} \xi_{-}}
\frac{1}{\epsilon^2_n + (\xi_{+} + \xi_{-})^2}, \label{fk} \\
g^{\pm}_{\bm k} &\equiv \pm \frac{1}{2\xi_{\mp}}
\frac{i\epsilon_n}{\epsilon^2_n+(\xi_{+} + \xi_{-})^2}, \label{gk} \\
h_{\bm k} &\equiv -\frac{1}{2}
\frac{\xi_{+} + \xi_{-}}{\epsilon^2_n+(\xi_{+} + \xi_{-})^2},
\label{hk} \\
f^0_{\bm k} &\equiv \frac{1}{4\xi^3}, \ \ h^0_{\bm k} \equiv
-\frac{1}{4\xi} \label{fk0}
\end{align}
with
\begin{align}
&{\bm \xi}_{+} {\bm \xi}_{-}
\equiv \sum^5_{j=2} a_{j,+}a_{j,-}, \nn\\
&\xi_{\pm} \equiv \sqrt{\sum^5_{j=2} a^2_{j,\pm}}, \ \ \
\xi \equiv \sqrt{\sum^5_{j=2} a^2_j}. \label{xi}
\end{align}

The vertex parts $\overline{\cal S}^{(1,1)}_{\mu;\mu}({\bm q},i\epsilon_n)$
($\mu=1,2,3$), defined in Eq.~(\ref{d1}),
are calculated as follows:
\begin{widetext}
\begin{eqnarray}
\overline{\cal S}^{(1,1)}_{3;3} \equiv
\bordermatrix{
& {\bm e}^3_{1} & {\bm e}^3_{2} &
{\bm e}^3_{3} & {\bm e}^3_{4} &
{\bm e}^3_{5} & {\bm e}^3_{6} &
{\bm e}^3_{7} & {\bm e}^3_{8} &
{\bm e}^3_{9} & {\bm e}^3_{10} &
{\bm e}^3_{11} & {\bm e}^3_{12} \cr
& \gamma_{3;1} & \gamma_{3;2} & \gamma_{3;3} & \gamma_{3;4}
& \gamma_{3;5} & \gamma_{3;6}
& 0 & 0 & \gamma_{3;9} & 0 & \gamma_{3;11} & 0 \cr }, \label{b0-1}
\end{eqnarray}
\end{widetext}
\begin{align}
\overline{\cal S}^{(1,1)}_{1;1}  &\equiv
\bordermatrix{
& {\bm e}^1_{1} & {\bm e}^1_{2} &
{\bm e}^1_{3} & {\bm e}^1_{4} \cr
& \gamma_{1;1} & \gamma_{1;2} & \gamma_{1;3} & 0 \cr }, \label{Sg11} \\
\overline{\cal S}^{(1,1)}_{2;2} &\equiv
\bordermatrix{
& {\bm e}^2_{1} & {\bm e}^2_{2} &
{\bm e}^2_{3} & {\bm e}^2_{4} \cr
& \gamma_{2;1} & \gamma_{2;2}
& \gamma_{2;3} & 0 \cr }, \label{Sg22}
\end{align}
where the coefficients take the forms
\begin{align}
\gamma_{3;1} &\equiv \frac{i}{N_{\Lambda}}
\sum_{\bm k} s_x (a_{3,+} g^{-}_{\bm k}
- a_{3,-} g^{+}_{\bm k}), \label{g31} \\
\gamma_{3;2} &\equiv \frac{i}{N_{\Lambda}}
\sum_{\bm k} s_y (a_{5,+} g^{-}_{\bm k}
- a_{5,-} g^{+}_{\bm k}), \label{g32}
\end{align}
\begin{align}
\gamma_{3;3} &\equiv \frac{1}{N_{\Lambda}}
\sum_{\bm k} s_y (a_{3,+} a_{4,-}
+ a_{3,-} a_{4,+}) f_{\bm k},  \label{g33} \\
\gamma_{3;4} &\equiv \frac{1}{N_{\Lambda}}
\sum_{\bm k} s_x (a_{5,+} a_{4,-}
+ a_{5,-} a_{4,+}) f_{\bm k},  \label{g34}
\end{align}
\begin{align}
\gamma_{3;5} &\equiv -\frac{1}{N_{\Lambda}}
\sum_{\bm k} s_y (a_{2,+} a_{3,-}
+ a_{3,-} a_{2,+}) f_{\bm k},  \label{g35} \\
\gamma_{3;6} &\equiv - \frac{1}{N_{\Lambda}}
\sum_{\bm k} s_x (a_{2,+} a_{5,-}
+ a_{2,-} a_{5,+}) f_{\bm k},  \label{g36}
\end{align}
where $\frac{J_1}{4}$ were omitted from the overall factors
in the right hand sides.
\begin{align}
\gamma_{3;9} &\equiv -
\frac{\sqrt{2} \!\ i}{N_{\Lambda}}  \sum_{\bm k}
s_x c_y (a_{3,+} a_{5,-} -
a_{3,-} a_{5,+}) f_{\bm k}, \label{g39} \\
\gamma_{3;11} &\equiv -
\frac{\sqrt{2} \!\ i}{N_{\Lambda}}  \sum_{\bm k}
c_x s_y (a_{3,+} a_{5,-} - a_{3,-} a_{5,+})
f_{\bm k}, \label{g311}
\end{align}
where $\frac{J_2}{4}$ were omitted from the overall factors
in the right hand sides.
\begin{align}
\gamma_{1;1} &\equiv -\frac{i}{N_{\Lambda}}
\sum_{\bm k} s_y (a_{5,+} g^{-}_{\bm k}
- a_{5,-} g^{+}_{\bm k}), \label{g11} \\
\gamma_{1;2} &\equiv -\frac{1}{N_{\Lambda}}
\sum_{\bm k} s_x (a_{4,+} a_{5,-}
+ a_{4,-} a_{5,+}) f_{\bm k}, \label{g12} \\
\gamma_{1;3} &\equiv \frac{1}{N_{\Lambda}}
\sum_{\bm k} s_x (a_{2,+} a_{5,-}
+ a_{2,-} a_{5,+}) f_{\bm k},  \label{g13}
\end{align}
where $\frac{J_1}{4}$ were omitted from the overall factors
in the right hand sides.
\begin{align}
\gamma_{2;1} &\equiv -\frac{i}{N_{\Lambda}}
\sum_{\bm k} s_x (a_{3,+} g^{-}_{\bm k}
- a_{3,-} g^{+}_{\bm k}), \label{g21} \\
\gamma_{2;2} &\equiv -\frac{1}{N_{\Lambda}}
\sum_{\bm k} s_y (a_{4,+} a_{3,-}
+ a_{4,-} a_{3,+}) f_{\bm k}, \label{g22} \\
\gamma_{2;3} &\equiv \frac{1}{N_{\Lambda}}
\sum_{\bm k} s_y (a_{2,+} a_{3,-}
+ a_{2,-} a_{3,+}) f_{\bm k}.  \label{g23}
\end{align}
where $\frac{J_1}{4}$ were omitted from the overall factors
in the right hand sides.

\section{Mass of low-energy modes in ${\rm Im}\chi_{zz}({\bm q},\epsilon)$
at ${\bm q}=(\pi,0)$ and $(0,\pi)$}
Low-energy excitations in the $Z_2$ planar phase
consist of not only spin-wave (director-wave)
modes at ${\bm q}=(0,0)$ but also another gapless
mode in ${\rm Im}\chi_{zz}({\bm q},\epsilon)$ at ${\bm q}=(\pi,0)$, i.e.
${\bm e}^3_{8}$-mode. Inside the $U(1)$
planar phase, the mass of the latter mode
becomes even negative, indicating an instability.
A direct evaluation of the PRA propagator
at ${\bm q}=(\pi,0)$ and $i\epsilon_n=0$
suggests that ${\bm e}^3_{8}$-mode becomes
decoupled from others and its
mass is given by $\alpha_8$;
\begin{eqnarray}
\Big[\overline{\cal S}^{(0,2)}_{33} \Big]_{{\bm q}=(\pi,0),\epsilon=0} \equiv
\bordermatrix{
& \cdots & {\bm e}^3_{8} & \cdots & \cr
\vdots & \ddots & 0  & \ddots & \cr
{\bm e}^3_8 & 0 &\alpha_8 & 0 & \cr
\vdots & \ddots & 0 & \ddots &  \cr
}   \label{2a0-1}
\end{eqnarray}
at ${\bm q}=(\pi,0)$ and $i\epsilon_n=0$, where $\alpha_8$ reads
\begin{align}
\alpha_8 &= - \frac{1}{N_{\Lambda}}
\sum_{\bm k} \Big(\frac{c^2_y}{\xi_{+}} - \frac{s^2_y}{\xi}\Big)
=- \frac{1}{N_{\Lambda}}
\sum_{\bm k} \frac{c^2_y-s^2_y}{\xi},   \label{alpha8} \\
\xi^2_{+} &= J^2_2 \eta^2 c^2_x s^2_y + J^2_2 \chi^2 s^2_x c^2_y
+ \frac{J^2_1 D^2}{4} (c^2_x + s^2_y), \nn \\
\xi^2 &= J^2_2 \eta^2 x^2_x s^2_y + J^2_2 \chi^2 c^2_x c^2_y
+ \frac{J^2_1 D^2}{4} (s^2_x + s^2_y). \nn
\end{align}
One can see that $\alpha_8$ given by Eq.~(\ref{alpha8}) is reduced
to zero in the
$Z_2$ planar phase, by noting that the mean-field
gap equation for the pariring fields, $D$,
$\chi$ and $\eta$, is given by the following
coupled equations
\begin{align}
J_1 D &= \frac{1}{N_{\Lambda}} \sum_{\bm k}
\frac{\partial \xi}{\partial D} \tanh\Big(\frac{\beta\xi}{2}\Big)
= \frac{1}{N_{\Lambda}} \frac{J^2_1 D}{4}
\sum_{\bm k} \frac{s^2_x + s^2_y}{\xi}, \nn \\
J_2 \chi &=  \frac{1}{N_{\Lambda}} \sum_{\bm k}
\frac{\partial \xi}{\partial \chi} \tanh\Big(\frac{\beta\xi}{2}\Big)
= \frac{1}{N_{\Lambda}} J^2_2 \chi\sum_{\bm k}
\frac{c^2_x c^2_y}{\xi}, \nn \\
J_2 \eta &=  \frac{1}{N_{\Lambda}} \sum_{\bm k}
\frac{\partial \xi}{\partial \eta} \tanh\Big(\frac{\beta\xi}{2}\Big)
= \frac{1}{N_{\Lambda}} J^2_2 \eta \sum_{\bm k}
\frac{s^2_x s^2_y}{\xi} \nn
\end{align}
at $\beta^{-1}=0$. Namely, in the $Z_2$ phase
 ($\eta\ne 0$ and $\chi \ne 0$), this gap equation leads to
\begin{eqnarray}
J^{-1}_2 = \frac{1}{N_{\Lambda}} \sum_{\bm k} \frac{c^2_x c^2_y}{\xi}, \ \ \
J^{-1}_2 =  \frac{1}{N_{\Lambda}} \sum_{\bm k} \frac{s^2_x s^2_y}{\xi}.
\label{gap-eq2}
\end{eqnarray}
Or,
\begin{align}
0 &= \frac{1}{N_{\Lambda}} \sum_{\bm k}
\frac{c^2_xc^2_y - s^2_x s^2_y}{\xi} \nn \\
&=  \frac{1}{N_{\Lambda}} \sum_{\bm k}
\frac{c^2_y - s^2_y - (s^2_x c^2_y - c^2_x s^2_y) }{\xi}. \nn
\end{align}
Since $\xi$ is symmetric under the exchange between
$k_x$ and $k_y$, the right hand side leads to
$\alpha_8=0$. In the $U(1)$
planar phase ($\eta=0$ and $\chi\ne 0$),  the gap
equation leads to
\begin{eqnarray}
J^{-1}_2 = \frac{1}{N_{\Lambda}} \sum_{\bm k} \frac{c^2_x c^2_y}{\xi}, \ \ \
J^{-1}_2 >  \frac{1}{N_{\Lambda}} \sum_{\bm k} \frac{s^2_x s^2_y}{\xi}.
\label{gap-eq3}
\end{eqnarray}
instead of Eqs.~(\ref{gap-eq2}).
This dictates that the mass of the ${\bm e}^3_8$ mode
becomes negative, $\alpha_8<0$. Similarly, one
can see from Eqs.~(\ref{a12}), (\ref{ad2}), and (\ref{add2})
that the mass of ${\bm e}^{1}_{2}$, ${\bm e}^{2}_{2}$ and
${\bm e}^3_{12}$ modes at ${\bm q}=(\pi,\pi)$ also
become negative in the $U(1)$ planar phase.


\begin{thebibliography}{99}
\bibitem{BW} This state was refered to as a two-dimensional
anologue of the Balian-Werthamer state\cite{bw1}
in Ref.\ \onlinecite{sm},
but it is usually called the planar state in the context of superfluid $^3$He.
We adopt the name ``planar state" in this paper.
\bibitem{comment2} The energy-momentum
dispersion for gapless spin-wave modes in a quasi-two-dimensional
system generally has a weak dispersion in the third
direction. The effective spatial dimension
`$d$' should be taken to be $2$, when the temperature is
sufficiently larger than a characteristic energy scale
associated with this weak dispersion, whereas $d$ should
be $3$, when the temperature is in the
same order of this energy scale.
\end{thebibliography}
\end{document}